\pdfoutput=1
\documentclass[12pt]{article}
\usepackage{xspace}
\usepackage{cite}
\usepackage{amsfonts}

\usepackage{comment}
\usepackage{gensymb}
\usepackage{booktabs}
\usepackage{placeins}
\usepackage[unicode=true,pdfusetitle,
 bookmarks=true,bookmarksnumbered=false,bookmarksopen=false,
 breaklinks=false,pdfborder={0 0 1},backref=false,colorlinks=false]
 {hyperref}
 
\usepackage[pdftex]{graphicx}
\usepackage{booktabs}
\usepackage[table,xcdraw]{xcolor}
\usepackage{multirow}
\usepackage{rotating,tabularx}
\usepackage{amsmath}
\usepackage[]{hepparticles}

\usepackage{lineno}
\usepackage[normalem]{ulem} 

\usepackage{color}

\newcommand{\bea}{\begin{eqnarray}}
\newcommand{\eea}{\end{eqnarray}}

\newcommand{\pythia}{\textsc{Pythia}\xspace}
\newcommand{\sherpa}{\textsc{Sherpa}\xspace}
\newcommand{\epos}{\textsc{Epos}\xspace}
\newcommand{\urqmd}{\textsc{U}r\textsc{QMD}\xspace}
\newcommand{\smashm}{\textsc{SMASH}\xspace}

\newcommand{\sqrts}{$\sqrt{s}$\xspace}
\newcommand{\cmse}{$\sqrt{s}$\xspace}
\newcommand{\cmsenn}{$\sqrt{s_{\rm NN}}$\xspace}

\newcommand{\pt}{\ensuremath{p_{\mathrm T}}\xspace}

\newcommand{\ulumi}{$cm^{-2} s^{-1}$\xspace}

\newcommand {\diff}{\mathrm{d}}

\begin{document}

\title{Performance of Monte Carlo event generators of pp collisions at NICA energies \date{}}
\author{Maxim Azarkin$^1$, Martin Kirakosyan$^1$
\\\\
{\it \small $^1$ P.N. Lebedev Physics Institute, }\\ {\small\it Moscow 119991, Russia}
}
\maketitle

\begin{abstract}

This paper presents an overview of Monte Carlo(MC) event generators for simulation of proton-proton 
collisions along with the results on hadron production at Nuclotron-based Ion Collider fAcility (NICA)
energies. Namely, mean multiplicities, mean transverse momenta, and rapidity distributions of p($\bar{\rm p}$), 
 $\pi^{\pm}$, $\mathrm{K}^{\pm}$ at different collision energies are presented. We also study two-particle angular
correlations for stable charged particles. Results of simulations with \pythia, \epos, and \urqmd 
event generators are compared to available data. Connections of studied quantities with physics mechanisms 
in MC generators are discussed.
We suggest a tuned set of parameters to address observed discrepancies between data and \pythia.

\end{abstract} 


\section{Introduction}
\label{Sec:intro}

The Nuclotron-based Ion Collider fAcility (NICA)~\cite{Kekelidze:2016hhw} is an accelerator complex at JINR
(Dubna) and is designed to operate at $4 \leq \sqrt{s} \leq 25$~GeV in proton-proton (pp) collision mode
at luminosities up to $L=10^{32}$ \ulumi and at $4\leq \sqrt{s_{\rm NN}} \leq 11$~GeV in heavy-ion collision
mode for luminosities up to $L = 10^{27}$ \ulumi.  The main objectives of NICA physics program
are experimental studies of a hot dense strongly interacting baryonic matter produced in heavy-ion collisions 
and spin physics, using polarized protons and light ions. Multi Purpose Detector MPD~\cite{Golovatyuk:2016zps}
is a primary tool to study the first part of NICA physics program while the second part is the main focus of 
Spin Physics Detector (SPD)~\cite{SPD_URL}. The complex is expected to start an operation for heavy-ion
programm in 2022~\cite{NICA_Status}.

One of the primary goals of NICA physics program are the studies of strongly-interacting dense matter formed 
in heavy ion (Au) collisions at \cmsenn in the range $4-11$ GeV.  These energies are of particular interest because
maximum baryon density at freeze-out is expected to be reached \cite{NICAwhitepaper, NICAbaryondens1} and
critical point (possibly even a triple point) of a QCD phase transition is expected to be observed \cite{NICAwhitepaper,QCDphase1}.
Needless to say, the physics of heavy ion collisions is a quite complex subject. For that reason the  Monte Carlo (MC) simulation of heavy ion collisions at these energies will facilitate the better understanding of physics of strong interaction and will help to untangle the knot of different physical mechanisms at play.
The proper simulation of heavy-ion physics is impossible without adequate understanding of proton-proton collisions. 
Characteristics of pp collisions are used to provide a reference for AA collisions allowing to distinguish effects
of phase transition of hot and dense baryonic matter. In this article we focus on physical mechanisms of the pp collisions 
and analyze the performances of different up-to-date event generators for simulation of pp collisions at NICA energies. 
Results of simulations are compared with the available experimental data. In particular, we use data of pp collisions at NICA energies, 
that were mostly obtained on experimental facilities from 60s-70s (see Sec.~\ref{Sec:Results}), ISR collider \cite{ISR1,ISR2}, and NA49 \cite{NA49}, NA61/SHINE experiments \cite{Aduszkiewicz:2017sei}
at SPS and data from two-particle correlations at NA61/SHINE. Nice compilation of some data on identified-particle yields is presented in~\cite{Kolesnikov:2019lrb}. 
One should stress that early papers did not provide full information on the uncertainties and phase space definitions for particle yields measured in various experiments and at different \cmse, some results even being inconsistent.
Thus, MC models of inelastic pp collisions at these energies have been suffering from that and are expected to give 
poor quantitative description of the forthcoming data. 

Nowadays typical MC generator for simulations of high-energy hadron collisions is a mixture of phenomenological models interleaved with theoretical calculations taking advantage of  asymptotic freedom of strong interaction, i.e. making use of perturbative QCD calculations (pQCD). 
Most phenomenological layers in the generator are constructed with limited amount of experimental data and are subjects of continuous studies with subsequent improvements.
In context of this paper the most relevant phenomenological parts are intial-state condition of colliding hadrons, string excitations and fragmentation, low-mass diffraction.  

At lower energies pQCD machinery breaks down since non-perturbative contribution from interaction through
hadronic resonances to pp total cross-section becomes non-negligible. 
Up to \mbox{\cmse $\sim$ $5$ GeV} these contributions are significant in total cross-section. 
In the energy range of $5-10$ GeV resonances vanish and pQCD processes gradually come into play 
in proton-proton collisions \cite{TotalCSbook}, however non-perturbative nature of strong interaction
still have significant contribution to a particle production. For that reason the study of generator 
performance at these energies appears to be crucial
for developing a phenomenological understanding of the physics of strong interaction.  
We will study mean multiplicities, mean transverse momentum, and rapidity densities and for pion, kaons, and protons. 
Two-particle correlations for all stable charged particles are of interest as well.
Following MC generators will be compared  with data: \pythia~\cite{Sjostrand:2006za, Sjostrand:2014zea}, \epos~\cite{epos1,epos2}, \urqmd~\cite{urqmd1,urqmd2}. 

The paper is organized as follows. Sec. \ref{Sec:MCmodels} contains a short review of the Monte Carlo generators that are being examined. 
A compilation of the experimental data on pp collisions at NICA energies and comparisons with MC models are presented in Sec. \ref{Sec:Results}, set of parameters for \pythia that give a better correspondence to data at 17.3 GeV are suggested. In Sec.~\ref{Sec:Summary} we summarize our finding on the 
data-MC comparison and features of particle production with the studied MC event generators.

\clearpage 
\section{Monte Carlo models}
\label{Sec:MCmodels}

The modern MC event generators for simulation of particle collisions in collider experiments have been developing for several decades and encompass numerous physics mechanisms to describe huge amount of data. Nevertheless, up-to-date and most complete models are tuned to describe quite high-energy data. 
Our intention here is to set out discussion of the simulation of pp collision at NICA energies.
We will examine the performance of MC models listed in Sec.~\ref{Sec:intro}, 
in particular their recent versions: \pythia~8.3, \epos~3.4, \urqmd~3.4, \smashm~2.0.1.
\pythia~8.306, \epos~3.4, \urqmd~3.4, and \smashm~2.0.1, can be used to simulate both pp and nucleus collisions (in some extent).
\pythia is a general-purpose event generators for simulation of wide range of physics processes 
in $pp$ collisions on both, soft and hard scales, while others are specialized on soft QCD processes. 
Another widely used general-purpose event generators are {\sc HERWIG}~\cite{Bahr:2008pv, Bellm:2015jjp} and \sherpa~\cite{Sherpawww}. 
However, they are not designed to run (there are no extrapolations of some key mechanisms) at NICA energies, resulting either in a crash or off-scale results. Therefore, we put aside these event generators for now.

\subsection{\pythia}
\label{Sec:MCmodels:pythia}

\pythia is a general-purpose MC event generator \cite{Sjostrand:2006za, Sjostrand:2014zea} and is extensively compared 
with the experimental data over the past three decades. Being a mixture of perturabtive and non-pertubative mechanisms, \pythia model is designated to describe a wide range of characteristics of pp collisions. Non-perturbative mechanisms
are phenomenological approximations to experimental data, while perturbative ones are computed using the corresponding mathematical approach.
\pythia is one of the most widely used generator in high energy physics and its components and physic's approaches 
(e. g. string fragmentation, approach to the total cross-section, elastic scattering and diffraction) are borrowed by other generators, therefore the description of \pythia physics is more detailed in this section compared to other generators.
Below, we describe models and mechanisms that are most relevant for minimum-bias physics in proton collisions at NICA energies and therefore are the most relevant for observables studied in this paper. In Sec.~\ref{Sec:Results}, we will try to re-evaluate free parameters of these mechanisms to achieve better agreement with data. 

In \pythia physics of total cross-section, diffractive and elastic scattering (TED, in short) is treated independently from the physics of inelastic cross-sections. 
Three approaches of TED physics are implemented in the latest version (8.3) of generator ~\cite{Bierlich:2022pfr}, 
namely Shuler-Sjostrand modification (SaS/DL) \cite{Schuler:1993wr} of Donachie-Landshoff model \cite{Donnachie:1992ny} (DL), 
modified (extended) \cite{Rasmussen:2018dgo} ABMST model \cite{Appleby:2016ask} and the Minimum Bias Rockefeller (MBR)~\cite{Rasmussen:2018dgo,MBR} parameterization. 
We will discuss SaS and ABMST models in this section. SaS is the standard one in \pythia while ABMST is an alternative, but as it will be seen, gives a better description of data at NICA energies (see Sec.\ref{Sec:Results}).
The MBR is qualitatively similar to SaS.

SaS model is based on a pomeron and reggion exchanges picture. In this approach, total cross-section is modeled by
a single cut pomeron and reggion exchange, which results in DL parametrization of a total cross-section:
\begin{equation}   
 \sigma_{tot} = X^{AB} s^{\epsilon} + Y^{AB} s^{-\eta}
\end{equation}
with $\epsilon = 0.0808$ and $\eta = 0.4525$, $X^{AB} = 21.7 $ and $Y^{AB} = 56.08$ for pp collisions.
While, elastic cross-section, single diffraction, double diffraction and central diffraction are represented by 2-pomeron,
3-pomerons and 4-pomeron diagrams respectively, giving a simple representation for differential cross-sections 
$\dfrac{d \sigma}{dt \prod_{i} d M_{X_i}}$ (number of $M_{X_i}$ depends on process type, i. e. 0 means for elastic scattering, $1$ means for single diffraction, $2$ means for double diffraction), see \cite{Rasmussen:2018dgo} and \cite{Schuler:1993wr} for details.
Central diffraction implementation is poorly tuned and switched off by default. While SaS model adequately describes diffraction at Tevatron energies, it significantly underestimates total elastic cross-section at LHC (not even taking into account its fundamental inability to predict differential elastic cross-section). And as it will be seen drawbacks of the model in description of high-rapidity gap events is visible at low energies as well.
 
ABMST model \cite{Appleby:2016ask} is also inspired by Donachie-Landshoff and originally was a model for elastic scattering and single diffraction. It is much more elaborate than SaS/DL model. For elastic scattering it includes Coulomb scattering amplitude and 5 other terms: triple gluon exchange amplitude, hard and soft pomerons and two reggion terms. For diffractive scattering ABMST model operates in 2 regimes: high and low-mass diffraction, separated by a cut:
 \begin{equation}
 	M_{\text{cut}} (s) = \left\{\begin{array}{lr}
 		3, & \text{for } s \leq 4000 \, \text{GeV}^{2}\\
 		3+ 0.6 \ln \left(\dfrac{s}{4000}\right)  & \text{for } s > 4000 \, \text{GeV}^{2}
 	\end{array}\right.
 \end{equation}	
 
In the low mass regime, diffraction is simulated by the exchange of 4 resonances parametrized by Breit-Wigner shapes with an added quadratic background from a high mass regime.  
 
In the high mass regime the model has a contribution from 5 components to the cross-section: from pion exchange in differential cross-section, and triple Reggion (Pomeron) terms, namely $\mathbb {P P P}$, $\mathbb {R P P}$, $\mathbb {R R P}$ and $\mathbb {RRR}$. 

Generally, in all \pythia models the diffractive event is split into two regimes: soft diffraction and hard diffraction. The soft diffraction, in turn, has two limits: low-mass diffraction high-mass diffraction , with a smooth (exponential) transition between them.  There are two regimes at low masses. At the very low masses, $M_x \leq m_B + 1$ ($2$ GeV for $p-p$ collisions) formed diffractive system decays isotropically into a two hadron state . Above $2$ GeV a collision process is viewed as either a pomeron kicking out of quark and  leaving the remnant diquark or pomeron kicking out a gluon of the incoming hadron, which results in a hairpin topology. The relative rate between this alternatives is:
\begin{equation}
	\dfrac{P_q}{P_g} = \dfrac{N}{M_{X}^{p}}
\end{equation} 
default values being $N = 5$ and $p = 1$.
In the high-mass ($M_{\rm X} \xspace \approx \xspace 10$~GeV or higher) limit of soft diffraction, the perturbative description is considered through the Pomeron PDFs $f_{i/\mathbb{P}}$ (parton inside Pomeron PDF) invoking the standard perturbative multiparton interactions framework.
The hard diffraction regime incorporates also Pomeron flux distribution functions $f_{\mathbb{P}/p}$ (i. e. Pomeron inside proton PDF). Hard diffraction appears to be not relevant at NICA energies, further details are omitted.

The extended ABMST model have been implemented in to \pythia 8.3, nevertheless, it can be easily reduced to the original ABMST model by the choice of relevant parameters.
Followed modification of the original model have been added:

a) parameters were adjusted to smoothen the dip between low-mass regime and high-mass regime of diffraction through the modification of a background from high-mass in a low-mass regime, see \cite{Rasmussen:2018dgo} for details. 

b) the standard ABMST shows an unobservable increase in cross-section at high masses around 
$\xi \sim 1$ (there is no observable rapidity gaps at this values of $\xi$ since $\delta y \sim \ln (\xi)$). 
This rise also results in an increase of cross-section with \cmse giving a significant contribution to diffractive cross-section. 
Therefore, the damping factor $1/\left(1 + (\xi \exp (y_{min}))^{p}\right)$ is introduced along with the scaling factor $k(s/m_{p}^{2})^p$. 

c) original ABMST-model does not include double diffractive events. Thus, double diffraction was added in \pythia \cite{Rasmussen:2018dgo}.

In \pythia model, a typical inelastic pp collision starts with multiple parton interactions 
(MPI)~\cite{Sjostrand1987, Sjostrand:2017cdm}, that encompass a wide range of 
phenomena on internal proton structure which can not be derived from the first principles. 
The basis of MPI is a fact that $2\rightarrow2$ partonic cross-section is dominated by $t$-channel single gluon exchange $\dfrac{d \sigma}{d p_{T}^{2}} \sim \dfrac{\alpha(p_{T}^{2})}{p_{T}^{4}}$. It is obvious that this cross-section is needed to be regularized. 
Currently, \pythia relies on the smooth damping factor $\dfrac{p_{T}^{2}}{(p_{T}^{2} + p_{T 0}^{2})^{2}}$.
Thus integrated inelastic cross-sections is:
\begin{equation}
\sigma_{\mathrm{int}} =
\int_{p_{\perp\mathrm{min}}^2}^{s/4} \dfrac{\alpha(p_{T}^{2} + p_{T 0}^{2})}{(p_{T}^{2} + p_{T 0}^{2})^{2}}
\, \mathrm{d}p_{\perp}^2~,
\label{tseq:sigint}
\end{equation} 
where $p_{\perp\mathrm{min}}$ is minimum transverse momentum of parton interaction considered.
Data shows that this regularization parameter is energy dependent, currently a power-like rise is assumed: 
\begin{equation}
	p_{T 0} (\sqrt{s}) = p_{T 0 \, \text{Ref.}} \left(\dfrac{\sqrt{s}}{7 \, \text{TeV}} \right)^{a}
	\label{ptdependence}
\end{equation} 
It is worth mentioning, that \pythia checks the integral of perturbative cross-section and readjusts $p_{T 0}$ 
so that $\sigma_{\rm int} \geq \sigma_{\rm tot} - \sigma_{\rm diff} - \sigma_{\rm el}$.
The behavior of the $2\rightarrow2$ partonic cross-section leads to the fact that multiple parton interactions mostly occur at low $p_{\rm T}$ (order of $p_{\rm T 0}$ ).  It is expected that the significance of MPI wanes with the decrease of collision energy. {
Nevertheless, the regularization scale $p_{T 0}$ is obviously reflected in kinematic characteristics of minimum bias physics at these collision energies. 
Parton interactions are then dressed with initial state radiation (ISR) and final state radiation (FSR) and followed by Lund string hadronization. However, role of these mechanisms is practically negligible in the studied collision energy range.

Let us elaborate on Lund fragmentation algorithm used in \pythia and other generators studied 
in this review. Lund string fragmentation model dates back to the beginning
of 80s \cite{Lund1,Lund2, Lund3}. It was motivated by results from the Regge theory and first
lattice calculations on linearity of QCD-potential at large distances, i.e. $V(r) = \kappa r$, 
$\kappa$ being a string tensions. This quasiclassical model was based on the assumption that fragmentation
of quark-aniquark pair to hadrons can be described by the breaking of a relativistic string stretched
between them through a quantum mechanical tunneling process. 
Probability of string breaking by producing oppositely oriented quarks with a mass $m_{\rm q}$ 
and transverse momentum with respect to string axis $p_{\perp}$ is given by a Gaussian distribution:
\begin{equation}
P(m_q,p_{\perp}) \sim \exp\left(-\frac{\pi m_{\mathrm q}^2}{\kappa}\right) \exp \left(-\frac{\pi p_{\perp}^2}{\kappa}\right)
\end{equation}
Value of $\kappa$ is obtained by comparison of simulations to experimental data.

Lund fragmentation is an iterative procedure. Strings start to break and hadrons are formed from one of endpoints to another. 
The choice of symmetric fragmentation function assures that the result does not depend on the endpoint chosen \cite{Lund2}:
\begin{equation}
	f(z) = N \dfrac{1}{z} (1-z)^{a}  \exp \left(-\frac{b ( p_{\perp h}^2 + m_{h}^2)}{z}\right)
\end{equation}
where $f$ is a probability for a hadron with a mass $m_h$ and transverse momenta $p_{\perp h}$ with respect to string axis to take away fraction $z$ of the sting energy-momentum, $N$ is a normalization. Parameters $a$ and $b$ are determined by comparison to experimental data. 

Production of different particle species in the string fragmentation process is regulated by dedicated free parameters. These parameters set relative rates. In this paper we study production of pions, protons, and kaons. The most relevant parameters in this context are suppression factors of diquarks and strange quarks. Namely, the suppression of the production of an $s$-quark during the breaking is given by the string parameter \texttt{StringFlav:probStoUD} and \texttt{StringFlav:probQQtoQ}. Also, the string fragmentation machinery in \pythia 8.306 allows to set relative rates of vector mesons with respect to pseudo scalar ones composed of same quarks, relative rates of excited mesons, as well as  factors for baryons and mesons in the popcorn models.

The picture of baryon production is more complicated \cite{LundBaryon}.Baryons are produced in \pythia through diquarks, when either string breaks by a production of diquark at a breaking point instead of a quark or (popcorn model), diquark is produced due to quark fluctuations before the breaking of a string forming at the and a baryon anti baryon pair and a meson.  If string breaks during the formation, effective diquark-antidiquark pair is produced that form color-neutral objects fragmenting into hadrons. 
Another mechanism of fragmentation to baryons implemented relatively recently (since version 6.3) is a mechanism of baryon production through junction topology, i. e. string is stretched between three {\it rgb} quarks with a junction in the middle. Gluon on a fragmentation stage in \pythia is considered to be a quark-antiquark pair. 

Very important constituents of the collision, especially for minimum-bias physics, are beam remnants. By definition, beam remnants are the leftovers of incoming protons after the partons that took part in hard interactions and ISR/FSR were removed. To conserve quantum numbers of beam particles, certain amount of partons are required to be added as follows. First, the number of added flavours are being determined. If there are valence quarks in a remnant, randomly selected diquark pairs are formed with probabilities based on $\mathbf{SU(3)_{fl.}\times SU(2)_{sp}}$ model. For all unresolved sea quarks kicked out during the MPI and showering a companion quark with opposite color and flavour is added into the remnant. At the end, gluon is added to carry a momentum of the remnant (for momentum conservation) and if necessary to balance the colour structure.
At the second stage colourless object is formed from remnant partons. Currently there are two models that realizes this procedure. Old model (used by default now) \cite{brold} adds initiator gluons at random to partons in a beam remnant while gluons are replaced by quark pairs as if splitting occurred. The new model \cite{brnew} is motivated by full $\mathbf{SU(3)}$ algebra and extends \pythia 's MPI machinery to beam remnants. First, colour configuration of initiator partons is considered, the weight of overall multiplets is chosen, so that larger multiplets would be suppressed compared to smaller ones: $\rho(M) ~ \exp(-\dfrac{M}{k_{sat}})$, saturation is regulated by $k_{sat} \equiv BeamRemnants:saturation$ which is a free parameter. After that, minimum number of gluons are added to the beam remnant to obtain a color singlet state. Then color connections and junction structures are assigned randomly, however two valence quarks, if they are present in the beam remnant,  are treated as parts of common junction. 

During a perturbative treatment of the event, only the longitudinal momenta is generated because calculations are based on collinear factorization. \pythia accounts to the transverse momentum generation by adding primordial $k_{T}$ in the model. For hard initiators primordial $k_{T}$ is added by a Gaussian distribution that has a width:
\begin{equation}
	\sigma(Q) = \dfrac{\sigma_{\text{soft}} Q_{1/2} +  \sigma_{\text{hard}} Q}{(Q_{1/2} + Q)} \dfrac{m}{m + m_{1/2}y_{\text{dump}}}
\end{equation}	
here there are two separate width parameters $\sigma_{\text{soft}}$ and $\sigma_{\text{hard}}$. $Q$ is a scale of the hardest process, parameter $Q_{1/2}$ controls the transition between hard and soft regimes (the bigger $Q$ is compared to $Q_{1/2}$ the less important is $\sigma_{\text{soft}}$ contribution and vice versa), $m$ is a mass of the system, $m_{1/2}$ is a parameter that controls dampening of small-mass and large-rapidity systems. Another parameter $\sigma_{\text{remn}}$ controls the primordial transverse momentum of partons in a remnant. The original transverse momentum of a beam is restored after rescaling $p_{i T}$ of all partons in a beam by a common factor.

\subsection{\epos}
\label{Sec:MCmodels:epos}

\epos is an event generator for both proton-proton and heavy ion collisions.
\epos~\cite{epos1,epos2} stands for Energy conserving quantum mechanical multiple scattering 
approach based  on (E), Parton ladders (P), Off-shell remnants (O), Splitting of parton ladders (S). 
Let us give a short summary of parts of \epos model relevant for proton-proton collisions. 

\epos is a parton based Gribov-Regge theory model where partons undergo multiple scattering through pQCD-inspired ladders,
 open ladders being reggeons (cut pomerons) while closed ladders representing pomerons.

However, despite being pQCD-inspired \epos approach is not perturbative, since ladder diagrams
have both hard and soft parts. These allows \epos to describe soft (diffractive, elastic) scattering, hard inelastic scattering and MPI in a single approach. Saturation like effects are incorporated by the means of ladder splitting. Parton distribution functions are not included ad hoc since they represent a dynamic process of parton emission that is being simulated in \epos framework.

At the final stage of the collision each parton ladder is translated into two color string (remnants being white objects) which ends up fragmenting into hadrons. 
Since \epos 3.0, the MC model is  extended to describe collective-like phenomena in pp collisions
invoking core-corona hydrodynamical approach to fragmenation. In this approach, before fragmentation each
string is either considered to be in a core spatial region or in a peripheral region (corona). 
Strings in the core are subject to additional hydrodynamic evolution, i. e. are hadronized with additional
longitudinal and radial flow effects, while usual routine is used for strings in the corona \cite{epos3}.

\subsection{\urqmd}
\label{Sec:MCmodels:urqmd}

UrQMD~\cite{urqmd1,urqmd2} ( Ultrarelativistic Quantum Molecular Dynamics) is a generator
of heavy ion collisions that incorporates:
\begin{itemize}
	\item microscopic transport theory (Boltzmann-Uehling-Uhlenbeck equation):
\begin{equation}
	\dfrac{\diff f_{i}(x,p)}{\diff t} \equiv \dfrac{\partial f_{i}(\mathbf{x},t,p)}{\partial t} +\dfrac{\diff \mathbf{x}}{\diff t}\dfrac{\partial f_{i}(\mathbf{x},t,p)}{\partial \mathbf{x}} + \dfrac{\diff p^{\mu}}{\diff t}\dfrac{\partial f_{i}(\mathbf{x},t,p)}{\partial p^{\mu}} = \mathrm {St} f_{i} (x,p)
\end{equation} 
of phase-space densities $f_{i} (x,p)$ for different particles produced in nuclear-nuclear collisions ($x$, $p$, being time-space coordinates and $4$ - momentum respectively, $\mathbf{x}$ -- space coordintes, $t$ -- time), 
where collision term $\mathrm{St} $ takes into account rescattering of different particles.

\item Soft interaction between nuclei is described by Skyrme-like potential (actual Quantum 
Mechanical Dynamics), which allows to calculate equations of state of many-body system. 
Such soft interactions are not relevant for pp collisions.

\item Partial elementary cross-sections of different hadron-hadron processes are hardcoded in UrQMD model. 
Most basic and abundant processes are experimentally measured, and fits of the experimental data are used.  
If no data were available, the additive quark model and detailed balance arguments were used (see~\cite{urqmd1}). 

\item Differential cross-section (i. e. angular distribution) for each elementary hadron-hadron $2-2$ process is described
by general formula obtained by the calculation in the scope of effective field theory for the system of
nucleons interacting through $\sigma$, $\omega$ and $\pi$ mesons \cite{MaoLiZhuo}.

\item For higher energy scatterings the Lund string model is used for calculations of string excitations and subsequent fragmentation.

\end{itemize}

\subsection{\smashm}
\label{Sec:MCmodels:smashm}

\smashm event generator (Simulating Many Accelerated Strongly-interacting Hadrons)~\cite{Weil:2016zrk,Mohs:2019iee} 
is designed to simulate nucleus collisions. As in the \urqmd  model, nuclear interactions in 
\smashm are treated within the transport model of solving relativistic Boltzmann equation, 
though a little different approach is used here. Key features of \smashm model relevant for hadron collisions are presented below: 
\begin{itemize}
\item Hadron-hadron collision cross-section is split into various mechanisms that contributes differently depending on collision energy:
\begin{itemize}
\item  At $\sqrt{s} < 4$ GeV (for $pp$ collision) inelastic collisions are described via resonance production, excitation and decays. The dependency of cross-sections on \cmse is a fit to data (same as for \urqmd, see section \ref{Sec:MCmodels:urqmd}). 
Angular distribution of $2-2$ are isotropic except for elastic processes $NN \rightarrow NN$, $NN \rightarrow N\Delta$ processes, 
where exponential ansatz \cite{Cugnon} is used $d \sigma/d t \sim \exp(b(s) t)$ ($b$ -- energy-dependent parameter, $t$ is a Mandelstam variable)
 and heavy resonance production $NN\rightarrow NR$ ($R = \Delta^{*} , N^{*}$) power function (for the portion of non) ansatz 
 $d \sigma/d t \sim t^{-a}$ is used. All formed resonances decay are supposed to be isotropic in the rest frame.  
\item At transition energies, 4$<$~\cmse$<$~5 GeV for proton-proton collisions, resonances gradually reduce their contribution to particle production 
(smooth weighting function for cross-sections is used) and inelastic collision is modeled with the soft-string subroutine. Similar to \urqmd, 
this model splits inelastic cross-section to single-diffractive, double-diffractive and non-diffractive parts. Soft-string excitations are
treated differently for each of them. Fragmentation of the formed excited strings is handled by \pythia 8.306 except for the fragmentation 
of leading baryons (see below).
\item At $\sqrt{s} > 5$ GeV, pQCD-based \pythia routines for parton scattering are also used though hard parton scattering starts to contribute 
only at much higher energies (see \cite{Mohs:2019iee}, fig. 1). Fragmentation of partons is also handled by \pythia.
\end{itemize} 
\item  Dependence of elastic collisions at $\sqrt{s} > 4$ GeV on $\sqrt{s}$ is a fit of experimental data. Simple exponential ansatz $d \sigma/d t \sim \exp(b(s) t)$ is used 
for angular distribution  .      
\item Dedicated fragmentation functions are used for leading baryons in soft (non-pertubative) non-diffractive string process.
It uses same functional form as for other particles, but it has parameters that lead to harder fragmentation \cite{Mohs:2019iee}.
\item Generally, free parameters of the string excitation and fragmentation are adjusted~\cite{Mohs:2019iee} to describe some
experimental data on proton-proton collisions measured in NA49 and NA61/SHINE experiments at $\sqrt{s}$~=~17.3~GeV.
\end{itemize}


\section{Results}
\label{Sec:Results}

\subsection{Particle yields}
\label{Sec:Res:PY}

Early data on hadron production mostly come from 60s and 70s from fixed-target experiments
on accelerator facilities in Brookhaven (BNL, fix target, incident proton energies: $2.4$, $2.85$, $5$, $6.9$, $8$ GeV, \cite{BNL1,BNL2,BNL3,BNL4,BNL5,BNL6}),
Princeton (PPA, i. e. Princeton-Pensilvania Accelerator, incident beam energies: $2.54$, $2.88$, $3.03$ GeV  \cite{PPA1}), 
CERN (PS, i. e. Proton Synchrotron Saclay 81-cm hydrogen bubble chamber, incident proton energy:
$5.5$, $19$, $12$, $24$ GeV \cite{Saclay1,ScandBubble1, PSCher1, PS2,PS3}; ISR, center of mass energies: 22, 31, 45,53~GeV \cite{ISR1,ISR2}; 
SPS, i. e. Super Proton Synchrotron,
$200$, $360$ GeV \cite{SPS1,SPS2}), Argonne National Laboratory (ANL, incident beam momentum $6$, $102$ GeV \cite{ANL1,ANL2},
ZGS, i. e Zero Gradient Synchrotron, incident proton momentum $12$, $12.4$ GeV \cite{ZGS1,ZGS2}), 
Serpukhov (Serpukhov accelerator, incident beam momentum $32$, $69$ GeV \cite{Serp1,Serp2}), 
Fermilab (Main Ring, incident beam momentum: $100$, $147$, $200$, $205$, $300$, $400$ GeV 
\cite{Fermilab1, Fermilab2, Fermilab3, Fermilab4,  Fermilab5,Fermilab6}).
Particle identification in the above-mentioned experiments was done by the means of either bubble chambers,
or time of flight cameras, Cherenkov detectors, or hybrid systems for particle identification. Combination of different
methods was used in \cite{Fermilab2,SPS1,SPS2, ISR1,ISR2}. Many of those  experiments had quite limited acceptance or low identification efficiencies. 
 
Relatively recent data on hadron production in proton-proton and nuclear collisions were obtained 
in fixed target experiments NA49 and NA61 at SPS in CERN. NA49 \cite{NA49exp} was a wide acceptance
spectrometer on SPS, it had an ability to measure particle momentum in a magnetic field and allowed
to identify particles via $\diff E / \diff x$ and time-of-flight (TOF) measurements. The experiment was active from 1994 to 2002. Being a direct descendant of NA49 experimental facility, NA61/SHINE experiment \cite{NA61exp} has a significantly improved system for precise $\diff E / \diff x$ and TOF measurements, and allows to measure energy of projectile fragments for heavy ion program.

This section presents predictions of four MC generators for 
mean multiplicity, mean transverse momentum, and rapidity distribution of various particle species produced in pp collisions at NICA energies. Namely, \epos 3.4, \pythia 8.306, \urqmd 3.4, \smashm 2.0.1 are chosen for this study for the reasons 
given in Sec.~\ref{Sec:MCmodels}. These predictions are compared to the experimental
data obtained in various experiments presented above. We also compare data and other generators to an adjusted version of \pythia 8.306 (see details in next subsection), that will be named \pythia 8.306 LE1 further in the text (here LE stands for Low Energy). Data and results of MC simulations
use inclusive selection of inelastic collisions that is widely referred to as minimum-bias one.

The particle multiplicity is one of the most basic observable to characterize high-energy particle collisions and includes in it all mechanisms of particle production. It is a basic observable to discriminate between models of particle production. We compare available data on multiplicity of pions, kaons, and protons with results of simulation with various event generators for a number of \cmse. Mean multiplicity of $\pi^{+}$  and  $\pi^{-}$ as a function of collision energy for data and MC models is shown in Fig.~\ref{Plot:s:pions}.  Data and all MC models show
a gradual increase of pion multiplicity with \cmse. The difference between $\pi^{+}$  and  $\pi^{-}$  multiplicities is apparently due to the noticeable role of charge conservation in collisions with low total multiplicity of produced particles. Relative differences between  $\pi^{+}$ and $\pi^{-}$ naturally decreases with \cmse. \epos 3.4 is in agreement with the data within
uncertainties.  \pythia 8.306 describes the data in the range  6.3~$\leq$\cmse$\leq$~12.3~GeV, showing higher pion multplicities up to 10\% at the highest studied energy \cmse~=~31GeV. \pythia 8.306 LE1 gives nearly the same pion multiplicities
in the range  6.3~$\leq$\cmse$\leq$~12.3~GeV as standard version, however it shows a better description of data at higher \cmse.
\smashm 2.0.1 and \urqmd 3.4 underestimate the pion
multiplicities. Same conclusions can be qualitatively applied to the mean kaon multiplicities as a function of collision
energy which is shown in Fig.~\ref{Plot:s:kaons}.  Completely different picture is observed 
for mean proton multiplicity which is shown in Fig.~\ref{Plot:s:protons} (left).
Existing data have large uncertainties, nevertheless, one can say that mean proton multiplicity
varies from 1 to 1.5 over the studied range  of \sqrts. MC models exhibit nearly 
flat dependencies with $\langle N_{\rm p} \rangle \approx$~1.2 for \epos 3.4, \pythia 8.306, 
and $\langle N_{\rm p} \rangle \approx$~1.5 for \urqmd 3.4 and \smashm 2.0.1. Dependence of mean antiproton
multiplicity ($\langle N_{\bar{\rm p}} \rangle $)  on \sqrts  is shown in Fig.~\ref{Plot:s:protons} (right).
It is qualitatively similar to the one for pions and koans. The difference between proton and antiprotons, apparently, comes from differences in production mechanisms that will be discussed along with their rapidity distributions below.

\begin{figure}[hbtp]
\begin{minipage}[h]{0.49\linewidth}
\center{\includegraphics[width=1\linewidth]{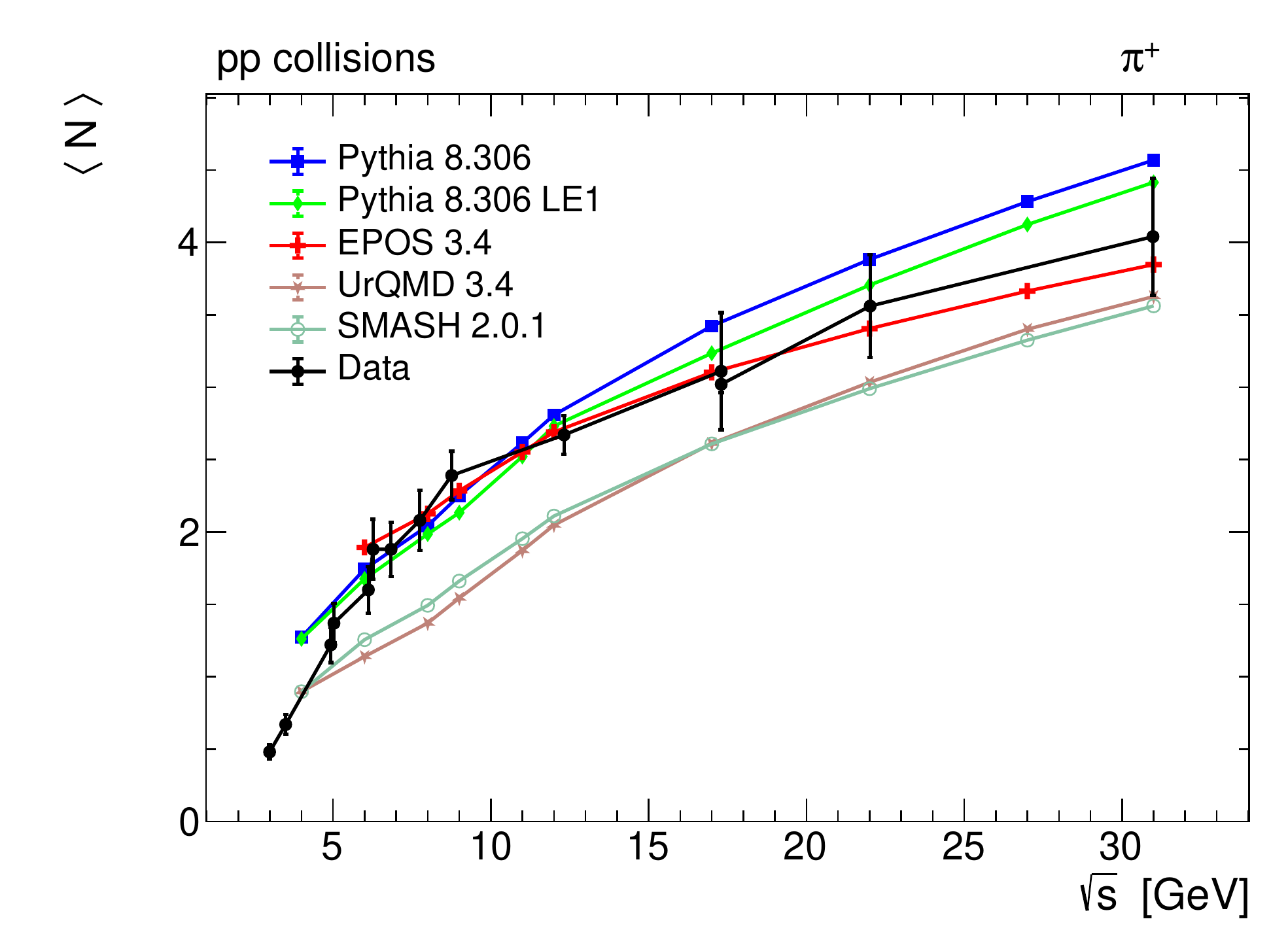} \\ }
\end{minipage}
\begin{minipage}[h]{0.49\linewidth}
\center{\includegraphics[width=1\linewidth]{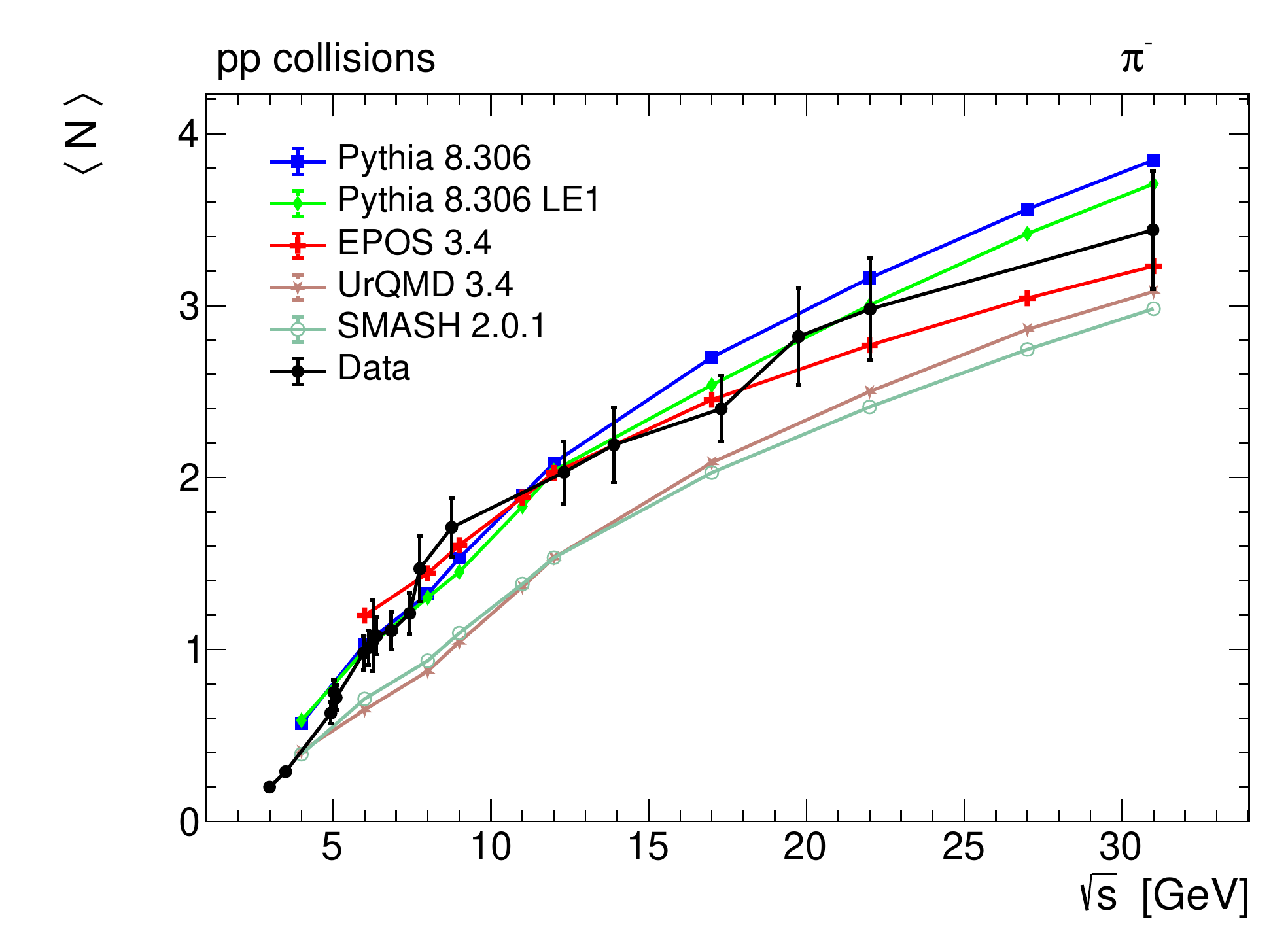} \\ }
\end{minipage}
\caption{Inclusive multiplicity of $\pi^{+}$(left) and $\pi^{-}$ (right) as a function of $\sqrt{s}$.}
\label{Plot:s:pions}
\end{figure}

\begin{figure}[hbtp]
\begin{minipage}[h]{0.49\linewidth}
\center{\includegraphics[width=1\linewidth]{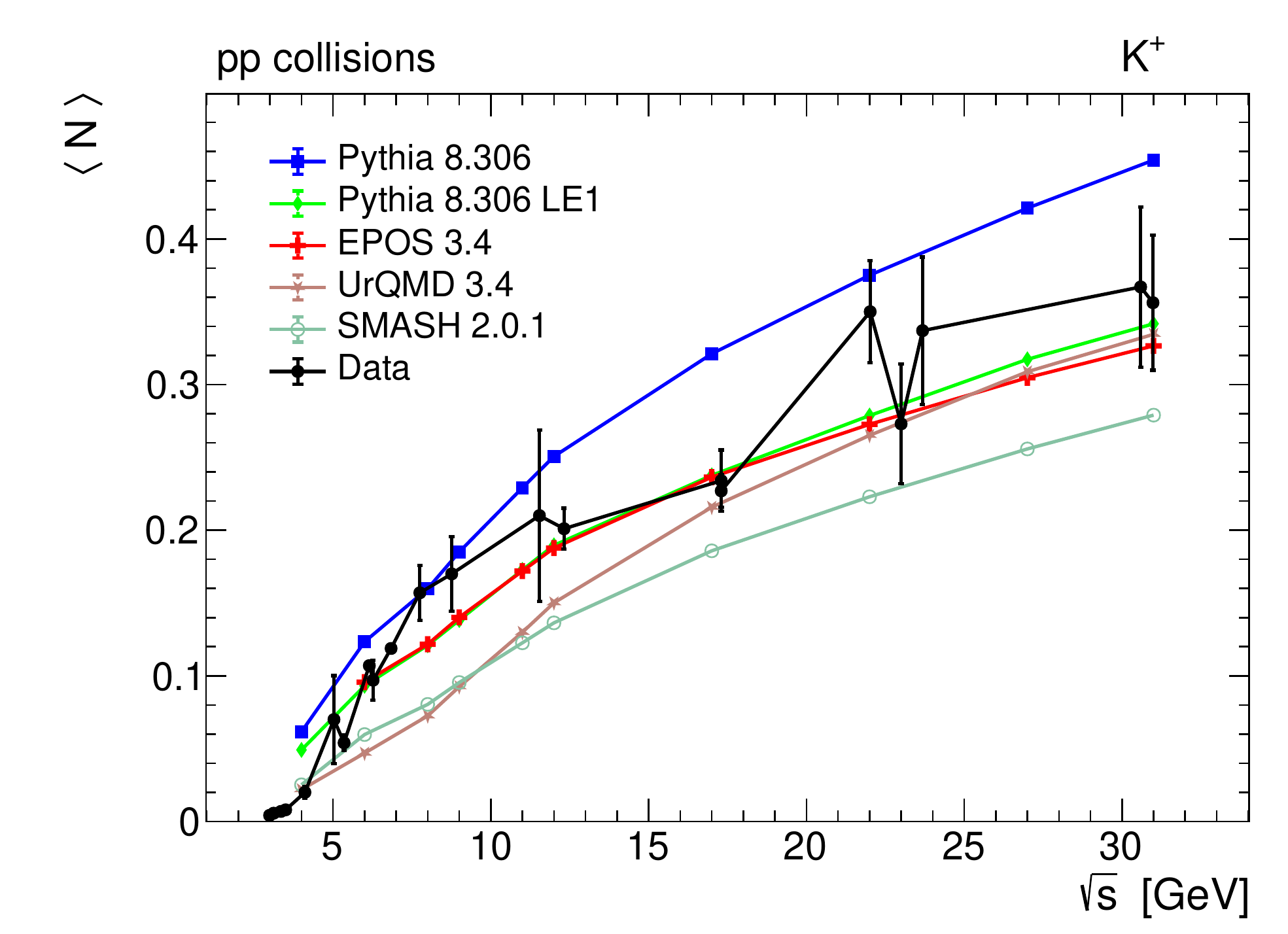} \\ }
\end{minipage}
\begin{minipage}[h]{0.49\linewidth}
\center{\includegraphics[width=1\linewidth]{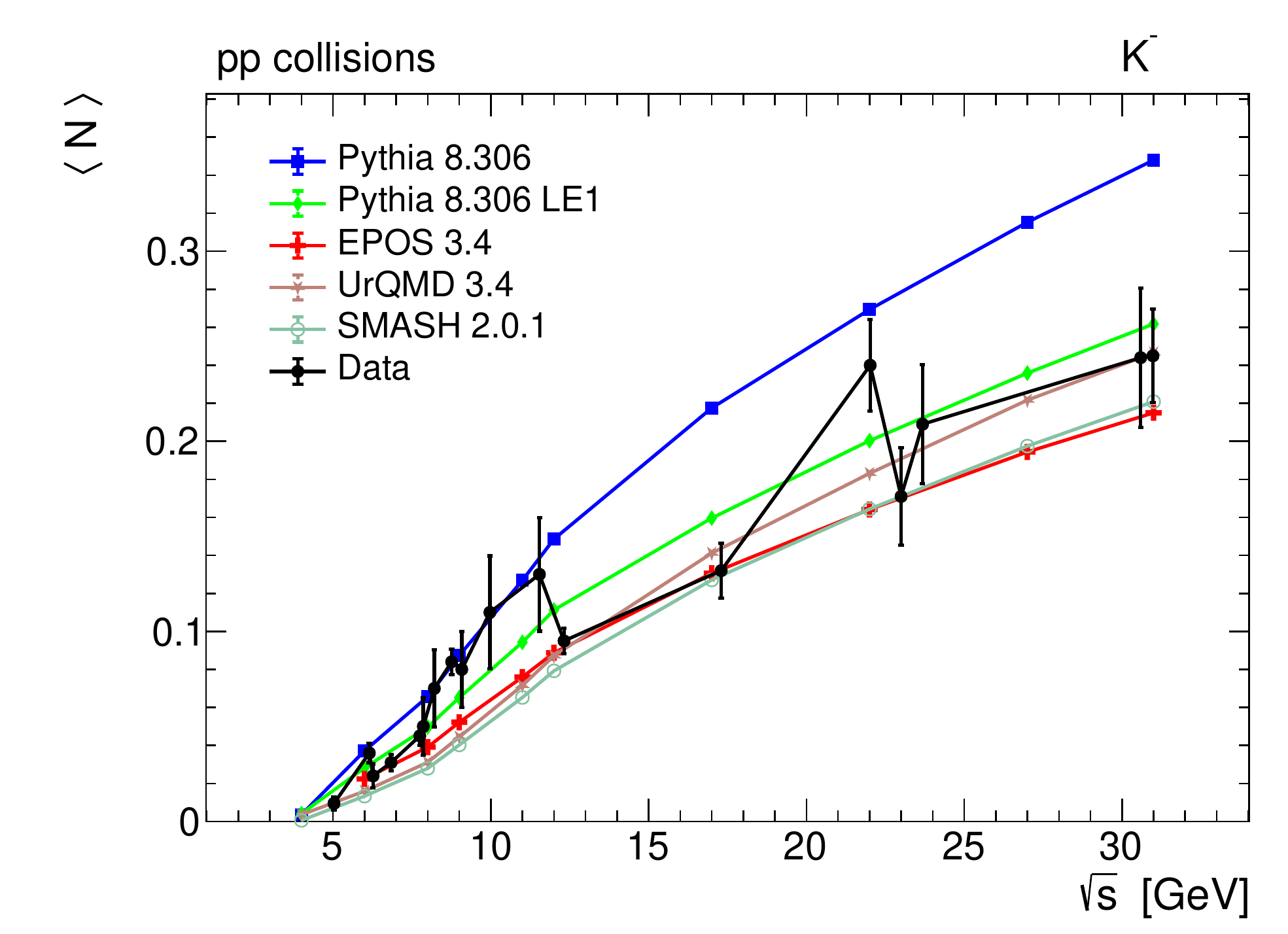} \\ }
\end{minipage}
\caption{Inclusive multiplicity of $\mathrm{K}^{+}$ (left) and $\mathrm{K}^{-}$ (right) as a function of $\sqrt{s}$.}
\label{Plot:s:kaons}
\end{figure}

\begin{figure}[hbtp]
\begin{minipage}[h]{0.49\linewidth}
\center{\includegraphics[width=1\linewidth]{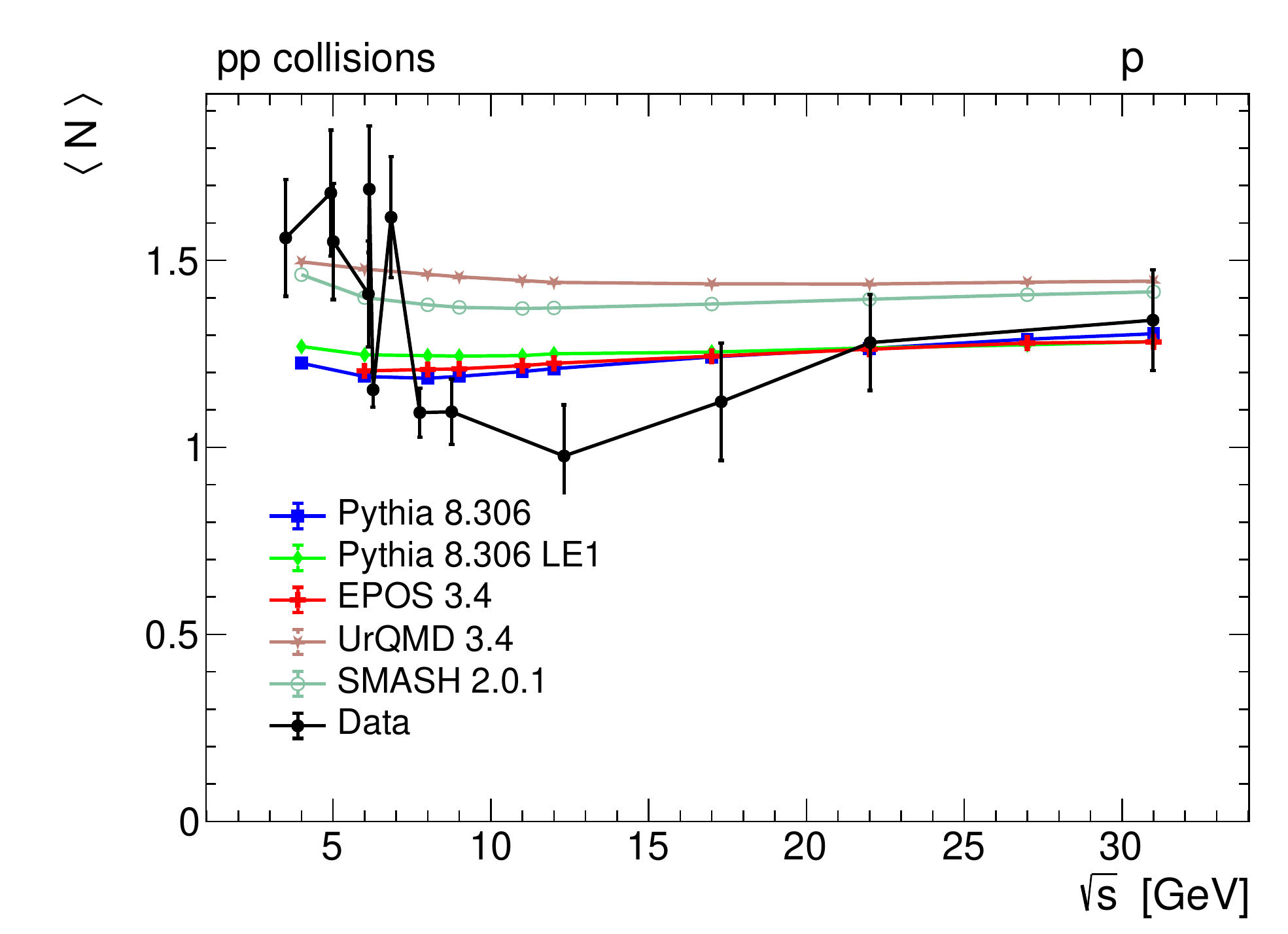} \\ }
\end{minipage}
\begin{minipage}[h]{0.49\linewidth}
\center{\includegraphics[width=1\linewidth]{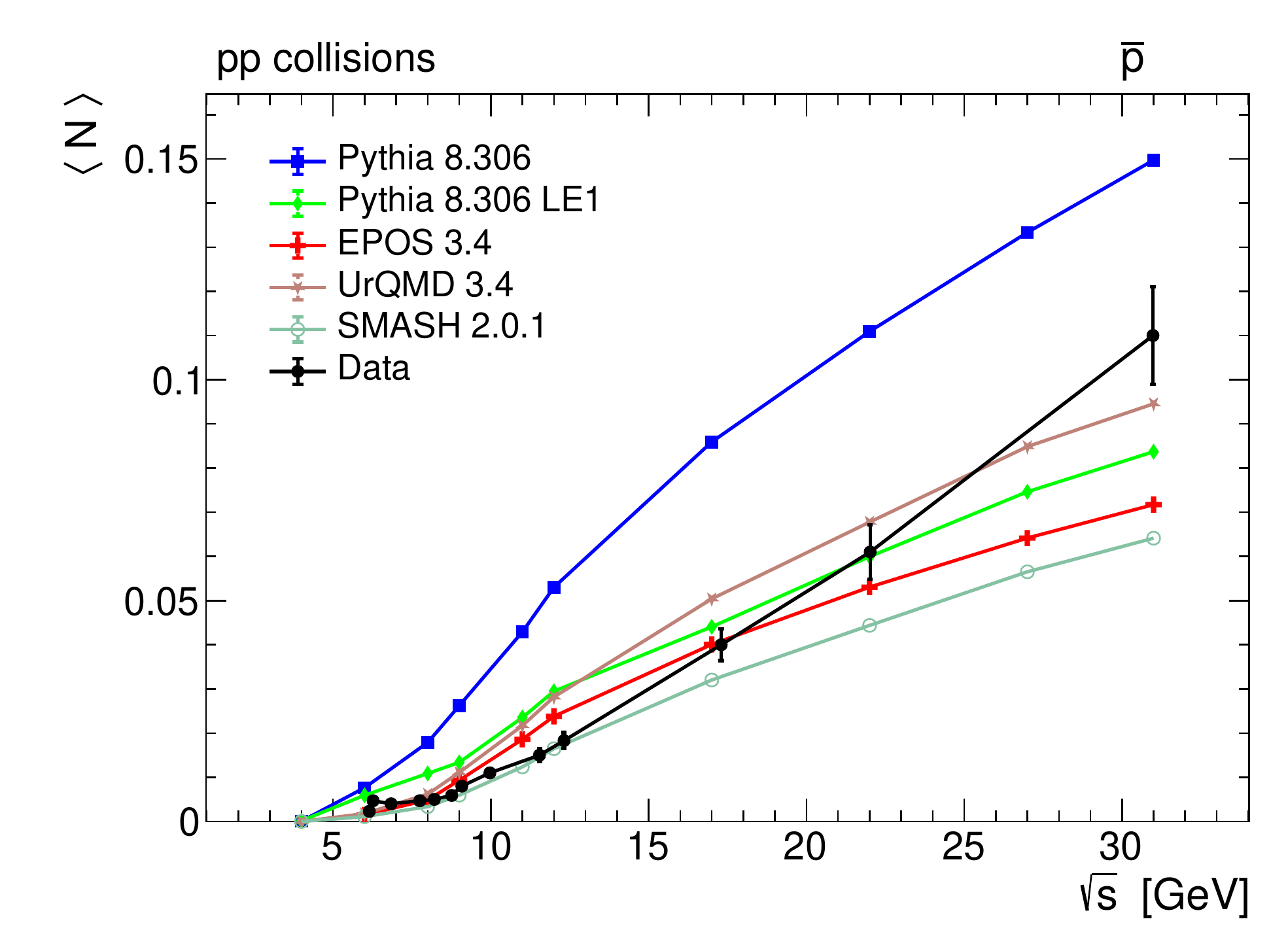} \\ }
\end{minipage}
\caption{Inclusive multiplicity of p(left) and $\mathrm{\bar{p}}$ (right) as a function of \cmse.}
\label{Plot:s:protons}
\end{figure}

It should be noted that NA61/SHINE and NA49 conducted high-quality measurements of other particle species 
at $\sqrt{s}$~=~17.3~GeV. 
The experimental data on the mean multiplicity were compiled and interpreted within statistical hadronization model~\cite{Matulewicz:2021smg}. It was found that the model can reasonably describe the data if canonical volumes of hadrons without and  with open strangeness were different. Comparison of these data with studied generators at $\sqrt{s}$~=~17.3~GeV are summarized in Table~\ref{table:AllParticleMult17GeV}.
One can clearly see that \epos 3.4 was tuned to describe the data, though $\Xi$ yields are not described as good as other particles. 
Default tune of \pythia 8.306 tends to overestimate most particle yields because it was tuned to describe data at LHC energies. \pythia 8.306 LE1 with adjusted parameters gives a significantly better agreement with data. 
\urqmd 3.4 and \smashm 2.0.1 noticeably  underestimate most particle multiplicities.
Generally, relative data--MC differences for the extended list of mesons and baryons are nearly universal, respectively.

\medskip

\begin{sidewaystable}
\caption{Mean multiplicity of various particles in pp collision at $\sqrt{s}$~=~17.3~GeV. Data are obtained in N49 and N61/SHINE experiments.
Particles not present in the standard generator output are marked  with "--". }
\label{table:AllParticleMult17GeV}
\centering
\begin{tabular}{|c|c|c|c|c|c|c|r} \hline
   Particle &  \multicolumn{6}{c|}{Mean Multiplicity}   \\ 
 \cline{2-7}
                                           & Data & \pythia 8.306& \pythia 8.306 LE1 & \epos 3.4 & \urqmd 3.4 & \smashm 2.0.1 \\ 
                                           
\hline

          $\pi^{+}$                        &  3.1 $\pm$ 0.3      &  3.41  & 3.22  &  3.11   & 2.62   &  2.60   \\
\hline
          $\pi^{-}$                        &  2.4 $\pm$ 0.3      &  2.68  & 2.52  &  2.45   & 2.09   & 2.03     \\
\hline
          $K^{+}$                          &  0.23 $\pm$ 0.02    & 0.321  & 0.237 & 0.27    &  0.216  & 0.186      \\
\hline
          $K^{-}$                          &  0.132 $\pm$ 0.014  & 0.217  & 0.156  & 0.132  &  0.141  & 0.127    \\
\hline
          $K^{0}_{\rm S}$                  &  0.18 $\pm$ 0.04    &  0.247  & 0.185   &  0.18  &  --  &  --    \\
\hline
          $K^{*}(892)$                     &  (7.6 $\pm$ 5)$10^{-2}$  & 0.082  &   0.063 &   0.077  &   0.081   &  --  \\
\hline
  &    &    &    &    &    &  \\[-1em]                
		  $\bar{K}^{*}(892)$               & (5.2 $\pm$ 5)$10^{-2}$   & 0.049  &   0.095  &  0.049  &  0.046  &  --  \\
\hline
          $\phi$                           &  (12.5 $\pm$ 0.4)$10^{-3}$  &  1.52 $\times 10^{-2}$ &   9.1$\times 10^{-3}$  &  13.3$\times 10^{-3}$  & 6.2$\times 10^{-3}$   &   --   \\
\hline
          $p$                              &  1.15 $\pm$ 0.03   &  1.24 &  1.26  & 1.24   & 1.43   & 1.38   \\
\hline
          $\bar{p}$                        &  (38.8 $\pm$ 1.2)$10^{-3}$  &  0.085 & 0.043   & 0.041   & 0.05   &  0.032   \\
\hline
           $n$                             &  0.67 $\pm$ 0.1  &  0.76  &  0.7  & 0.68   &  0.53  & 0.58   \\
\hline
          $\Lambda$                        &  0.12 $\pm$ 0.012  &  0.156  &   0.113 &  0.125  &  0.092  & 0.061    \\

\hline
          $\Xi^{-}$                        &  (8 $\pm$ 1)$10^{-4}$   &  1.5$\times10^{-3}$  &  8.7$\times10^{-4}$  &  6$\times10^{-4}$  &  1.6$\times10^{-3}$   &   4$\times10^{-4}$   \\
          \hline
            &    &    &    &    &    &  \\[-1em] 
          $\overline{\Xi}{}^{-}$                        &  (3.4 $\pm$ 0.6)$10^{-3}$  &  2.6$\times10^{-3}$  &   2.8$\times10^{-4}$  &  2.7$\times10^{-3}$   &  2$\times10^{-3}$   & 2.6$\times10^{-3}$     \\
\hline
\end{tabular}

\end{sidewaystable}

\medskip
\medskip
\medskip

Another basic characteristic is the transverse momentum (\pt) of produced particles. It
is important for both understanding of particle production mechanisms and optimization of detector design. 
Figs.~\ref{Plot:pt:pions},~\ref{Plot:pt:kaons},~\ref{Plot:pt:protons} show mean transverse momentum 
($\langle \pt \rangle$) of pions, kaons, and protons as a function of collision energy for MC models.
The striking feature of these dependencies is the fact that mean transverse momentum $\langle \pt \rangle $
is nearly flat at \cmse~$>$~10~GeV and almost equal for particles and antiparticles. 
Nevertheless, kaon and proton have higher $\langle \pt \rangle $ than pions.
There is a significant spread (up to $25\%$) of $\langle \pt \rangle$ across MC models, nevertheless 
\epos~3.4 and \pythia~8.306 LE1 converge at \cmse~$>$~10~GeV. 
These results are in the first approximation consistent with the data obtained recently by NA61/SHINE experiment \cite{Aduszkiewicz:2017sei}, though measured two-dimensional distributions $\frac{{\rm d}N}{{\rm d}y{\rm d}\pt}$ do not allow to extract exact values of $\langle \pt \rangle $ because of blind spots in the detector acceptance.

\begin{figure}[hbtp]
\begin{minipage}[h]{0.49\linewidth}
\center{\includegraphics[width=1\linewidth]{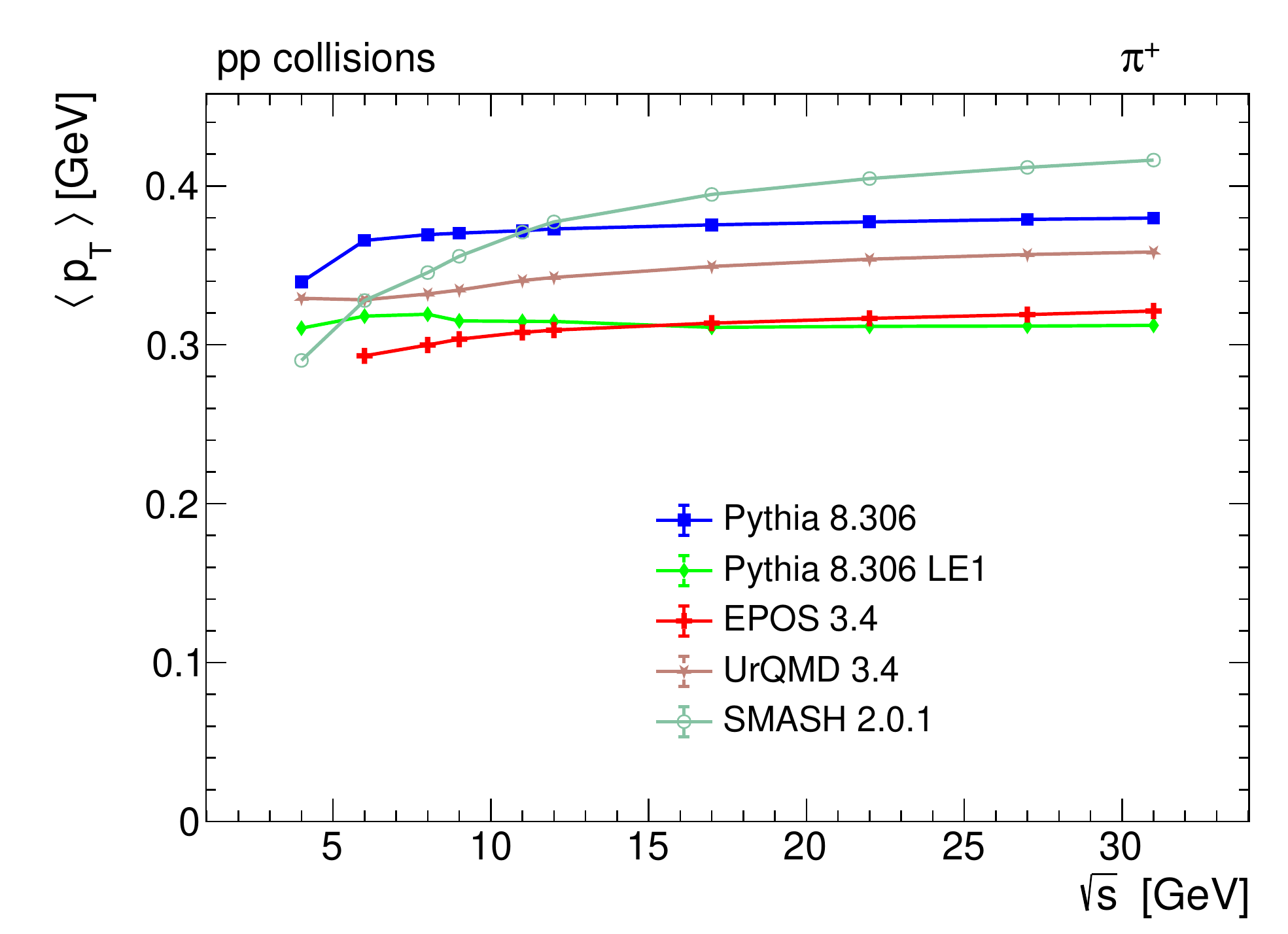} \\ }
\end{minipage}
\begin{minipage}[h]{0.49\linewidth}
\center{\includegraphics[width=1\linewidth]{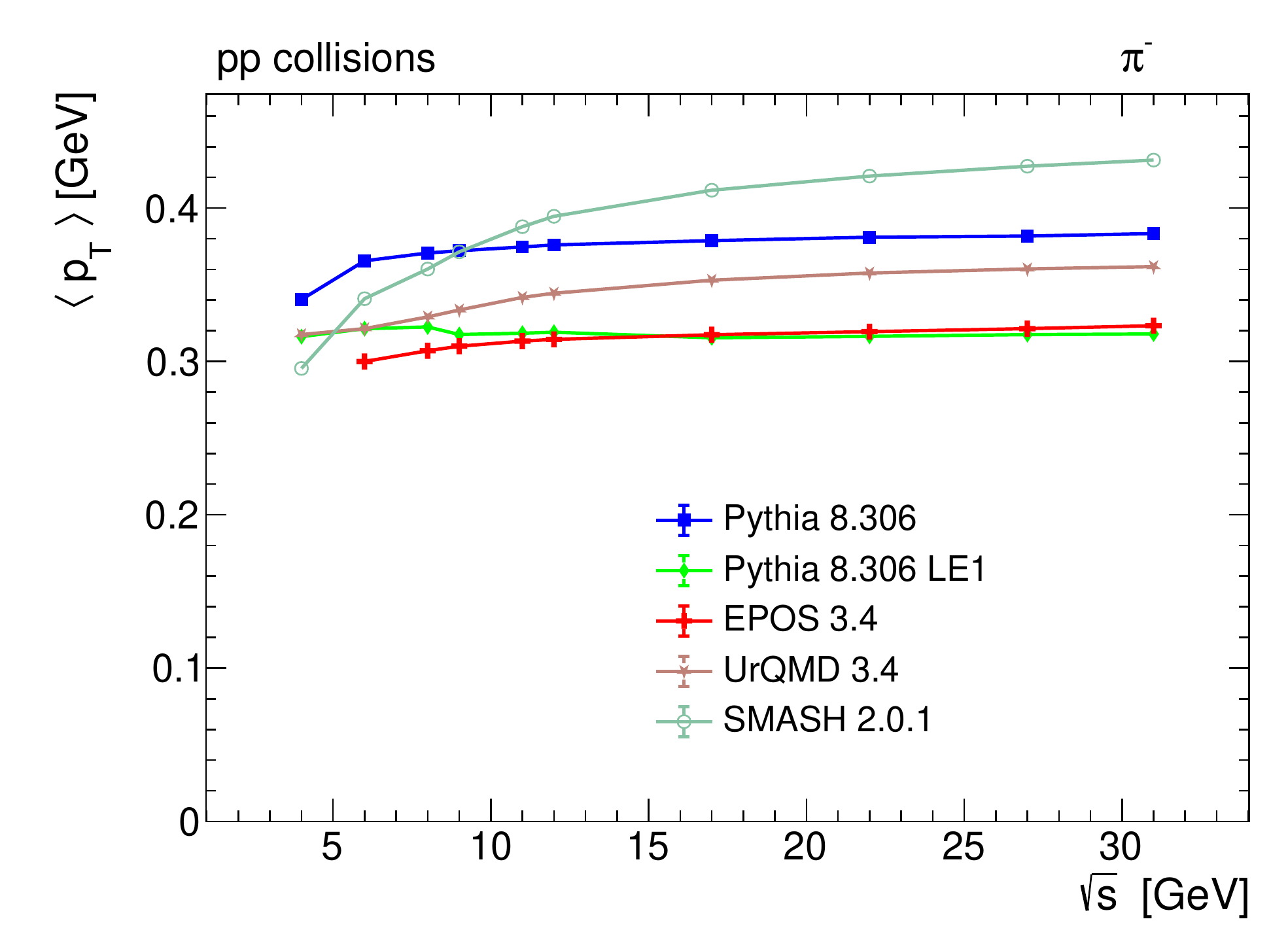} \\ }
\end{minipage}
\caption{Mean \pt of  $\pi^{+}$ (left) and  $\pi^{-}$ (right) as a function of $\sqrt{s}$.}
\label{Plot:pt:pions}
\end{figure}

\begin{figure}[hbtp]
\begin{minipage}[h]{0.49\linewidth}
\center{\includegraphics[width=1\linewidth]{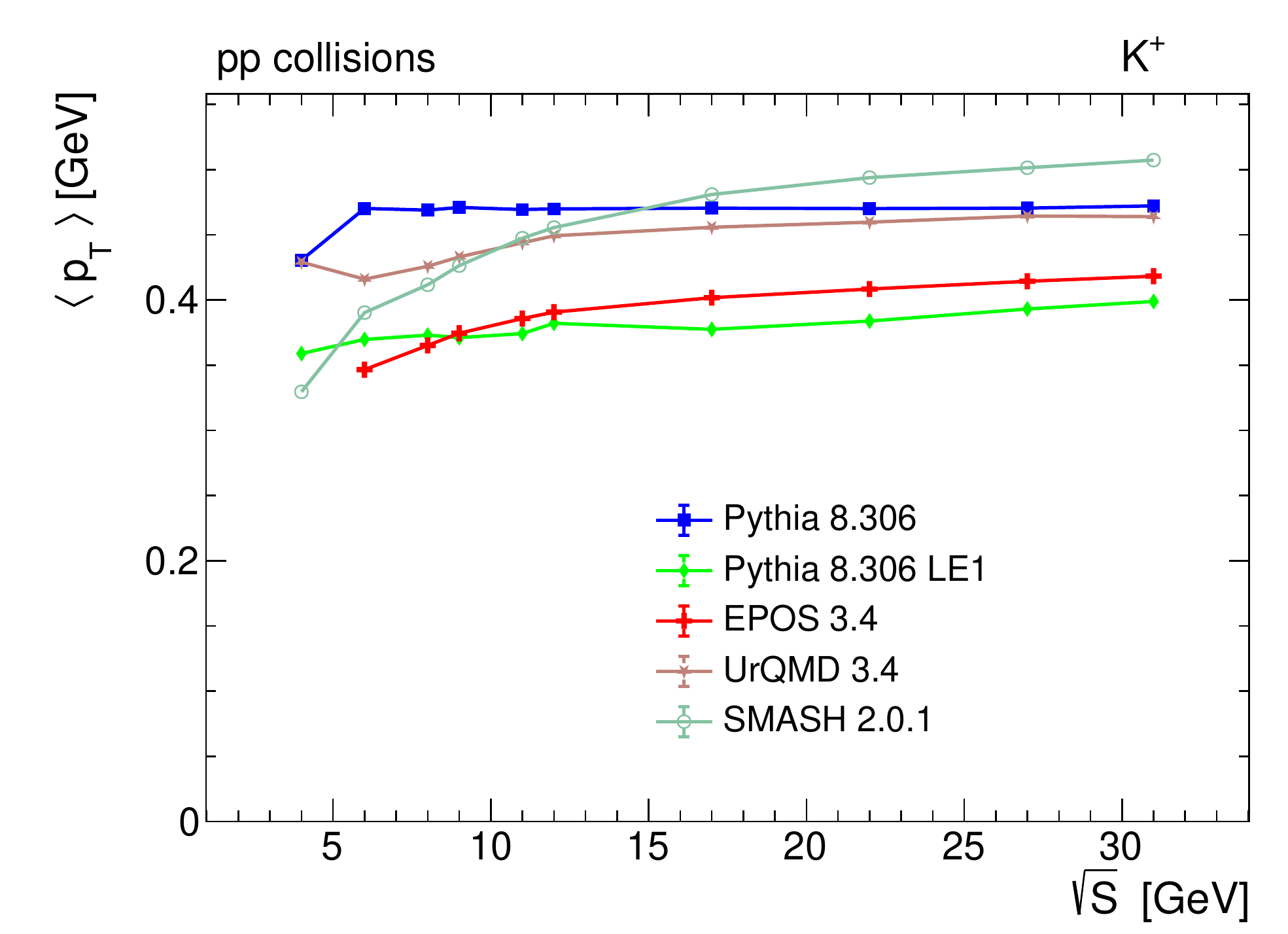} \\ }
\end{minipage}
\begin{minipage}[h]{0.49\linewidth}
\center{\includegraphics[width=1\linewidth]{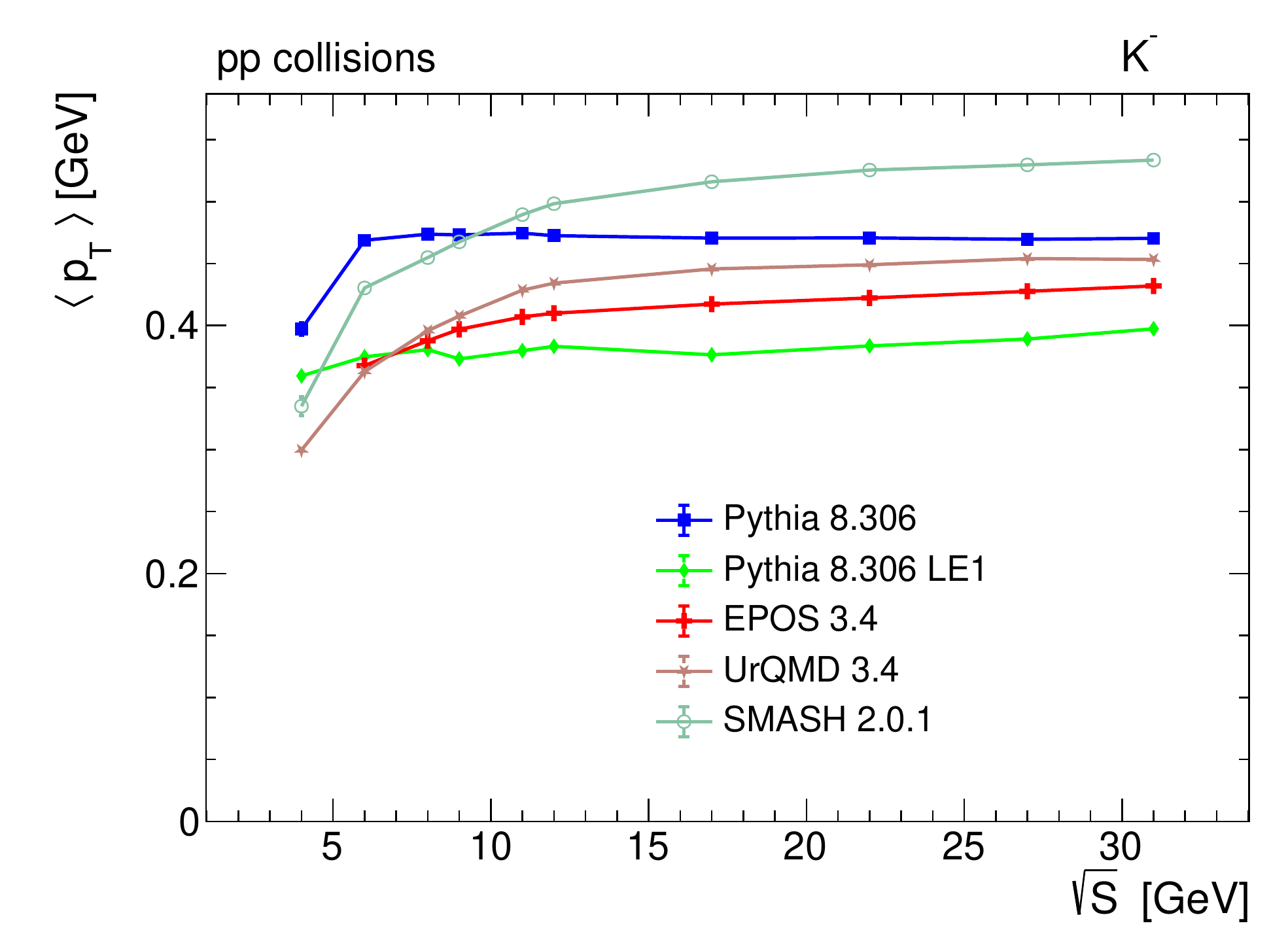} \\ }
\end{minipage}
\caption{Mean \pt of $\mathrm{K}^{+}$ (left) and $\mathrm{K}^{-}$ (right) as a function of $\sqrt{s}$.}
\label{Plot:pt:kaons}
\end{figure}

\begin{figure}[hbtp]
\begin{minipage}[h]{0.49\linewidth}
\center{\includegraphics[width=1\linewidth]{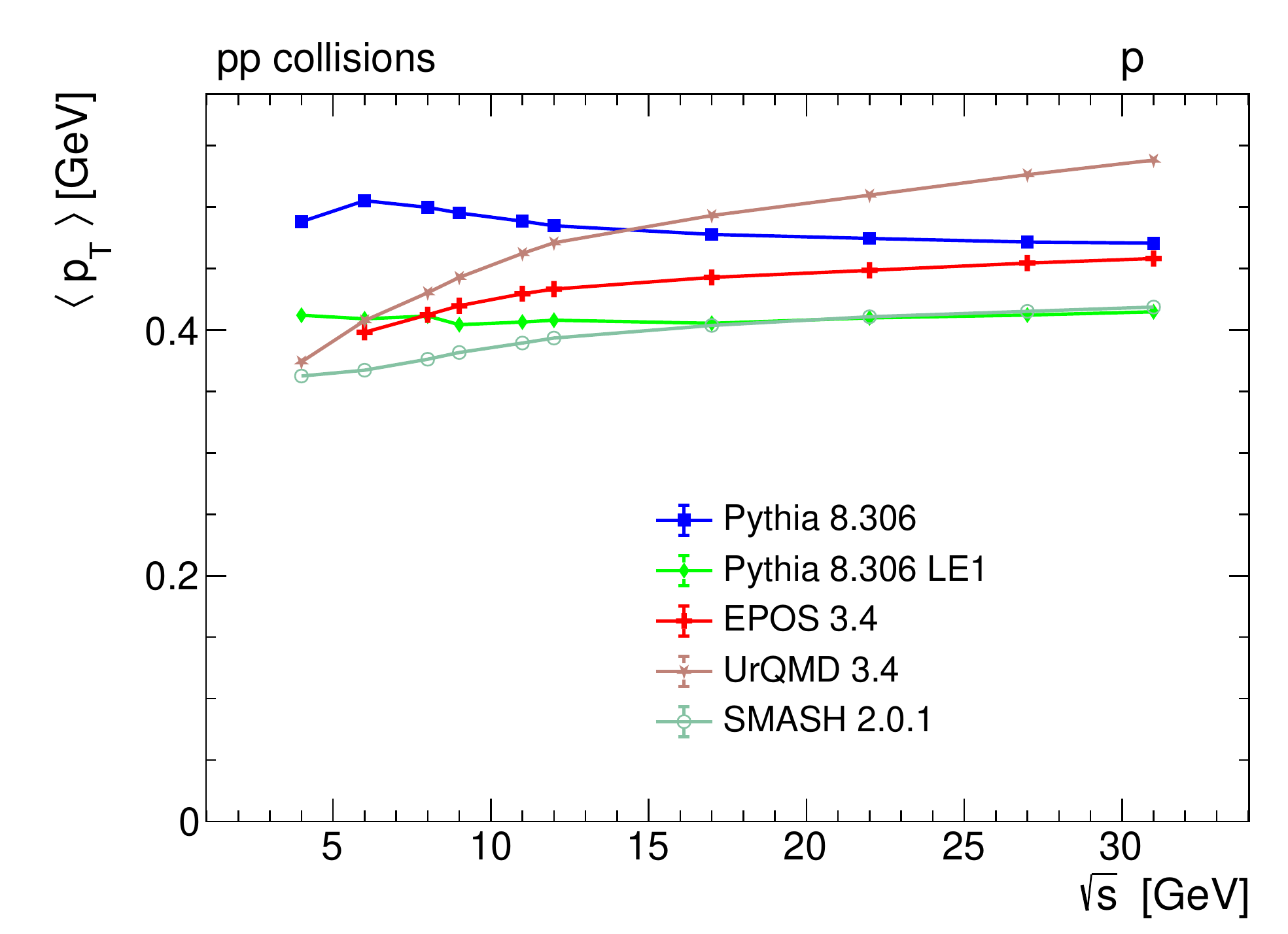} \\ }
\end{minipage}
\begin{minipage}[h]{0.49\linewidth}
\center{\includegraphics[width=1\linewidth]{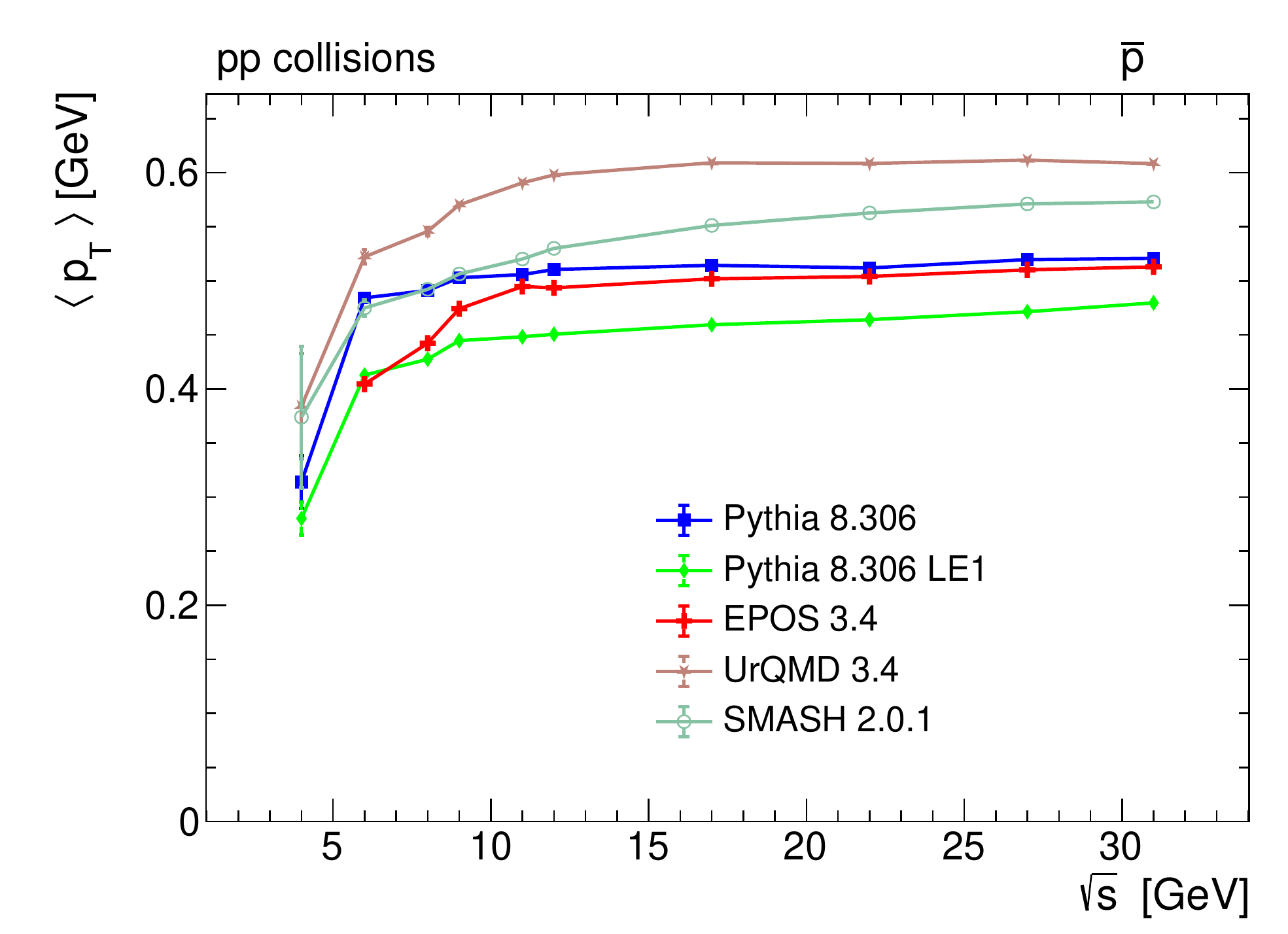} \\ }
\end{minipage}
\caption{Mean \pt of  p (left) and $\mathrm{\bar{p}}$ (right) as a function of $\sqrt{s}$.}
\label{Plot:pt:protons}
\end{figure}


Rapidity distribution of the produced particles obviously carries imprints of almost all physical mechanisms used in generators. Below we compare rapidity distributions of $\pi^{\pm}$, $K^{\pm}$ and $p(\bar{p})$ 
measured by NA61/SHINE experiment~\cite{Aduszkiewicz:2017sei} with the studied MC models. The comparison is performed at
$\sqrt{s}$~=~6.3, 12.3, and 17.3~GeV that covers major part of NICA energy ranges for pp collisions. However, there are no high quality data measured for higher energies achievable at NICA (i. e. \cmse up to 25 GeV).

Fig.~\ref{Plot:y:pions} shows rapidity distributions of $\pi^{+}$ and $\pi^{-}$ for the data and MC models of interest 
at $\sqrt{s}$~=~6.3, 12.3, and 17.3~GeV. All models and data show that 
the pion production rates have a maximum at $y$~=~0, an artifact of a distribution for perturbative $2->2$ parton scattering, and decline to a few percent level of maxima at  $y$~=~3. The main difference between $\pi^{+}$ and $\pi^{-}$ is 
reduced yield of the latter. 
Same conclusions apply to rapidity distributions of $\mathrm{K}^{+}$ 
and $\mathrm{K}^{-}$ which are shown in Fig.~\ref{Plot:y:kaons}.

Fig.~\ref{Plot:y:protons}\, shows rapidity distributions of protons and antiprotons for the data
measured and MC models of interest, at $\sqrt{s}$~=~6.3, 12.3, and 17.3~GeV. The proton distributions have 
broad maxima in rapidity range $1<\vert y \vert <2.5$ for data and all MC models. MC models show another narrower maximum at the edge of the rapidity distributions. Antiprotons
have a single maximum at $y$~=~0. Proton multiplicity is higher than the one for antiprotons and lighter kaons. 
Baryon number conservation plays a major role forcing beam remnants (which in the first approximation is considered to be a diquark) to fragment back to protons, caring large momentum fraction of the incoming protons. In contrast, antiproton are produced during the fragmentation of string stretched in hard scattering of valence quarks. 
Regarding MC models for proton distributions, \epos~3.4 slightly overshoots the height of the first maximum at all collision energies, while \pythia~8.306, \smashm ~2.0.1, and \urqmd~3.4 do not reflect data at very central rapidities and incorrectly describes the broad peaks at all studied collision energies. 
It should be noted that description of the longitudinal momentum of protons proved to be challenging for Fritiof-like models~\cite{Uzhinsky:2014kxa, Uzhinsky:2014jba}.

\begin{figure}[hbtp]
\begin{minipage}[h]{0.49\linewidth}
\center{\includegraphics[width=1\linewidth]{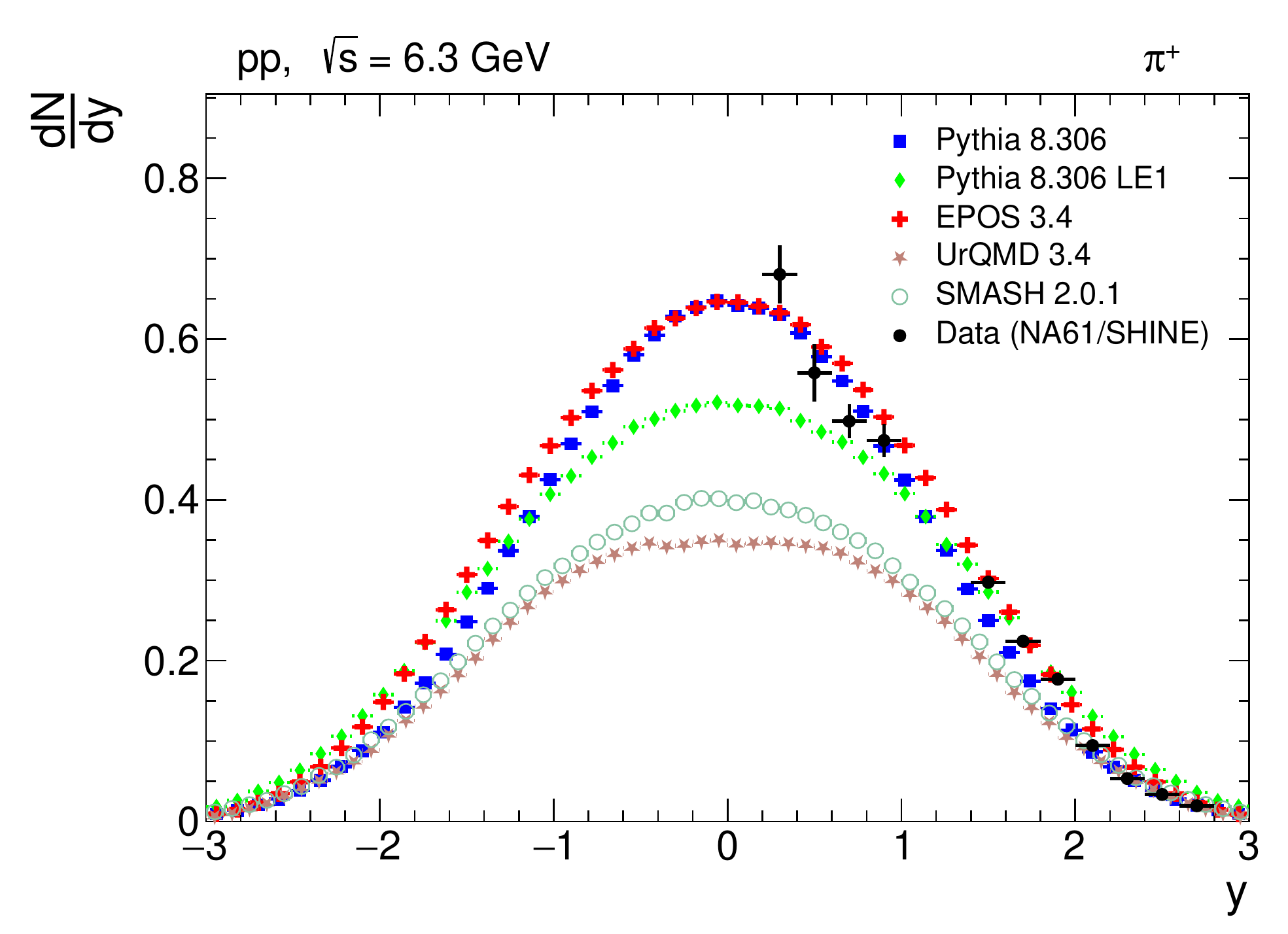} \\ }
\end{minipage}
\begin{minipage}[h]{0.49\linewidth}
\center{\includegraphics[width=1\linewidth]{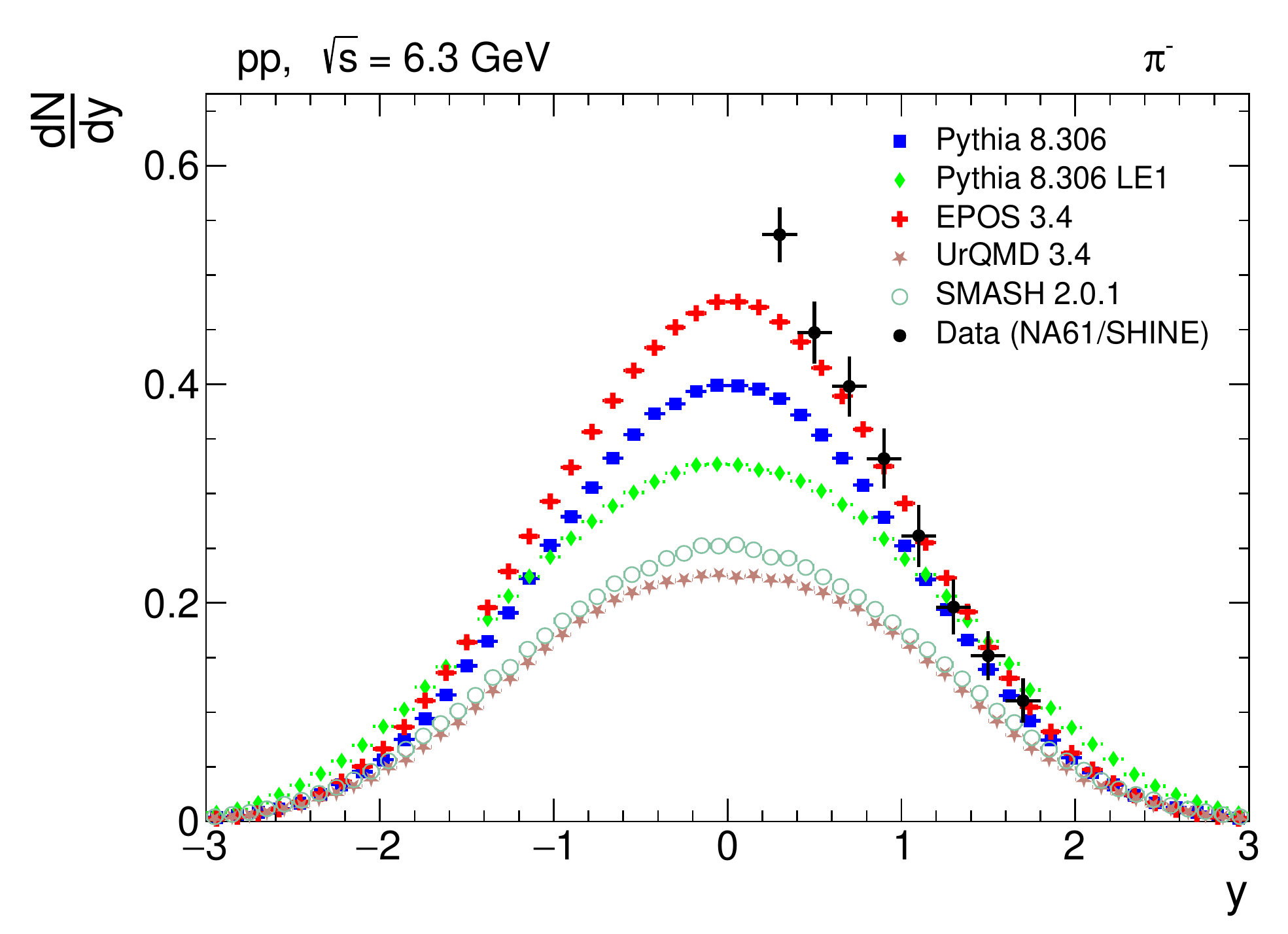} \\ }
\end{minipage}
\begin{minipage}[h]{0.49\linewidth}
\center{\includegraphics[width=1\linewidth]{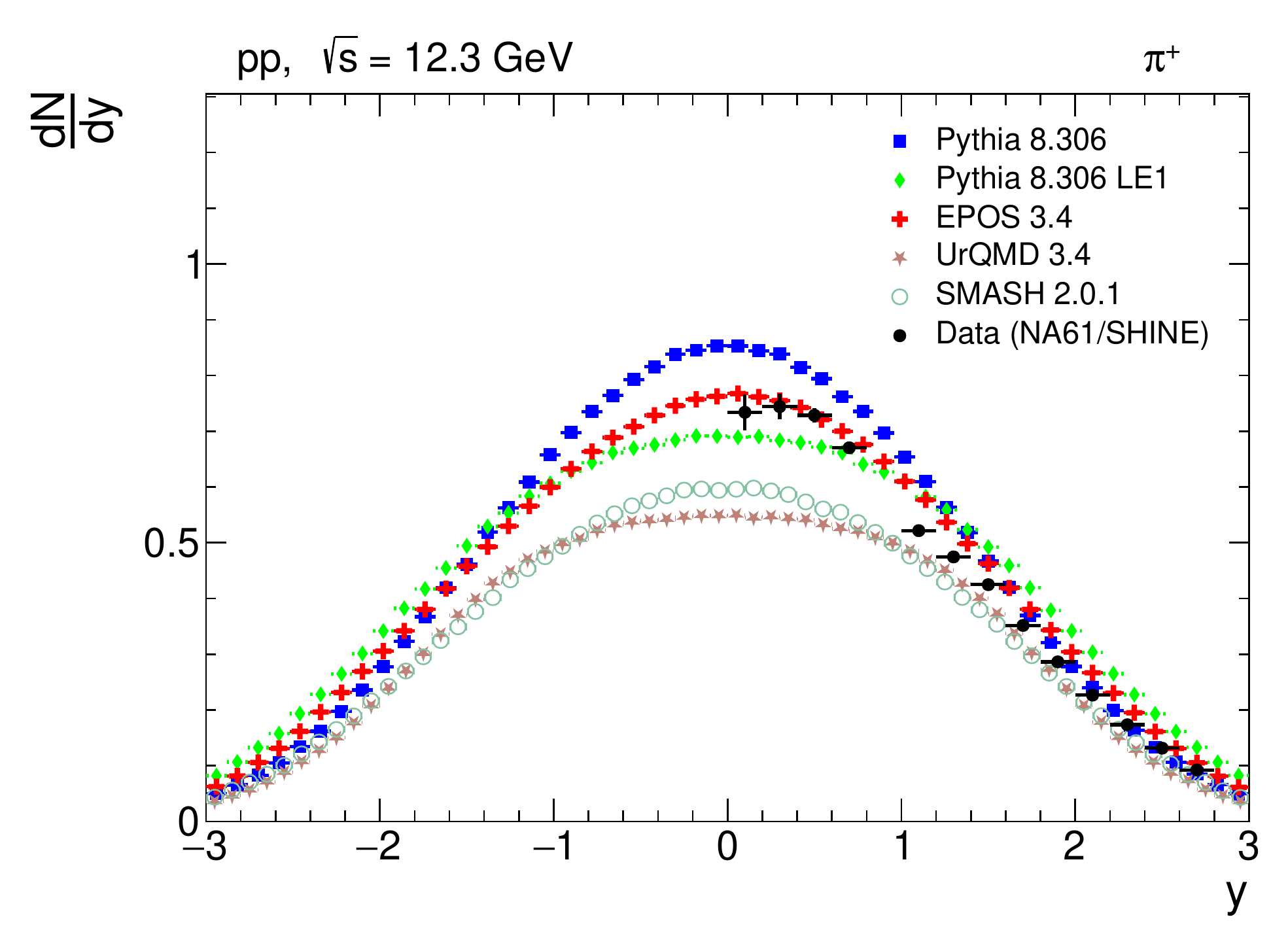} \\ }
\end{minipage}
\begin{minipage}[h]{0.49\linewidth}
\center{\includegraphics[width=1\linewidth]{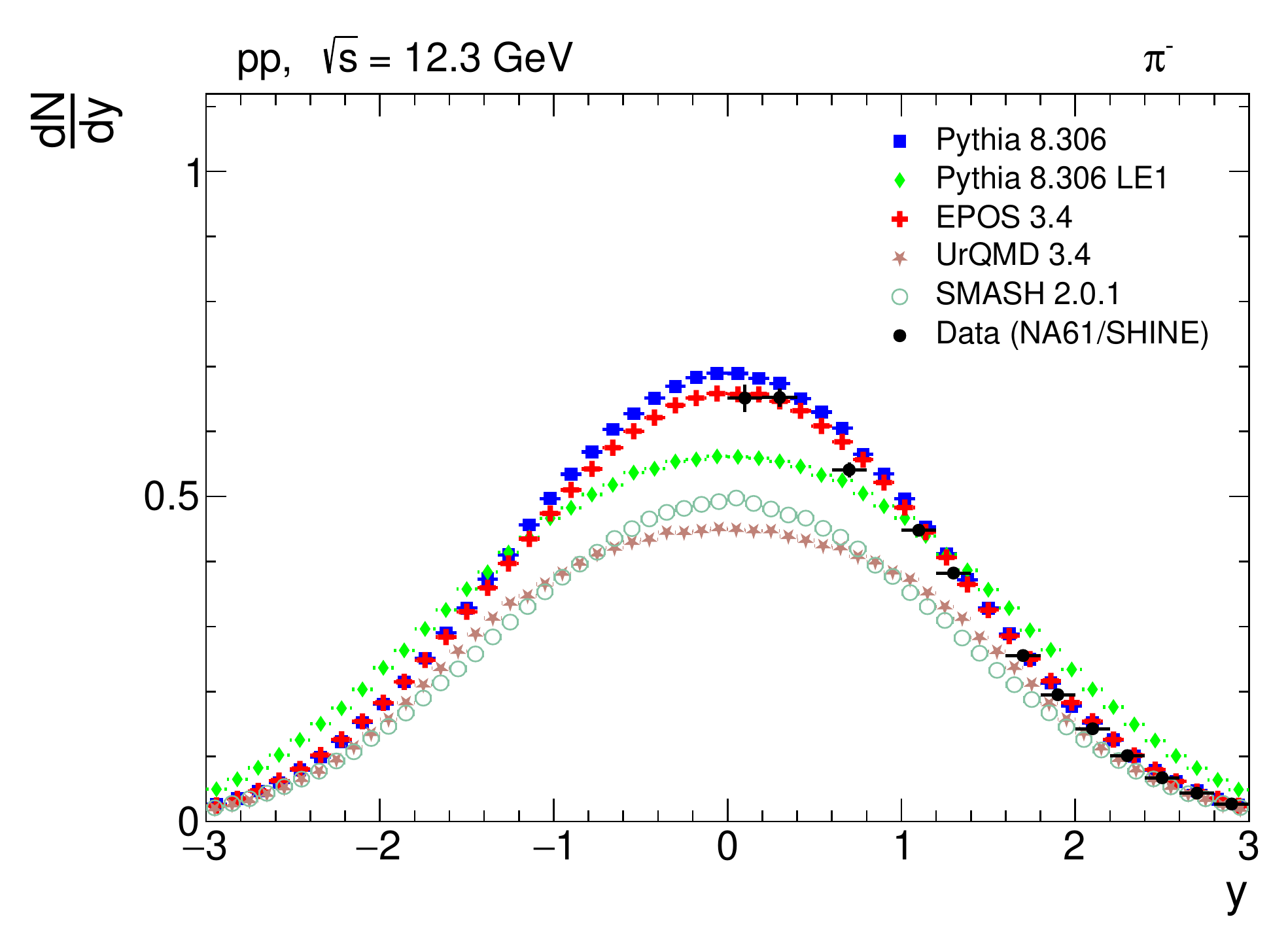} \\ }
\end{minipage}
\begin{minipage}[h]{0.49\linewidth}
\center{\includegraphics[width=1\linewidth]{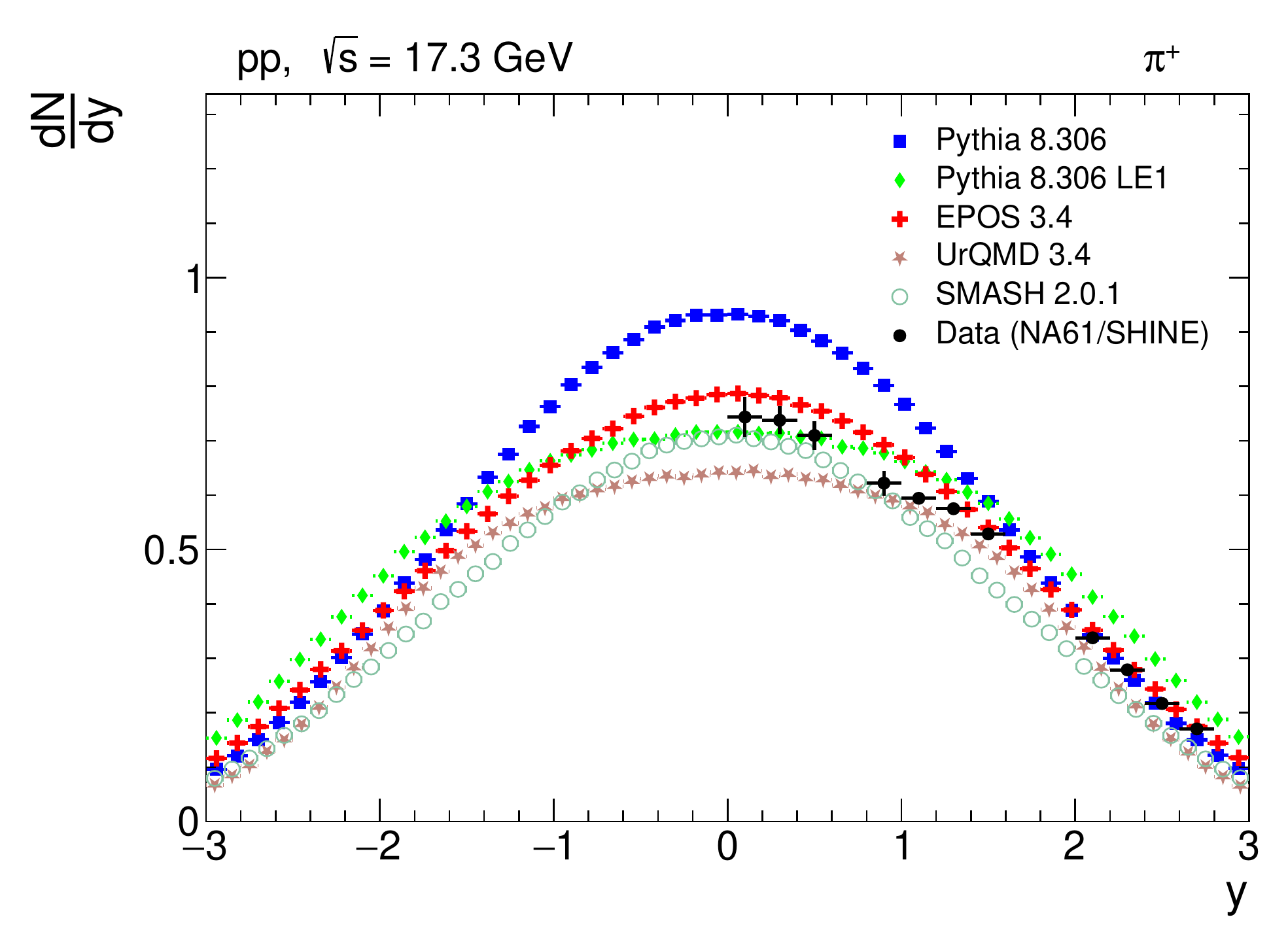} \\ }
\end{minipage}
\begin{minipage}[h]{0.49\linewidth}
\center{\includegraphics[width=1\linewidth]{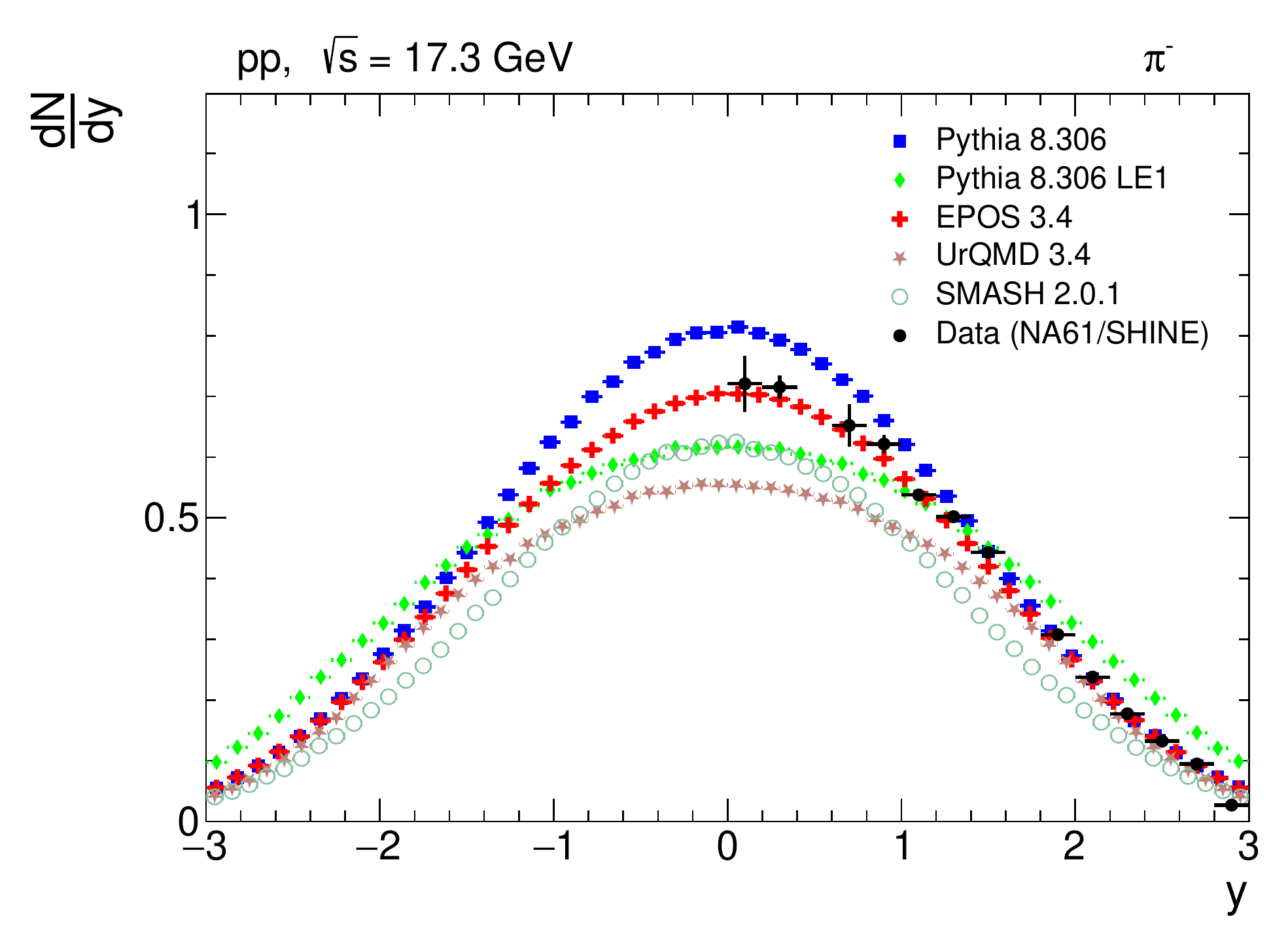} \\ }
\end{minipage}
\caption{Rapidity distributions of $\pi^{+}$ (left) and  $\pi^{-}$ at $\sqrt{s}$~=~6.3~GeV (top row), 
$\sqrt{s}$~=~12.3~GeV (middle row), and $\sqrt{s}$~=~17.3~GeV (bottom row).}
\label{Plot:y:pions}
\end{figure}

\begin{figure}[hbtp]
\begin{minipage}[h]{0.49\linewidth}
\center{\includegraphics[width=1\linewidth]{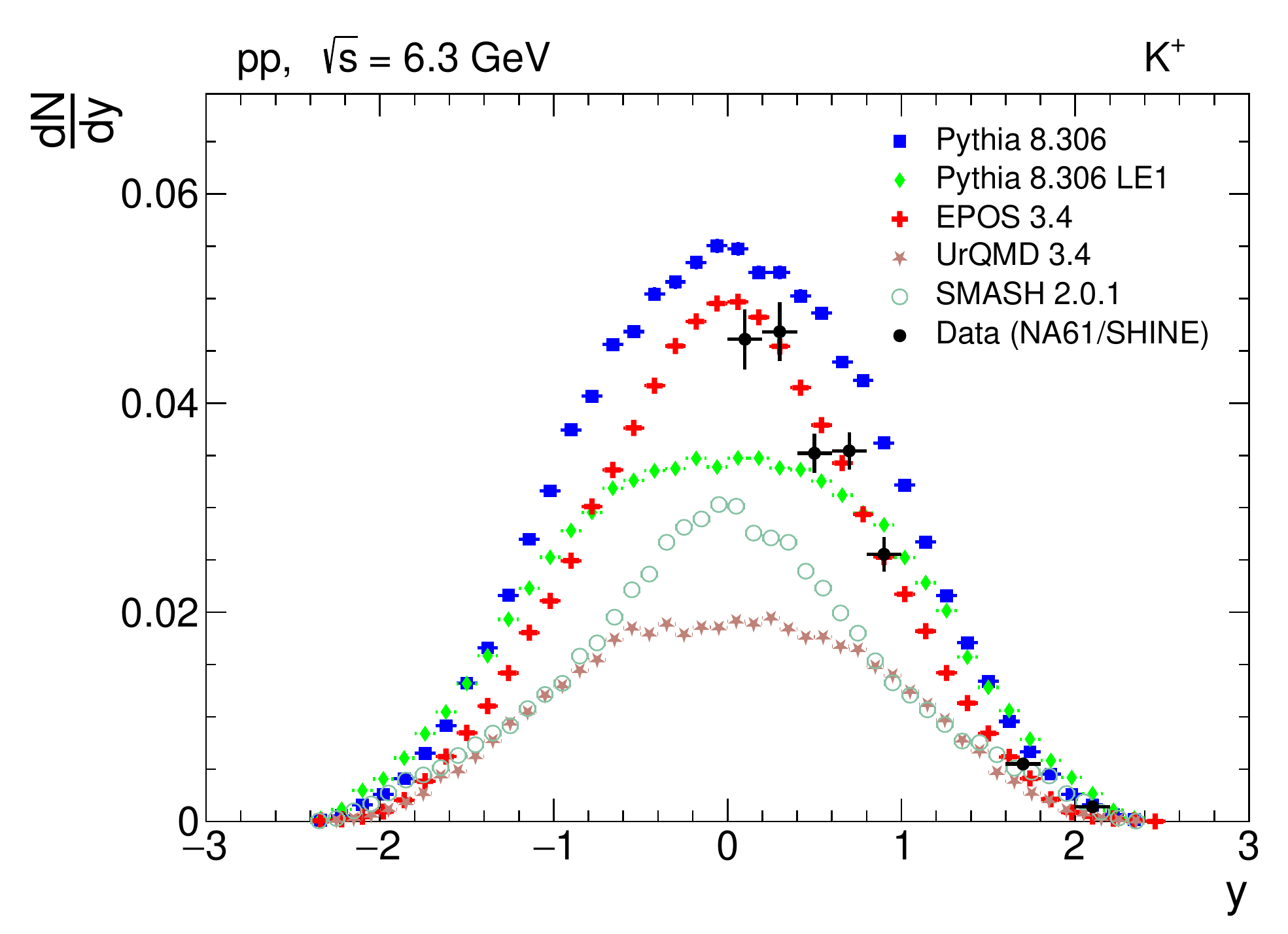} \\ }
\end{minipage}
\begin{minipage}[h]{0.49\linewidth}
\center{\includegraphics[width=1\linewidth]{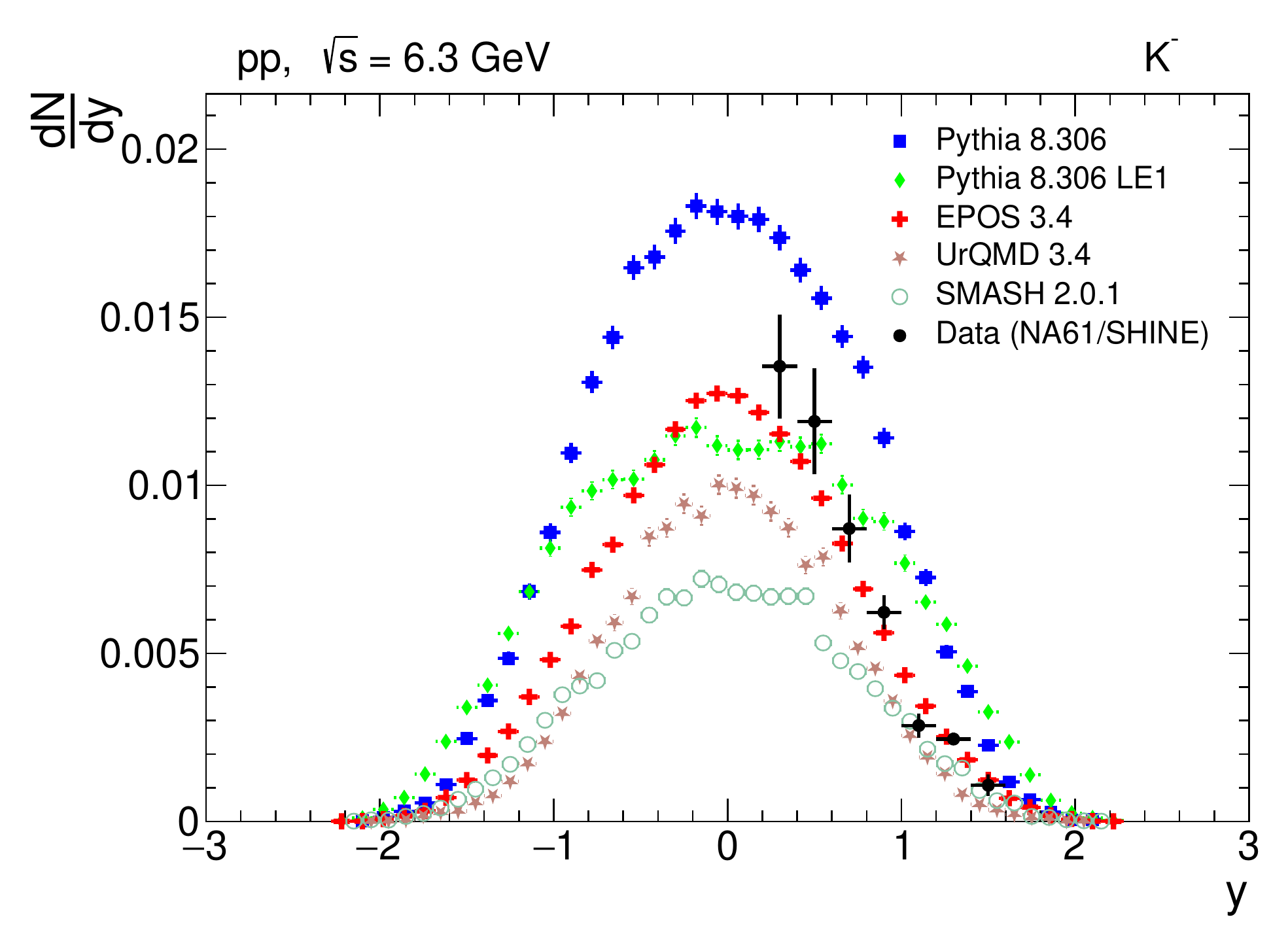} \\ }
\end{minipage}
\begin{minipage}[h]{0.49\linewidth}
\center{\includegraphics[width=1\linewidth]{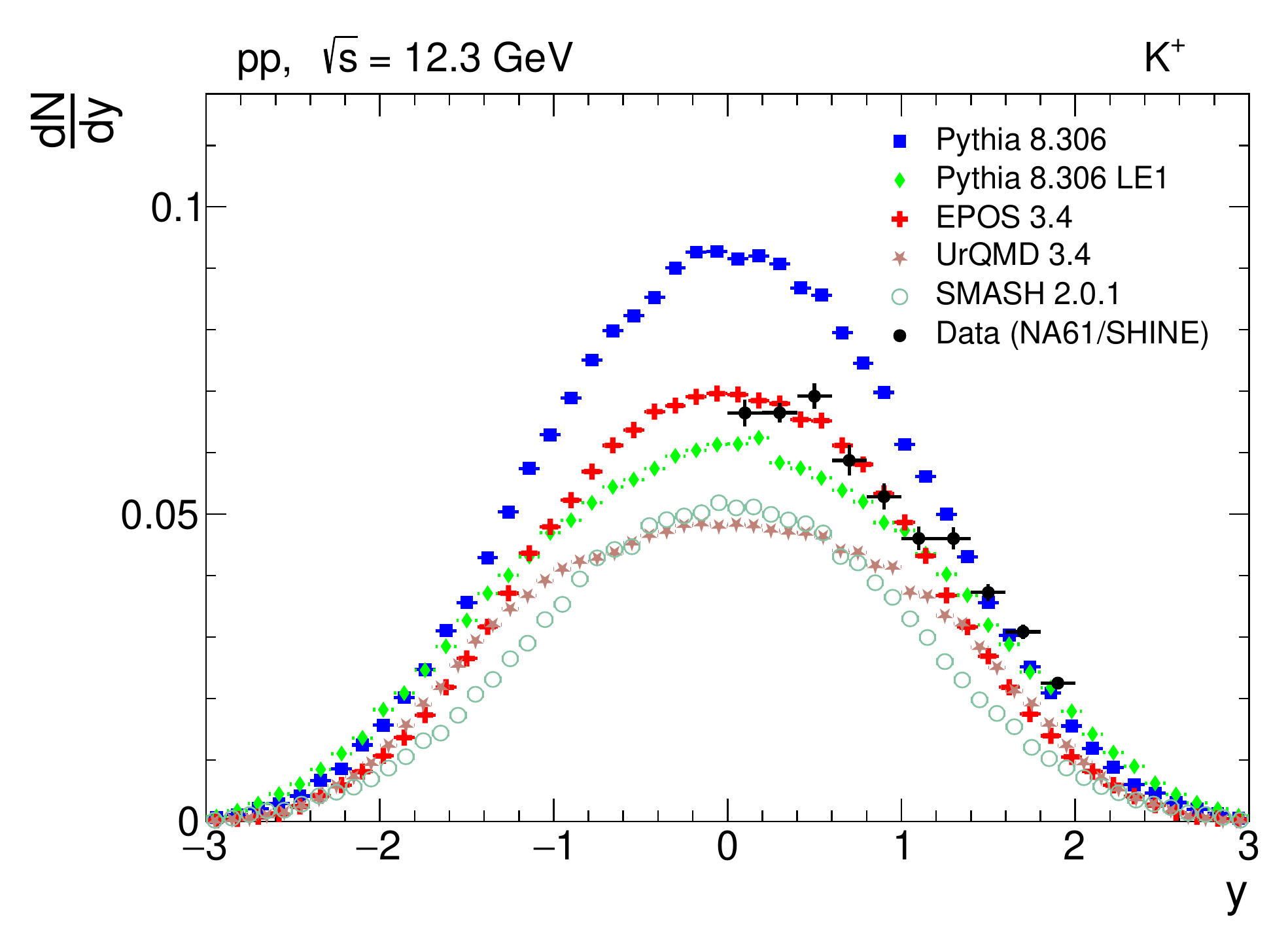} \\ }
\end{minipage}
\begin{minipage}[h]{0.49\linewidth}
\center{\includegraphics[width=1\linewidth]{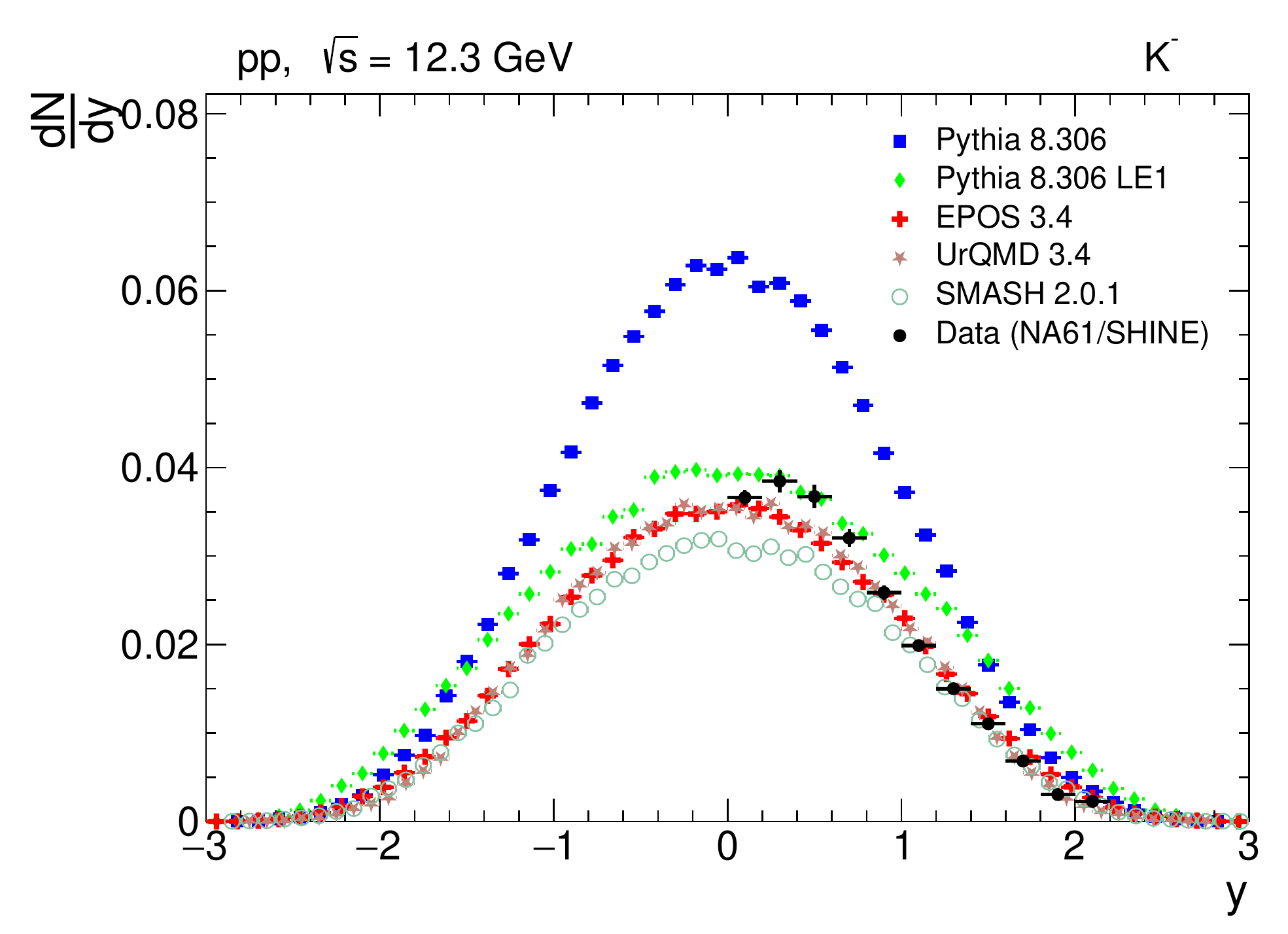} \\ }
\end{minipage}
\begin{minipage}[h]{0.49\linewidth}
\center{\includegraphics[width=1\linewidth]{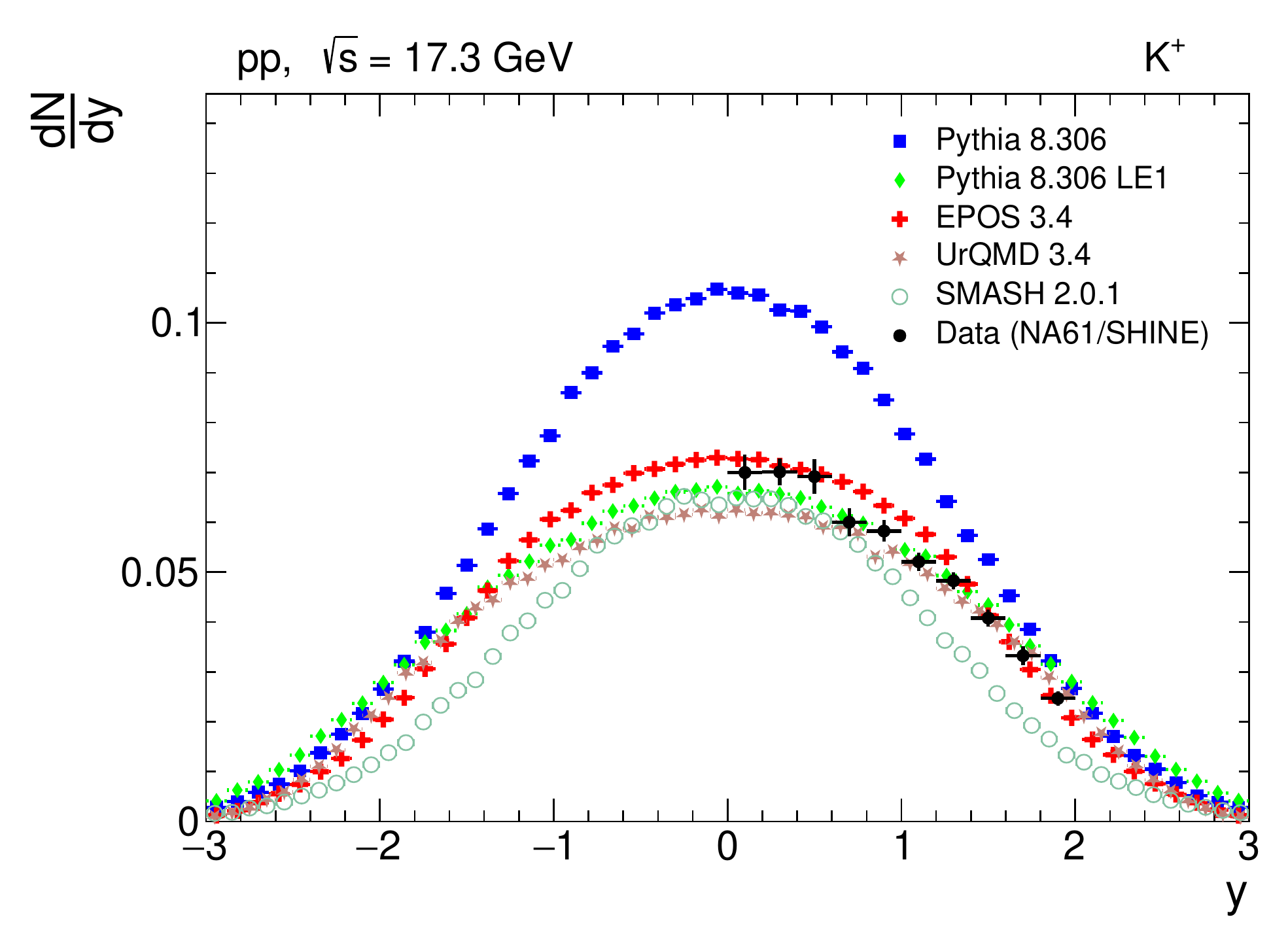} \\ }
\end{minipage}
\begin{minipage}[h]{0.49\linewidth}
\center{\includegraphics[width=1\linewidth]{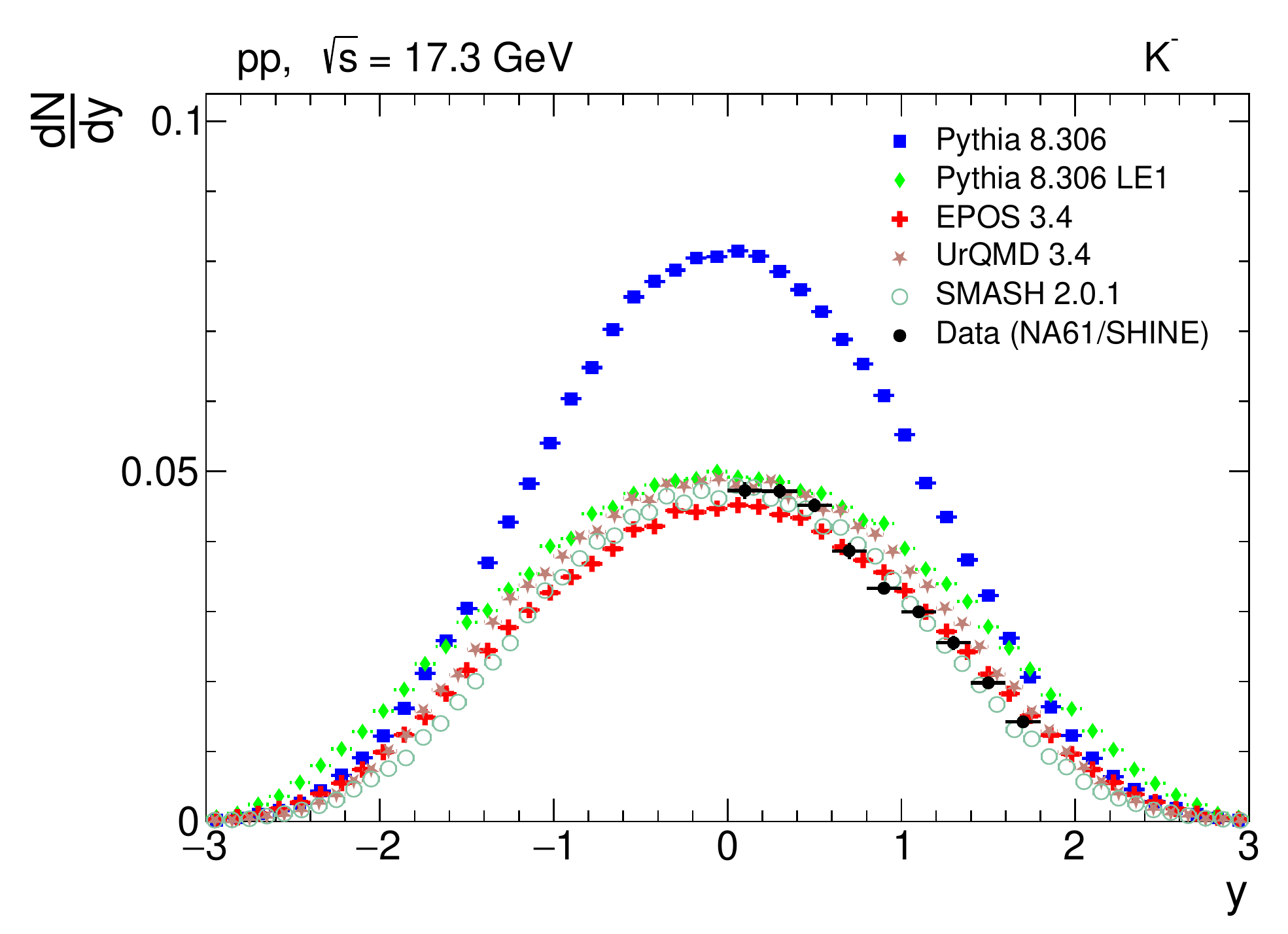} \\ }
\end{minipage}
\caption{Rapidity distributions of  $\mathrm{K}^{+}$ and  $\mathrm{K}^{-}$ at $\sqrt{s}$~=~6.3~GeV (top row), 
$\sqrt{s}$~=~12.3~GeV (middle row), and $\sqrt{s}$~=~17.3~GeV (bottom row).}
\label{Plot:y:kaons}
\end{figure}

\begin{figure}[hbtp]
\begin{minipage}[h]{0.49\linewidth}
\center{\includegraphics[width=1\linewidth]{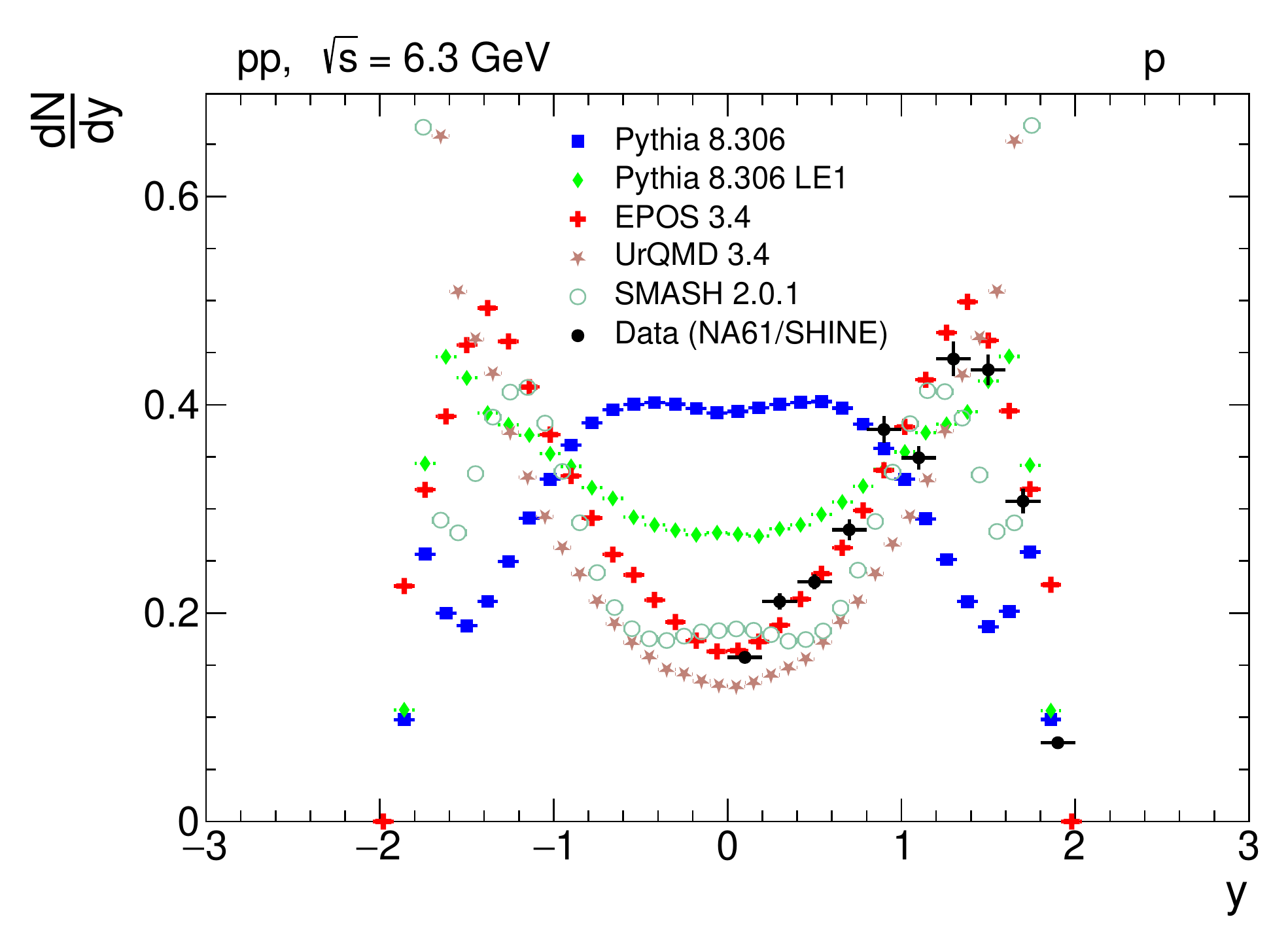} \\ }
\end{minipage}
\begin{minipage}[h]{0.49\linewidth}
\center{\includegraphics[width=1\linewidth]{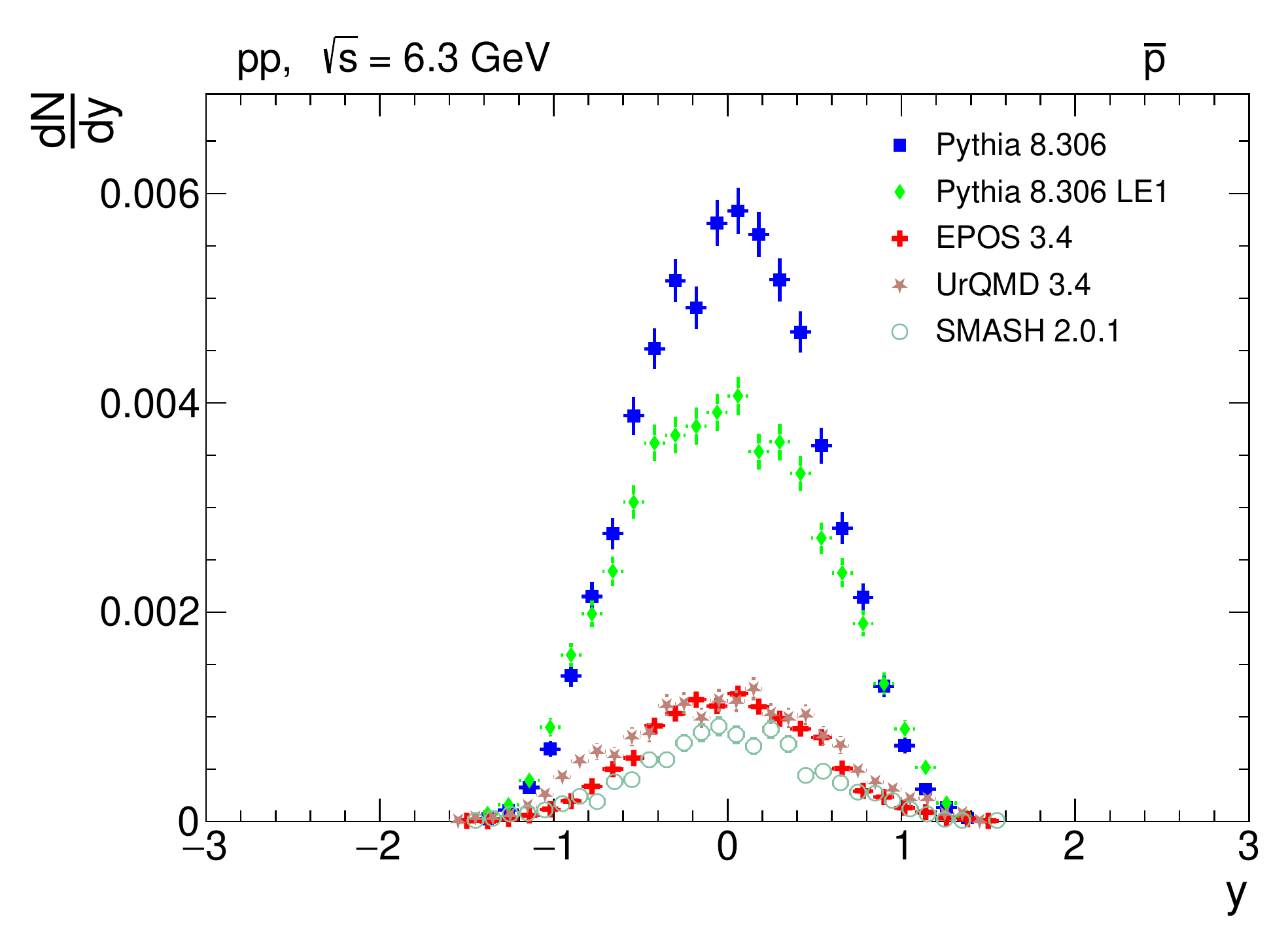} \\ }
\end{minipage}
\begin{minipage}[h]{0.49\linewidth}
\center{\includegraphics[width=1\linewidth]{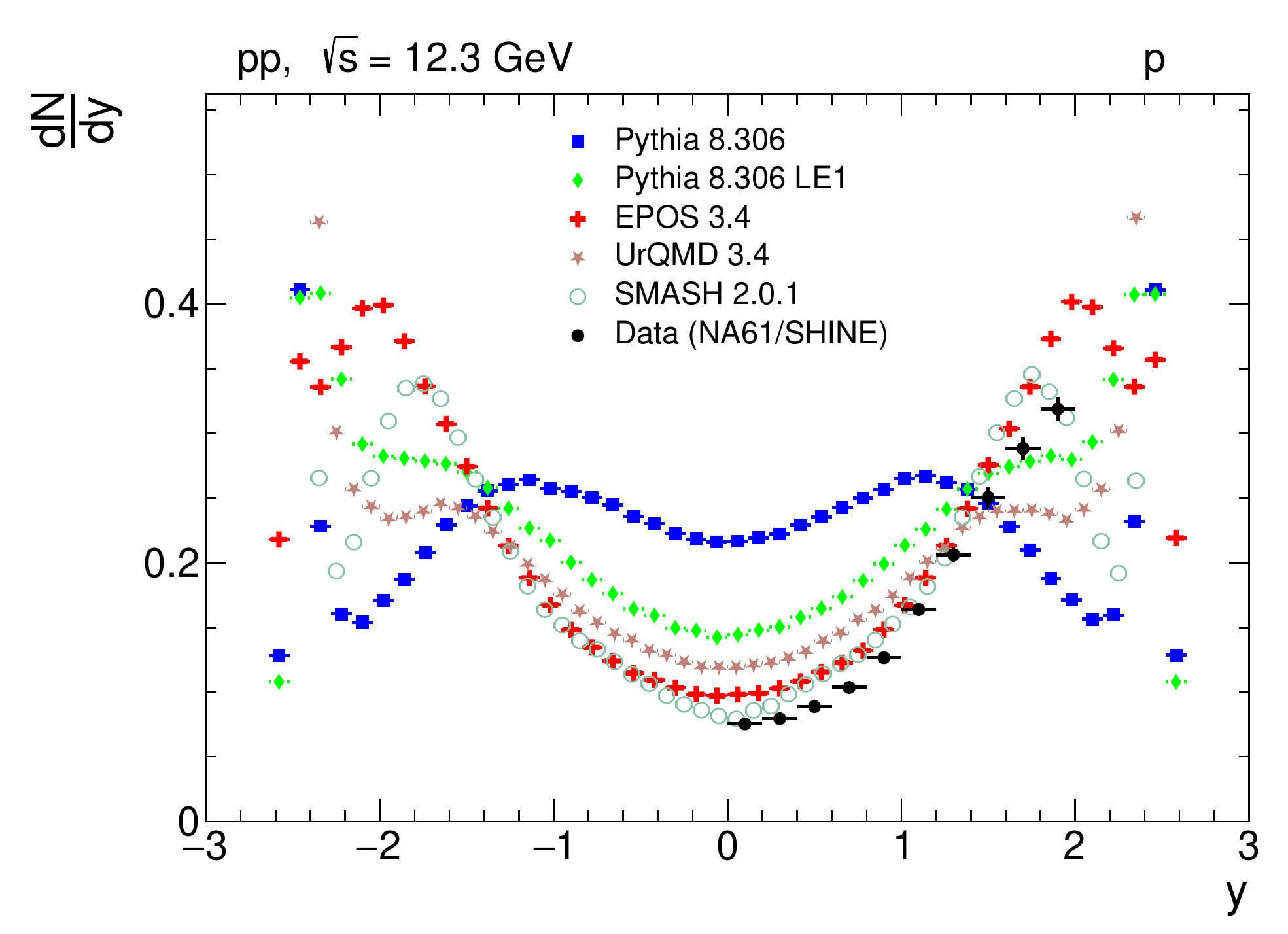} \\ }
\end{minipage}
\begin{minipage}[h]{0.49\linewidth}
\center{\includegraphics[width=1\linewidth]{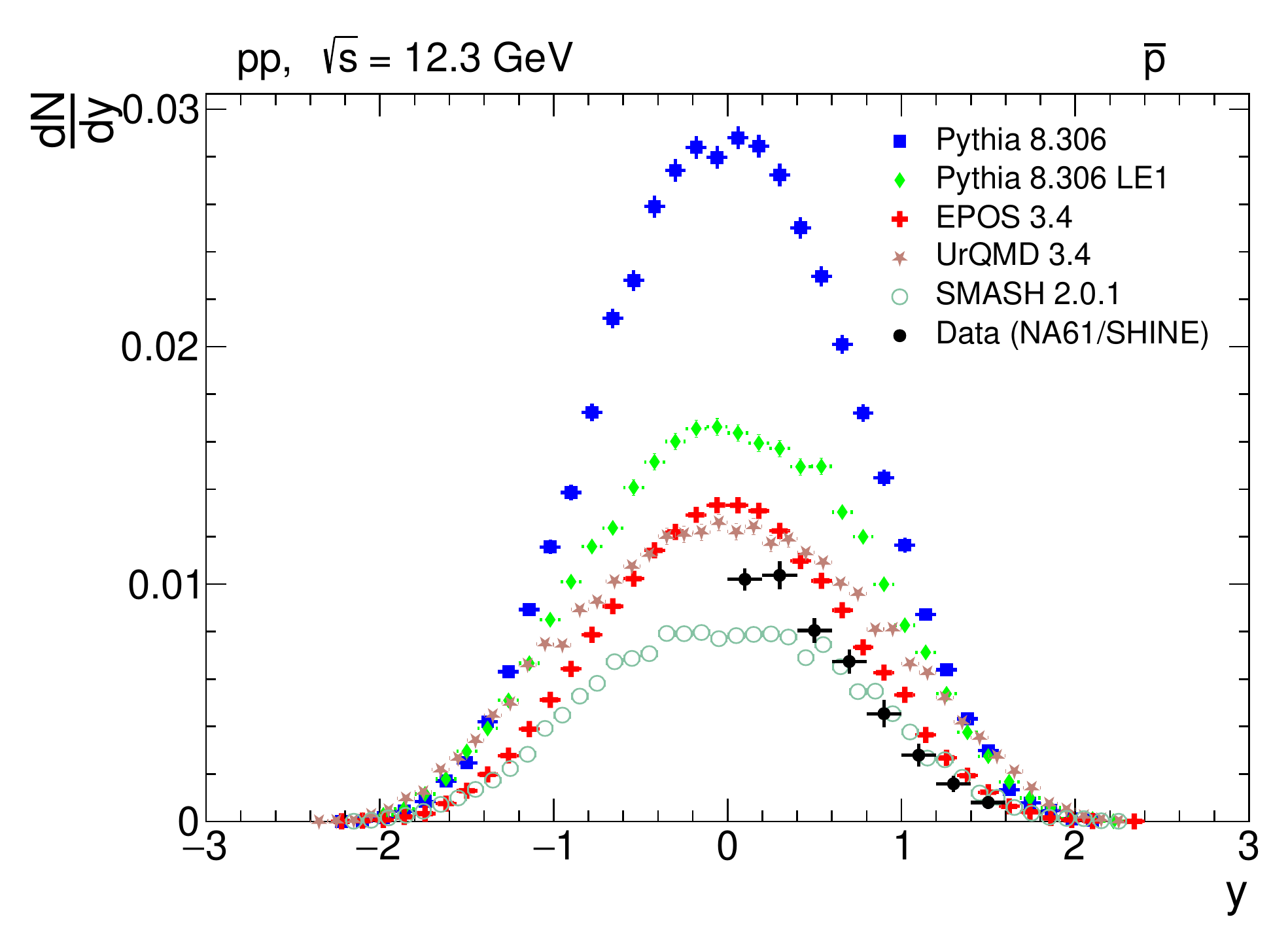} \\ }
\end{minipage}
\begin{minipage}[h]{0.49\linewidth}
\center{\includegraphics[width=1\linewidth]{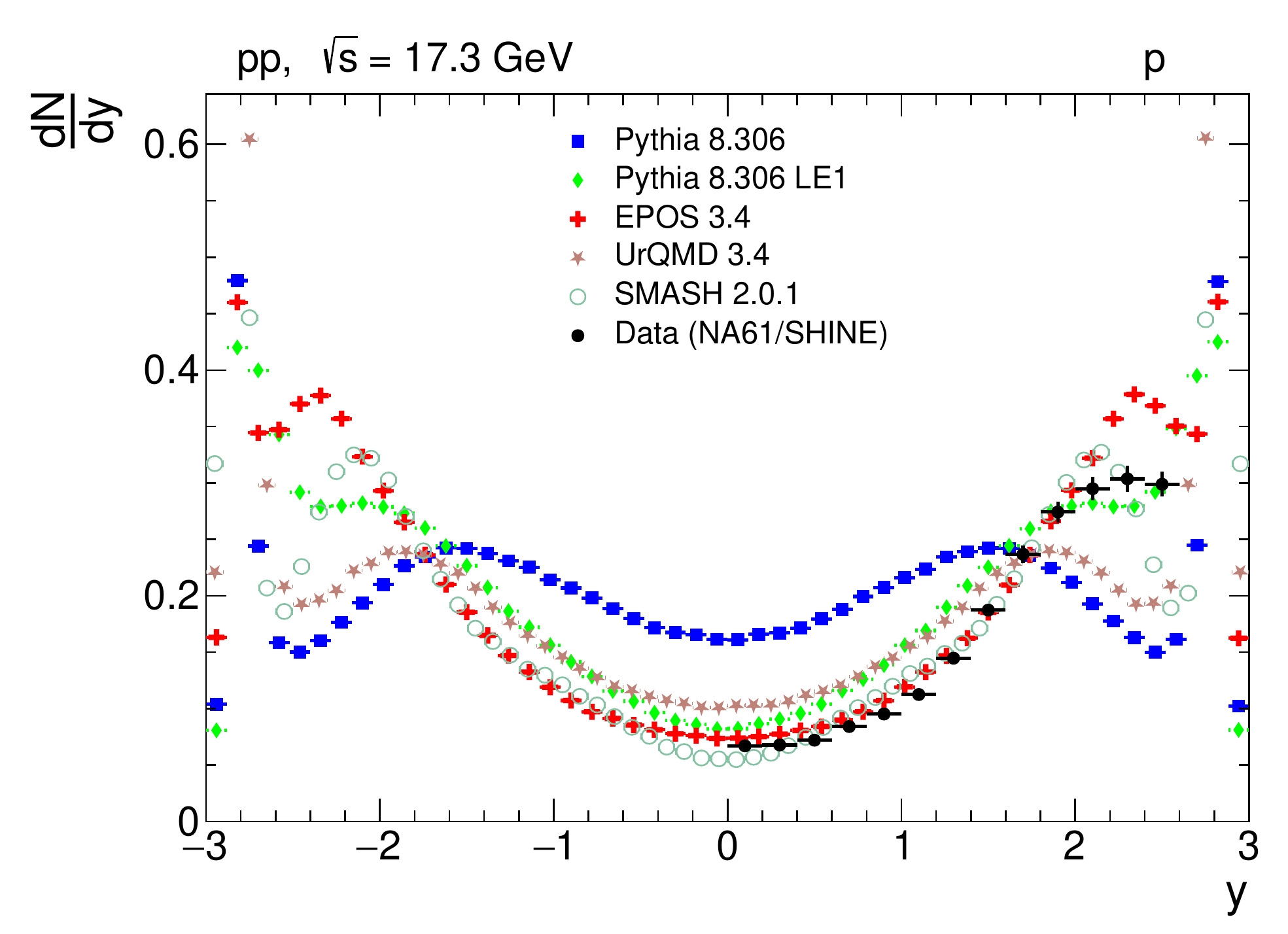} \\ }
\end{minipage}
\begin{minipage}[h]{0.49\linewidth}
\center{\includegraphics[width=1\linewidth]{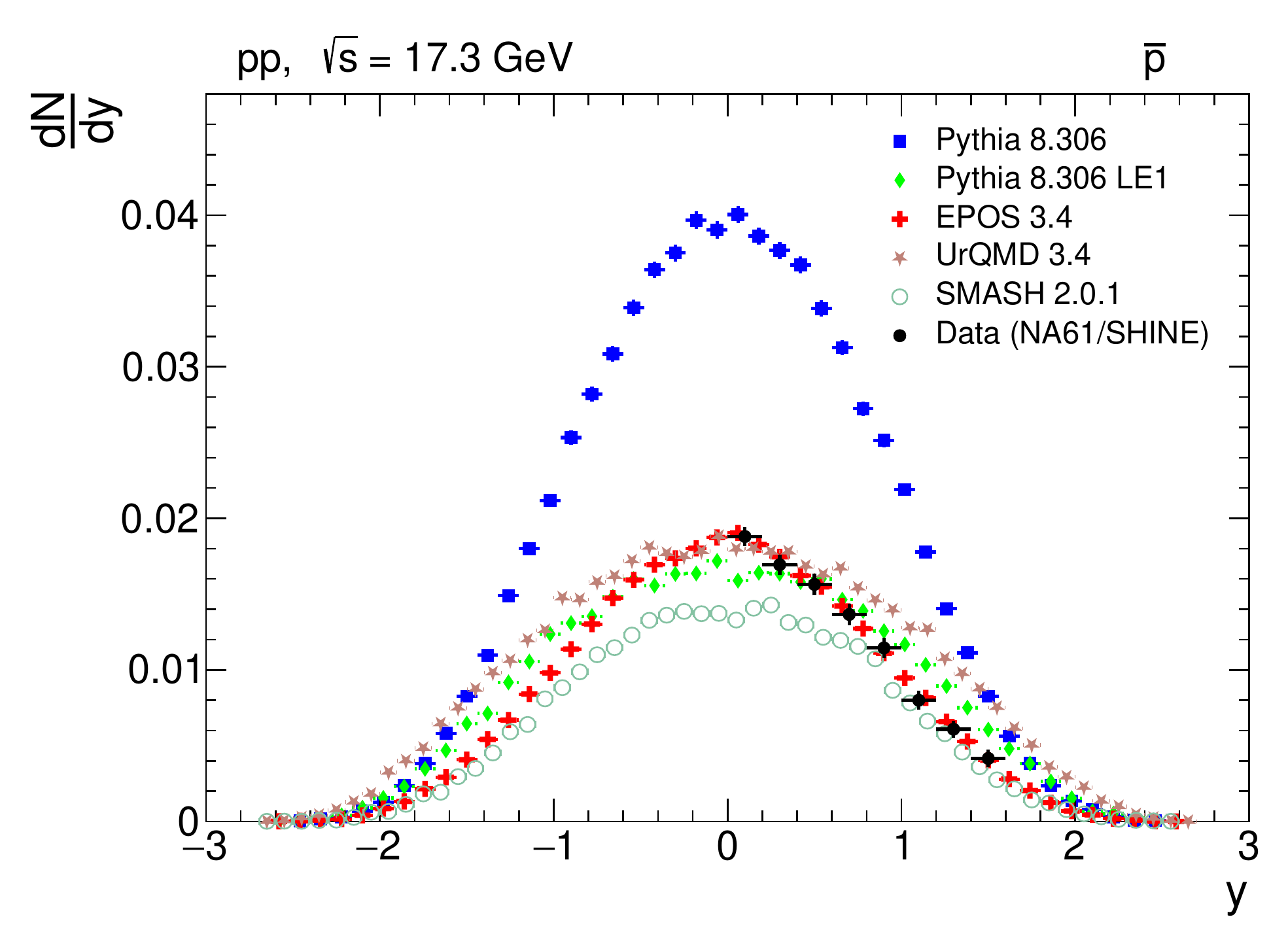} \\ }
\end{minipage}
\caption{Rapidity distributions of protons and antiprotons at $\sqrt{s}$~=~6.3~GeV (top row), 
$\sqrt{s}$~=~12.3~GeV (middle row), and $\sqrt{s}$~=~17.3~GeV (bottom row). 
Antiproton rapidity distribution for  \pythia~8.306 at $\sqrt{s}$~=~6.3~GeV is not shown since it is completely off scale.}
\label{Plot:y:protons}
\end{figure}

Let us now outline here what these data tell us about issues with mechanisms of hadron production in different generators and parameters of these models that may be a subject for tuning to improve it.

\begin{itemize}
		
\item Fig.~\ref{Plot:y:protons} reveals issues with the description of peaks in rapidity distribution for all generators as it was mentioned above. The broad peak is connected to the handling beam remnants. As we see the model for beam remnants in \pythia, \smashm, \urqmd in present form is not able to adequately describe data. The second sharp peak at the very edge of rapidity distributions at $\sqrt{s}$~=~6.3, 12.3, and 17.3~GeV is presented in all 4 generators (i. e. also present in \epos) and is connected to the mechanism of diffraction. 

\item In contrast, antiproton are produced in the fragmentation of string stretched in hard scattering of valence quarks.  In \pythia, discrepancy in antiproton multiplicities are most likely connected to the \pythia's fragmentation to baryons since the overshooting for mesons is not that dramatic. 
			
\item Both \smashm and \urqmd generators underestimate particle production out of resonance region, 
while at the same time overestimate diffractive cross-section. \urqmd also overestimates
antiproton production in energy range $6 - 20$ GeV.
Most likely tuning the parameters of soft-string production (see Sec.~\ref{Sec:MCmodels:urqmd} 
and Sec.~\ref{Sec:MCmodels:smashm}) will allow to better describe data at these energies.  
		
\item  As it can be seen from Fig.~\ref{Plot:y:protons} for protons \smashm provides a reasonable description of measured rapidity distributions except for few highest-$y$ points (i. e. diffraction region). This advantage is probably due to the dedicated fragmentation function for leading baryons which are usually protons at these energies.

\end{itemize}

\subsubsection{Pre-tuning \pythia 8.306 for NICA energies}	
\label{Sec:Res:Tuning}
	
As it was mentioned earlier, the standard version of  8.306 for simulation of minimum bias physics was tuned at high energies (Tevatron, LHC). 
Here, we show that varying key parameters can significantly improve the generator performance at \cmse = 17.3 GeV.  	

As it can be seen in Sec.~\ref{Sec:Res:PY} there is a significant discrepancy between the standard version of \pythia 8.306 and data. 
Most notably, \pythia~8.306 with default parameter values shows higher particle density at central rapidities and 
a lack of longitudinal momentum for protons. Therefore, it is natural to start with adjusting parameters 
for physics of TED. The standard configuration of \pythia~8.306 relies on SaS/DL model, however it  unsatisfactory describes the data
in this energy range. \pythia 8.306 offers an option to use ABMST model, which has been fitted to data in the energy range 17.2~$<$~\cmse~$<$~546~GeV,
and its extended version (see Sec.\ref{Sec:MCmodels:pythia}) that addresses known issues and is more flexible for tuning.
Therefore, we have chosen the extended ABMST model as a starting point in our attempt to bring \pythia 8.306 closer to data at NICA energies. 
In order to better describe the data, we tuned some diffraction parameters. 
We reduced single-diffraction cross-section by setting $k$-factor (i.e. parameter \texttt{SigmaDiffractive:ABMSTmultSD}) to $0.75$
and narrowed range for damping of small-rapidity gap event, i.e. partially rolling back to the original ABMST model). 
Values of the tuned parameters (parameter \texttt{SigmaDiffractive:ABMSTmultSD})) along with the default values are listed in Tab.~\ref{tab:Tune1}. 

As it is mentioned above, \pythia 8.306 overestimates yields of most particles in central rapidities (i.e. pions, kaons, antiprotons), indicating that \pythia overestimates the average number of pertubative parton interactions.
To address this issue, we adjust $p_{\rm T0}$, value that regularizes infrared divergence of parton $2 \rightarrow 2$ scattering (see (\ref{ptdependence})).
It is fair to say that it is a pretty standard step for a tuning procedure in minimum-bias physics of hadron collisions.
In order to decrease a number of interactions, one have to increase $p_{\rm T0}$.
We adjust $p_{T 0}$ so that it would be as large as possible at \cmse = $17.3$ GeV, 
i.e. $\sigma_{\rm int} \approx \sigma_{\rm in \, ND}$. Thus, it minimizes a bias in generation of events.
In practice, we reduce parameter ``a'' (see formula (\ref{ptdependence})) to increase $p_{\rm T0}$ and make dependence 
$\sigma_{\rm int}$ on \cmse flatter, trying to mimic the dependence of $\sigma_{\rm in \, ND}$ on \cmse. 
However, it should be noted that $p_{T 0}$ value should be readjusted for all energies since power-like dependence on $\sqrt{s}$  is not fundamental and does not reflect the behavior of $\sigma_{\rm in \, ND}$ on \cmse at low energies.

As we can see from multiplicity and pseudorapidity distribution \pythia predicts higher values of multiplicities of both strange
particles and antiprotons. For that reason two string parameters were readjusted, namely \texttt{:probStoUD} and
\texttt{StringFlav:probQQtoQ}. The former is a suppression factor of production of $s$-quark relatively to $u$ and $d$ during the breakup of strings,
the latter is a suppression factor of diquark production. 

Another important parameter in context of this study is a transverse momentum acquired by quark and antiquark at each string breaking.
By default \pythia 8.306 uses Gaussian distribution with width given by parameter \texttt{StringPT:SigmaPt}.
We find that moderate decreasing the values of this parameter improves agreement with data.
To further improve the rapidity distributions for protons we changed the model for beam remnants to the new  model, 
which uses colour rules (see Sec.~\ref{Sec:MCmodels:pythia}), i. e. we put parameters of \texttt{BeamRemnants:remnantMode} 
and \texttt{ColourReconnection:mode} to 1. Also, it was found out that decreasing primordial $k_{\rm T}$ also improves the distributions. 
Summary of tuned parameters is listed in the table \ref{tab:Tune1}. 

\begin{table}[tp]
	\centering\small 
	\begin{tabular}{llll}
		\bf Parameter & \bf LE1 & \bf (Default) & \bf Comment
		\\\toprule 
		\texttt{SigmaTotal:mode}   & 3 &  1 & The ABMST parametrizations 
		\\                  &   &    & for pp and p$\bar{p}$
		\\\hline
		
		\texttt{SigmaDiffractive:ABMSTmodeSD}   & 3   &  1      & The modified ansatz diffraction 
		\\                                &     &         & model for he ABMST parametrizations 
		\\\hline
		
		\texttt{SigmaDiffractive:ABMSTmultSD} &  0.75   &  1.0      & The scaling factor for SD cross sections
		\\\hline
		
		\texttt{SigmaDiffractive:ABMSTdampenGap} &  on   &  on      & It allow to dampen small rapidity-gap
		\\                                  &   &            & diffractive events
		\\\hline
		
		\texttt{SigmaDiffractive:ABMSTygap} & 1.0   & 2.0           & Damping diffractive events 
		\\                                  &   &            & with rapidity gap $\vert y \vert < ABMSTygap$
		\\\hline
		\texttt{MultipartonInteractions:pT0Ref} & 2.28   & 2.28      &$p_{\perp 0}$ at \cmse~=~7 TeV
		\\\hline
		\texttt{MultipartonInteractions:ecmPow} & 0.152   & 0.215      & This gives $p_{\perp 0}$~=~0.92~GeV 
		\\                                &        &            & at \cmse~=~17.3 GeV
		\\\hline
		
		\texttt{StringFlav:probStoUD}  & 0.165 & 0.217  & the suppression of s quark production 
		\\                       &       &        & relative to ordinary u or d one.
		\\\hline
		
		\texttt{StringFlav:probQQtoQ}   & 0.034   & 0.081  & The suppression of diquark production
	    \\                       &       &          & with respect to quarks.
		\\\hline
		\texttt{StringPT:sigma} & 0.29   & 0.335   & The width of \pt acquired in the 
		\\                      &        &            & fragmentation process
		\\\hline
		\texttt{BeamRemnants:remnantMode}    & 1 & 0  & New model using colour rules from QCD
		\\\hline
		
		\texttt{ColourReconnection:mode}  & 1   & 0      & New model using colour rules from QCD
		\\\hline
		\texttt{BeamRemnants:primordialKTsoft} &  0.6  & 0.9      & $\sigma$ of primordial $k_{\mathrm T}$ in the soft-interaction
		\\                       &       &         &  limit.
		\\\hline
		\texttt{BeamRemnants:primordialKTremnant} & 0.3   & 0.4      & $\sigma$ of primordial $k_{\mathrm T}$ of beam-remnant
		\\                       &       &         & partons.
		\\\bottomrule
	\end{tabular}
	\caption{Tuned parameters \label{tab:Tune1}}
\end{table} 

\subsection{Two-particle angular correlations}
\label{Sec:Res:Cor}

Two-particle angular correlation functions have been serving as an indispensable tool for studying particle production in both pp and relativistic heavy-ion collisions.
In this paper we study angular correlations between stable ($c\tau \xspace >$~1~cm) 
charged particles in pp collisions for different MC models at NICA energies and compare them with data from NA61/SHINE.
Four distinct sources of correlations at NICA energies may be highlighted. The first one is due to the string fragmentation. These processes mostly result in relatively short-range rapidity correlations. The second one is due to decay of an intermediate state. The third one comes from the fragmentation of beam remnants to hadrons.
The fourth one is a diffraction that results in long-range rapidity correlations. At highest NICA energies jet-induced correlations, MPI correlations, and ISR or FSR-induced correlations may manifest.

We follow a standard approach~\cite{Eggert:1974ek} (and also used in NA61/SHINE) to compute two-particle angular correlation, which is defined as follows:

\vspace{-0.4cm}
\begin{equation}
\label{2pcorr_incl}
C(\Delta\eta,\Delta\phi) =
\left<\left(\frac{S_{N}(\Delta\eta,\Delta\phi)}
{B_{N}(\Delta\eta,\Delta\phi)}\right)\right>_{events}
\end{equation}

\noindent where $S_N$ and $B_N$ are the signal and
random background distributions defined in Eqs.~(\ref{2pcorr_signal}) and
(\ref{2pcorr_background}) respectively,
$\Delta\eta=(\eta_1-\eta_2)$ and $\Delta\phi=(\phi_1-\phi_2)$ are the
differences in pseudorapidity and azimuthal angle between the two particles, respectively.
The final $C(\Delta\eta,\Delta\phi)$ is found by averaging over events.

The signal distribution:

\vspace{-0.4cm}
\begin{equation}
\label{2pcorr_signal}
S_{N}(\Delta\eta,\Delta\phi)=\frac{1}{N^{\rm pair}}
\frac{d^{2}N^{\rm pair}}{d\Delta\eta d\Delta\phi}
\end{equation}

\noindent is calculated by counting all charged-particle pairs ($N^{\rm pair}$) within each event with
charged-particle multiplicity $N$.

The background distribution:

\vspace{-0.4cm}
\begin{equation}
\label{2pcorr_background}
B_{N}(\Delta \eta,\Delta \phi)=\frac{1}{N^{\rm mixed}}
\frac{d^{2}N^{\rm mixed}}{d\Delta\eta d\Delta\phi}
\end{equation}

\noindent is calculated by counting number of pairs of particles from two different randomly-selected events ($N^{\rm mixed}$) with the same $N$.

We study correlations of stable-charged particles in MC models (\pythia~8.306, \epos~3.4, \urqmd~3.4, and \smashm~2.0.1) 
and compare them to the data provided by NA61/SHINE collaboration~\cite{Aduszkiewicz:2016mww}. 
The study is done at \cmse~=~6.3, 12.3, and 17.3~GeV. 
The data were collected with a minimum-bias trigger. Simulated collisions were generated with minimum-bias configurations as well. 
The data presented by NA61/SHINE experiment were not corrected for blind sports of NA61/SHINE acceptance in the studied momentum and rapidity ranges, instead the collaboration
provided the corresponding systematic uncertainty of 3\% -- 7\%. Moreover, the experiment measured 
particles with \pt~$<$~1.5~GeV.

Fig.~\ref{Plot:Corr2d:6GeV} shows two-particle correlation functions for stable charged particles
with \pt~$<$~1.5~GeV for the data  and studied MC models  at \cmse~=~6.3~GeV. 
Both data and all studied MC models show a convex structure at $\Delta\phi = \pi$ (away-side region) and $\Delta\eta = 0$. 
However, all MC models also show a distinct increase of the amount of correlations at $\Delta\phi = \pi$ 
and $\Delta\eta > 3$, while this structure can not be observed in data due to the limited rapidity range 
of NA61/SHINE experiment available for this measurement. 
The away-side increase at $\Delta\eta > 3$ is mostly associated with the presence of at least one proton in the collision
for all studied MC models.  It is worth noting  that data show ridge-like structure along $\Delta\eta = 0$ that is not reproduced by all MC models.
Fig.~\ref{Plot:Corr2d:6GeV} shows  that \smashm 2.0.1, \urqmd 3.4 miss the qualitative
description of data for the hill at $\Delta\phi = \pi$ and $\Delta\eta = 0$ at \cmse~=~6.3~GeV, producing a double-bump structure. 
Nevertheless,  \epos 3.4 and \pythia (both) produce single-bump away-side correlation.

Fig.~\ref{Plot:Corr2d:12GeV} shows two-particle correlation functions for 
stable charged particles with \pt~$<$~1.5~GeV for the data and studied MC models at \cmse~=~12.3~GeV. 
The key change with respect to \cmse~=~6.3~GeV is that data, \epos 3.4 and \urqmd 3.4 show ridge-like
structure at $\Delta\eta = 0$, which means that at least three particles are produced in the string fragmentation.
This is expected as strings may have higher mass at higher \cmse.

Fig.~\ref{Plot:Corr2d:17GeV} shows two-particle correlation functions for 
stable charged particles with \pt~$<$~1.5~GeV for studied MC models
(\epos 3.4, \pythia 8.306, \urqmd 3.4, \smashm 2.0.1) at \cmse~=~17.3~GeV. 
This collision energy expectedly moves farther the increase of ridge-like structure at $\Delta\eta = 0$
and the away-side high-$\Delta\eta$ structure is out of $\eta$-range of the plot. It is worth mentioning that
MC models do not produce any convex structure around $\Delta\phi = 0$ and $\Delta\eta = 0$ that means absence of hadron jet production at this energy.

\begin{figure}[hbtp]

\begin{minipage}[h]{0.45\linewidth}
\center{data (NA61/SHINE)}
\center{\includegraphics[width=1\linewidth]{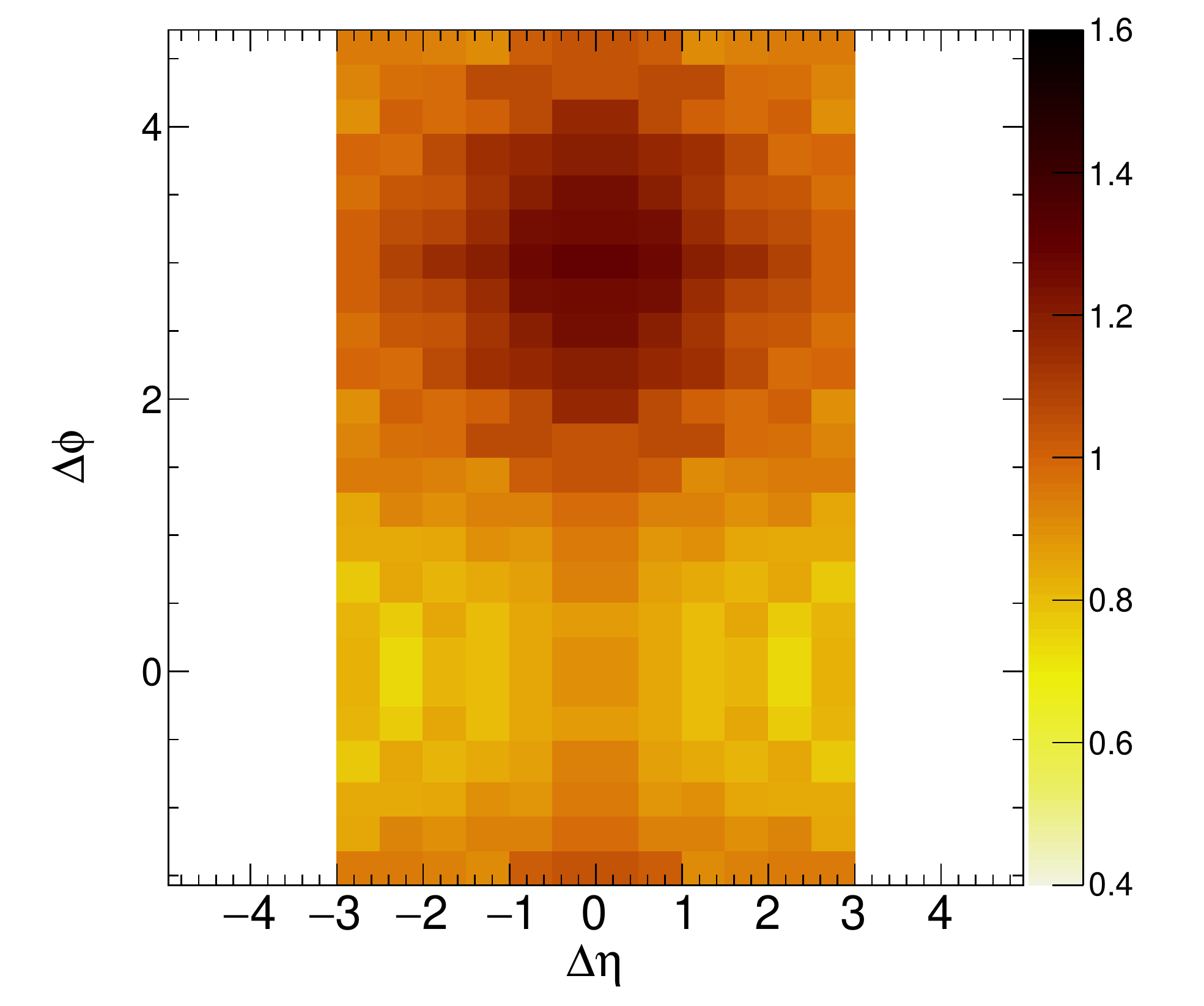} \\ }
\end{minipage}
\begin{minipage}[h]{0.45\linewidth}
\center{\epos 3.4}
\center{\includegraphics[width=1\linewidth]{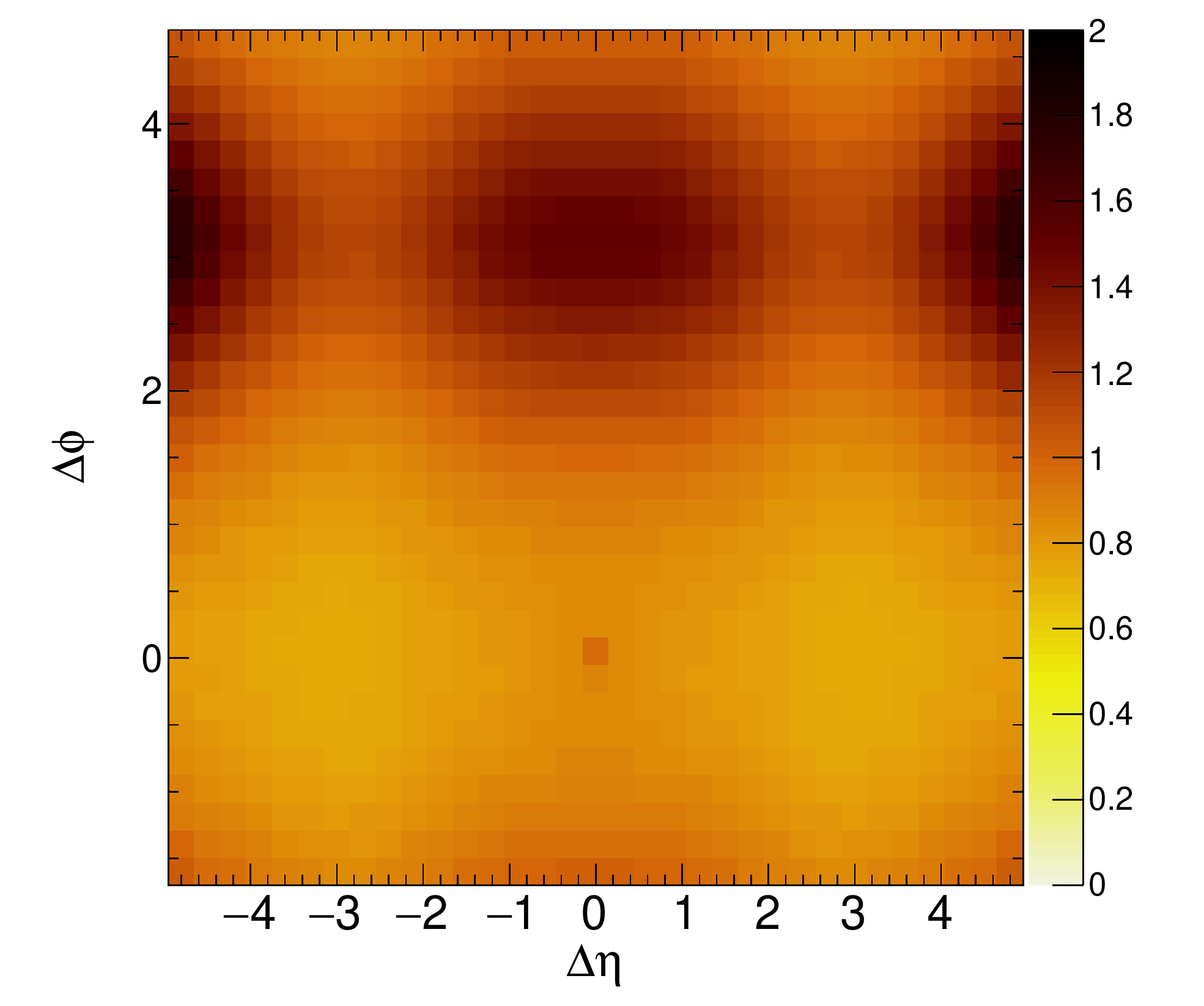} \\ }
\end{minipage}

\begin{minipage}[h]{0.45\linewidth}
\vspace{+0.3cm}
\center{\pythia 8.306}
\center{\includegraphics[width=1\linewidth]{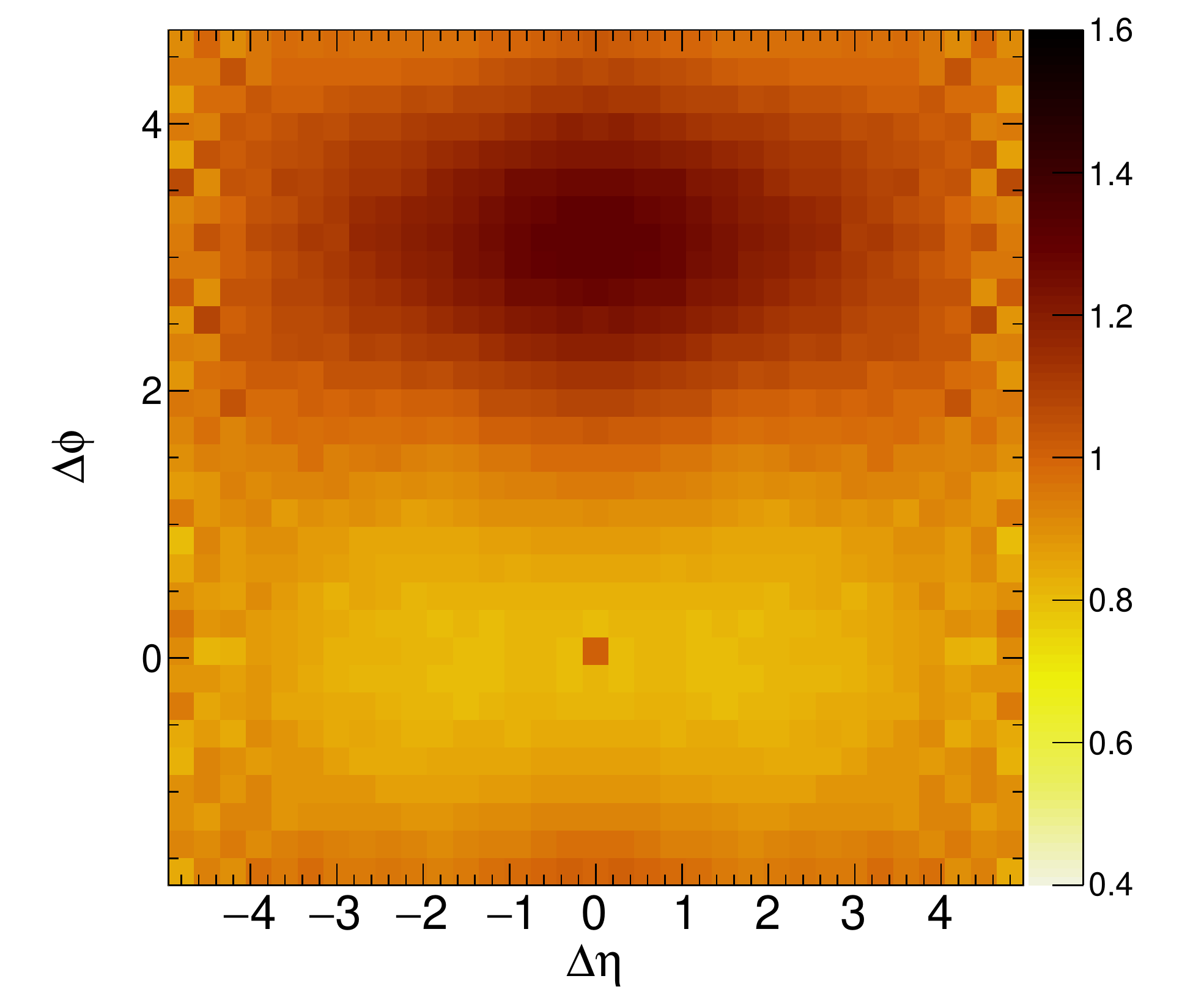} \\ }
\end{minipage}
\begin{minipage}[h]{0.45\linewidth}
\vspace{+0.3cm}
\center{\pythia 8.306 LE1}
\center{\includegraphics[width=1\linewidth]{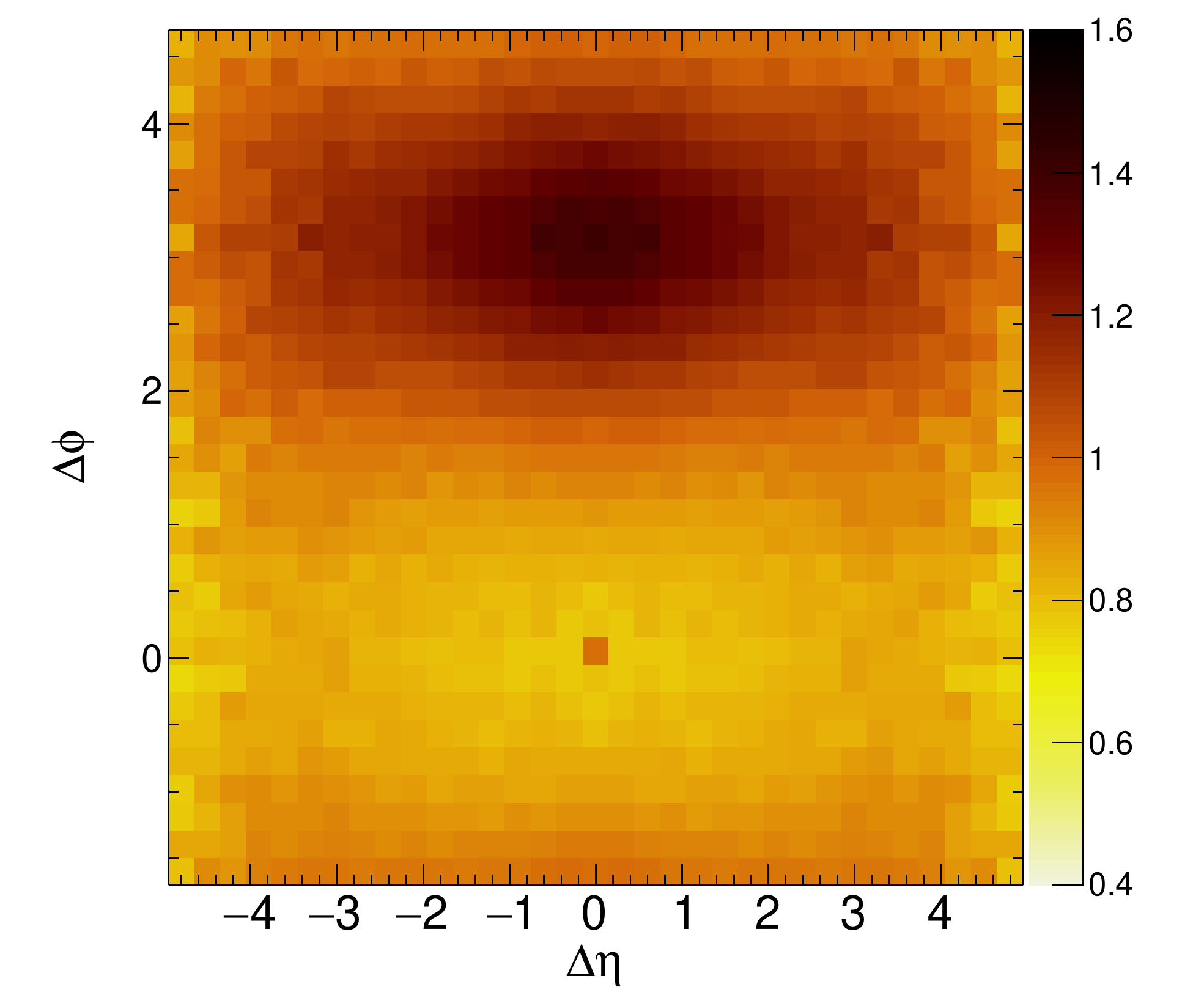} \\ }
\end{minipage}

\begin{minipage}[h]{0.45\linewidth}
\vspace{+0.3cm}
\center{\smashm 2.0.1}
\center{\includegraphics[width=1\linewidth]{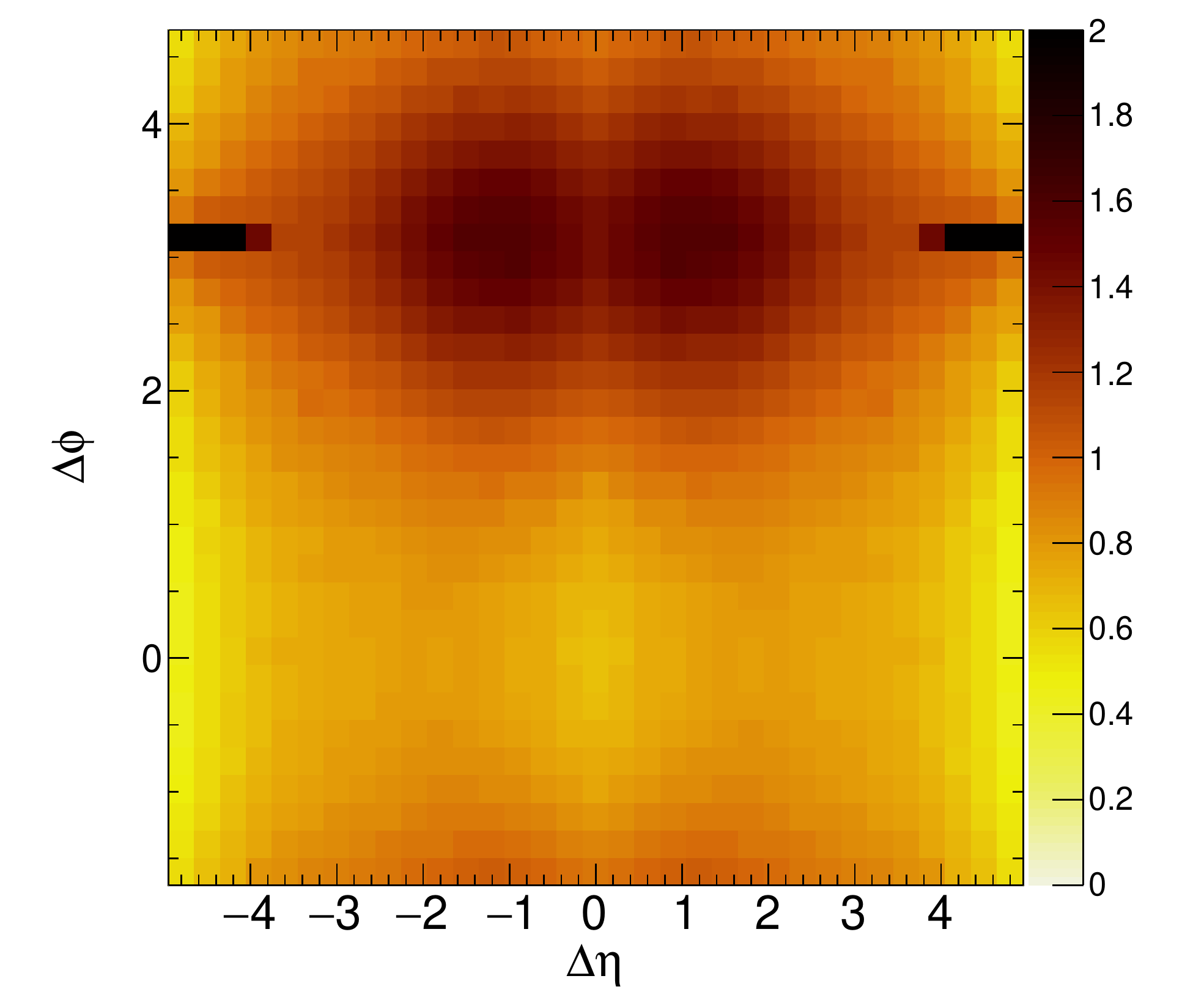} \\ }
\end{minipage}
\begin{minipage}[h]{0.45\linewidth}
\vspace{+0.3cm}
\center{\urqmd 3.4}
\center{\includegraphics[width=1\linewidth]{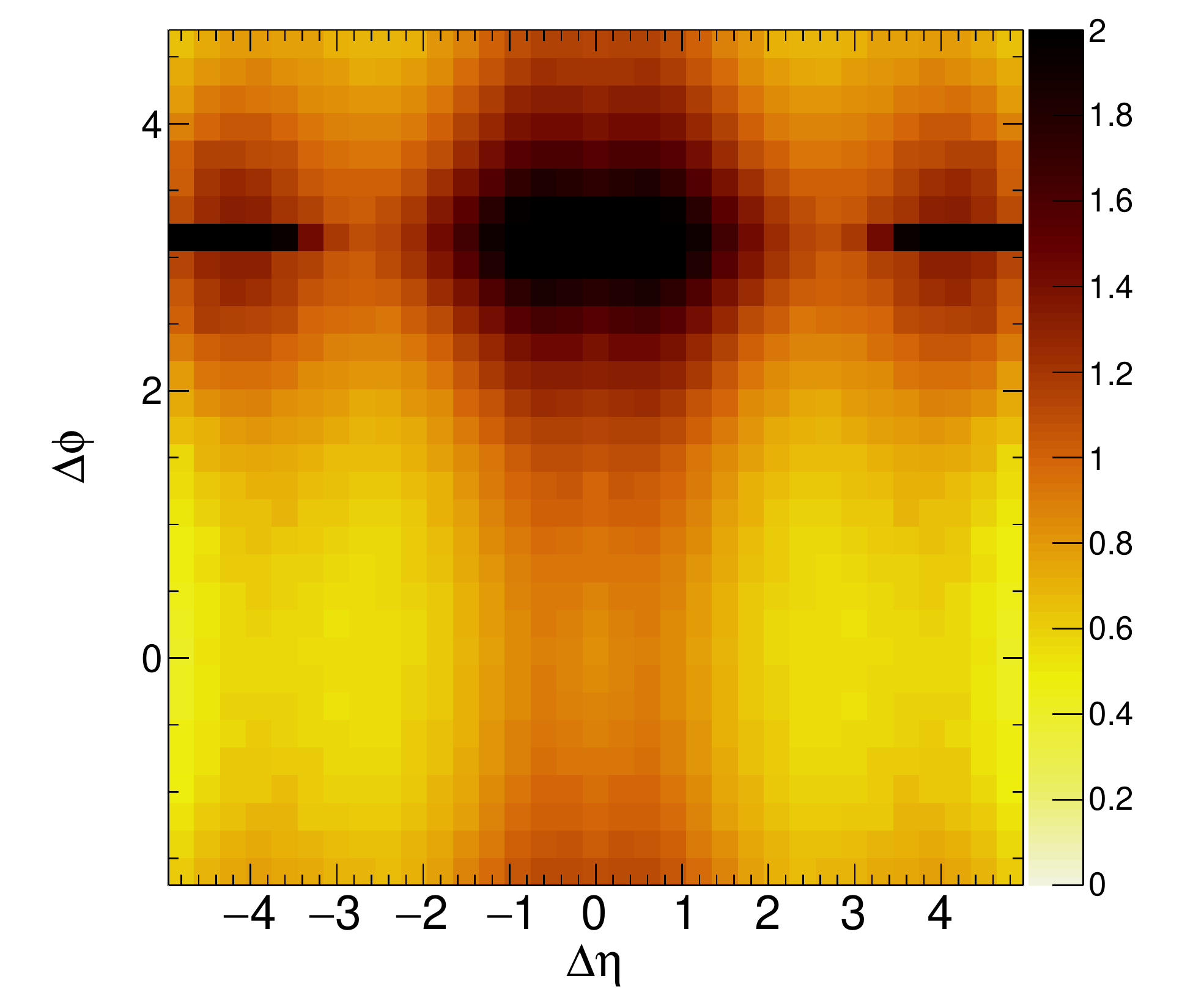} \\ }
\end{minipage}

\caption{ Two-particle correlation functions $C(\Delta\eta,\Delta\phi)$ for different MC models and data at \cmse~=~6.3~GeV. Both particles are required to have \pt~$<$~1.5~GeV. Values of $C(\Delta\eta,\Delta\phi)$ are  color-coded.}
\label{Plot:Corr2d:6GeV}
\end{figure}

\begin{figure}[hbtp]

\begin{minipage}[h]{0.45\linewidth}
\center{data (NA61/SHINE)}
\center{\includegraphics[width=1\linewidth]{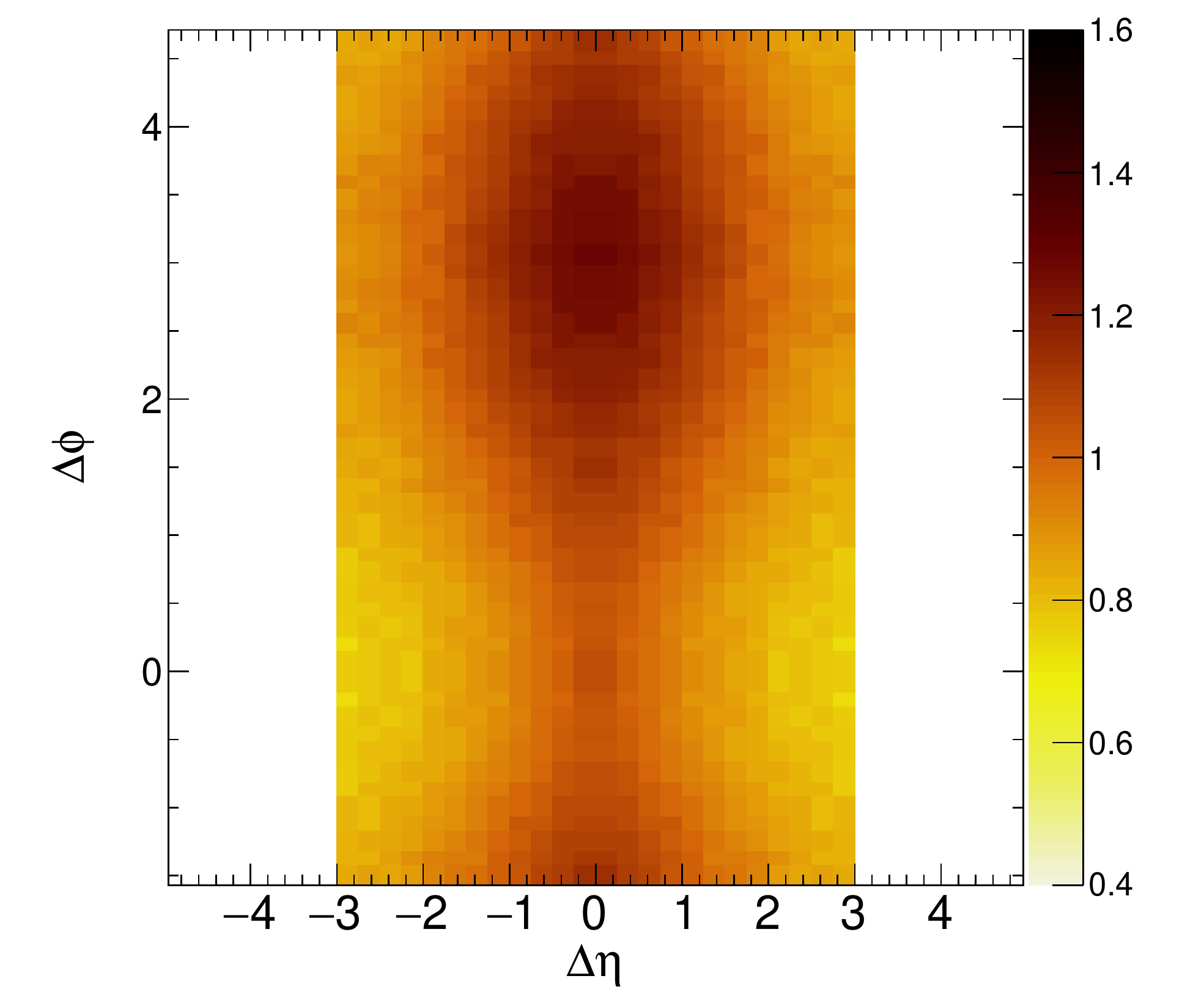} \\ }
\end{minipage}
\begin{minipage}[h]{0.45\linewidth}
\center{\epos 3.4}
\center{\includegraphics[width=1\linewidth]{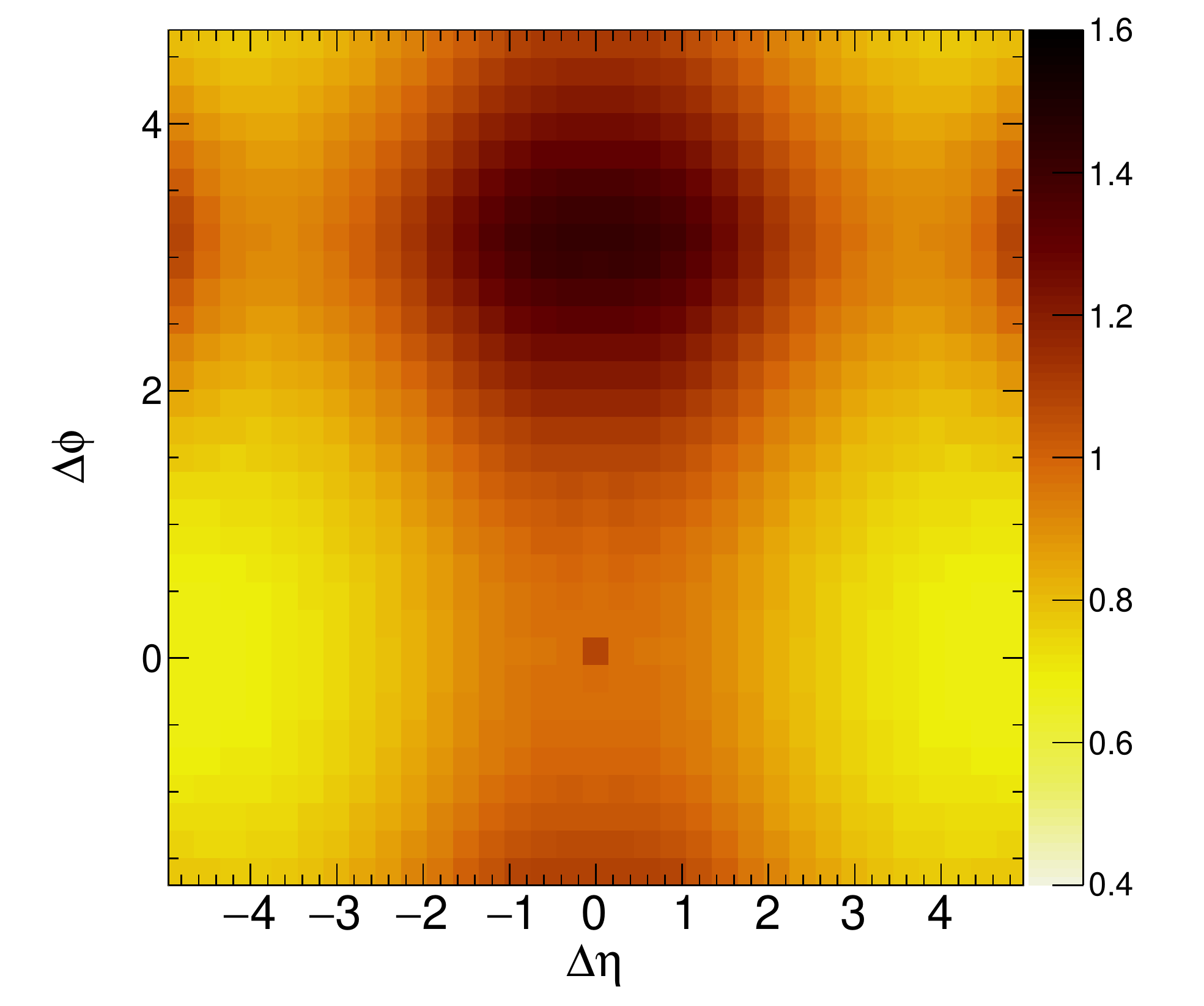} \\ }
\end{minipage}

\begin{minipage}[h]{0.45\linewidth}
\vspace{+0.3cm}
\center{\pythia 8.306}
\center{\includegraphics[width=1\linewidth]{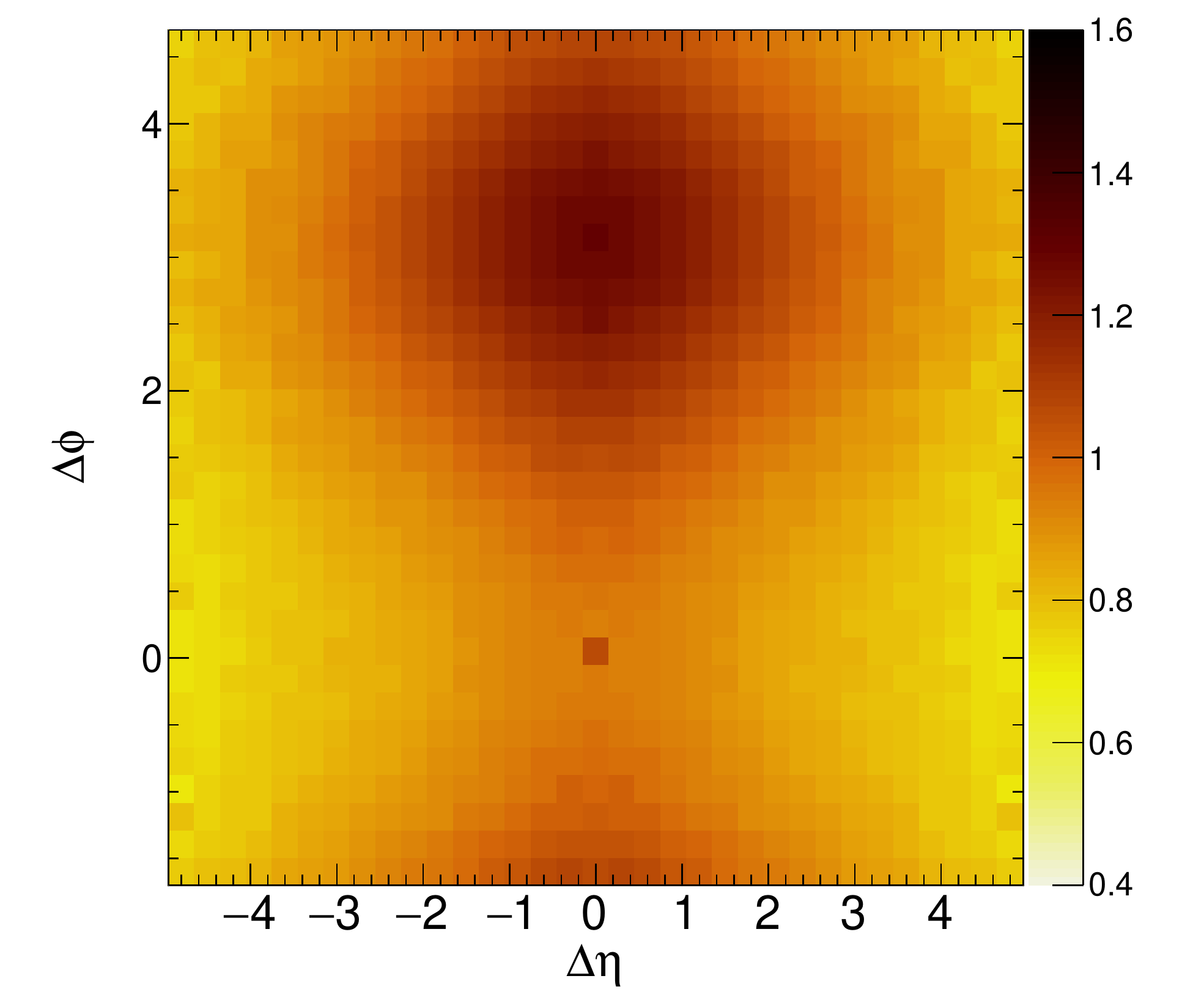} \\ }
\end{minipage}
\begin{minipage}[h]{0.45\linewidth}
\vspace{+0.3cm}
\center{\pythia 8.306 LE1}
\center{\includegraphics[width=1\linewidth]{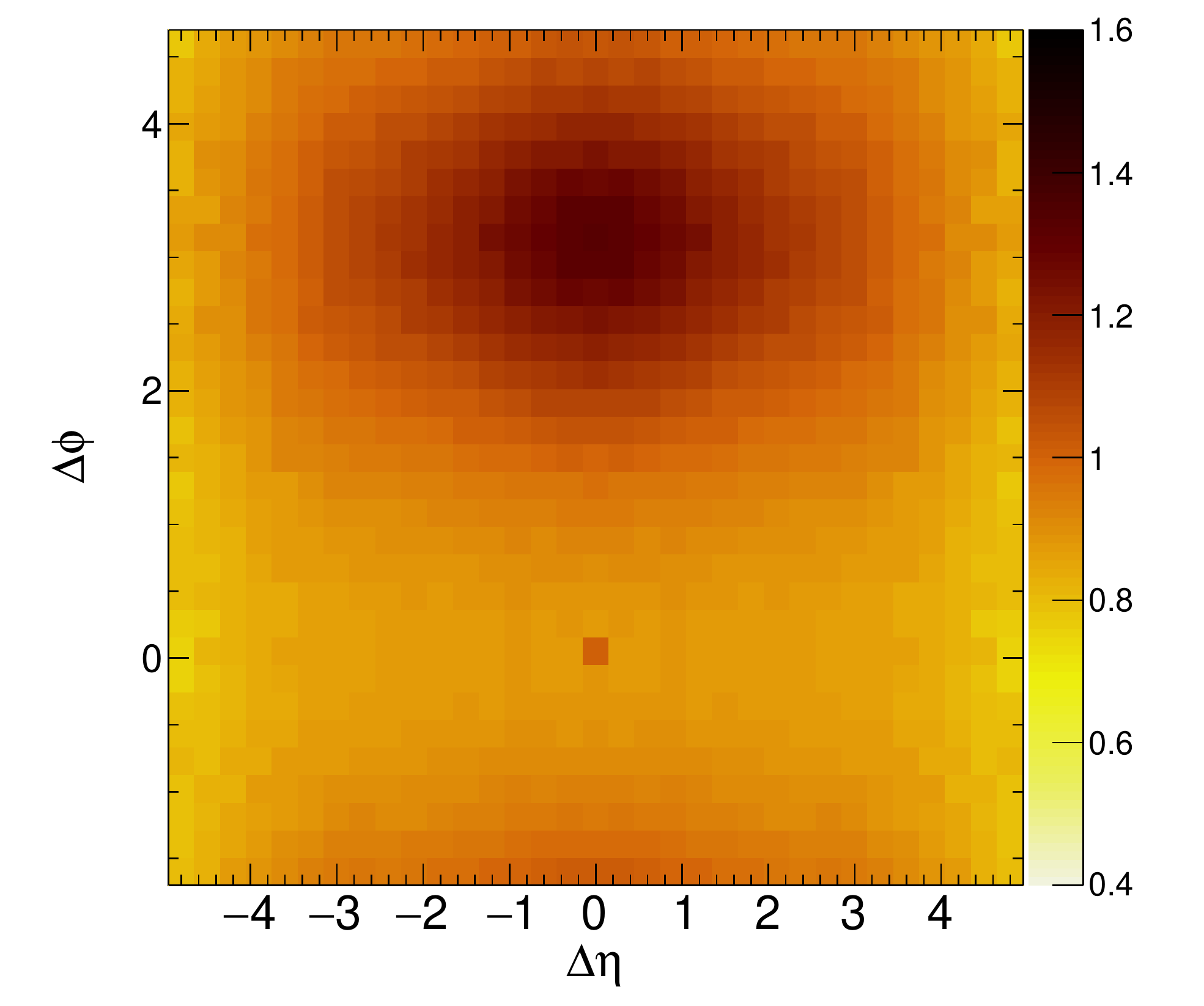} \\ }
\end{minipage}

\begin{minipage}[h]{0.45\linewidth}
\vspace{+0.3cm}
\center{\smashm 2.0.1}
\center{\includegraphics[width=1\linewidth]{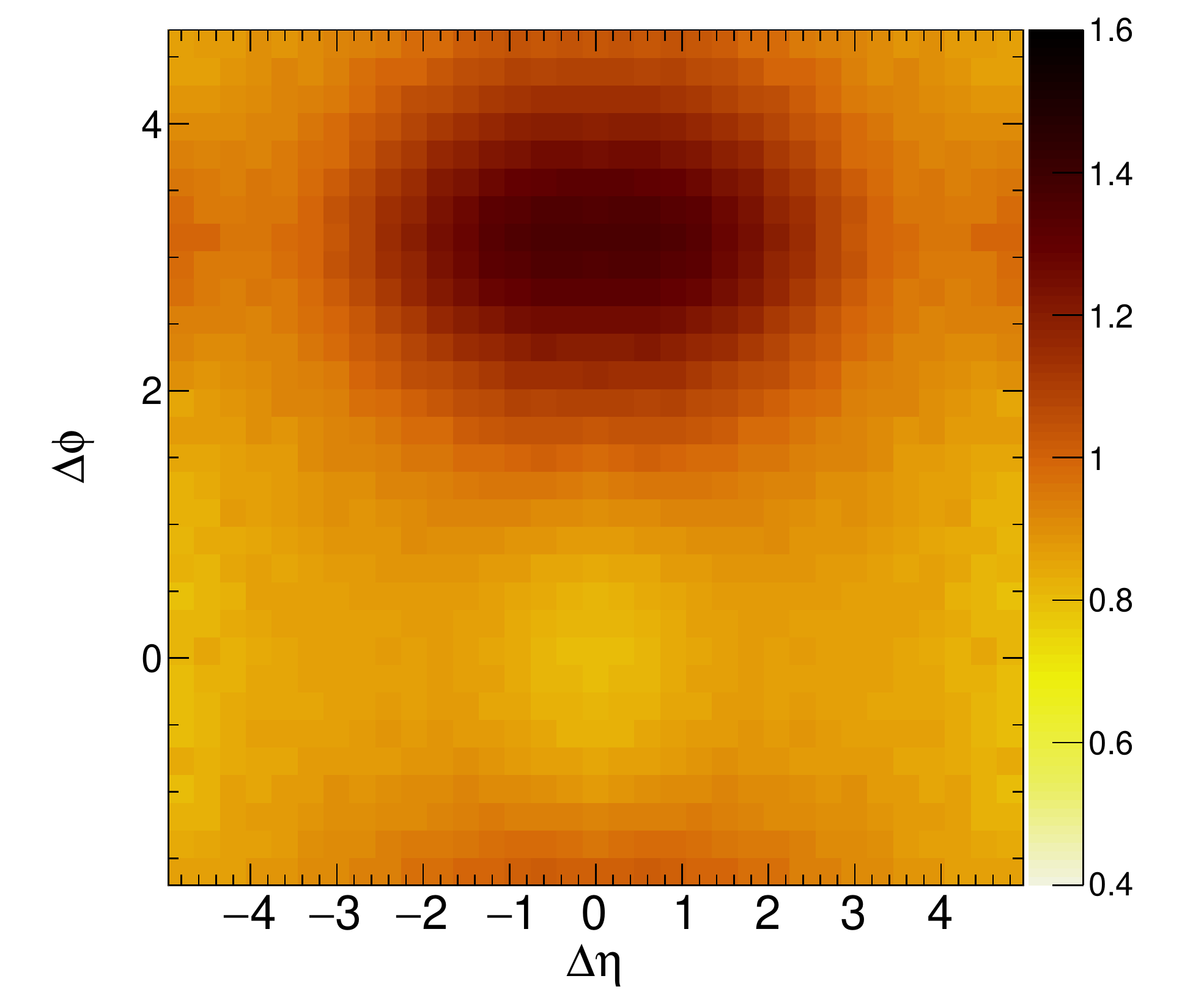} \\ }
\end{minipage}
\begin{minipage}[h]{0.45\linewidth}
\vspace{+0.3cm}
\center{\urqmd 3.4}
\center{\includegraphics[width=1\linewidth]{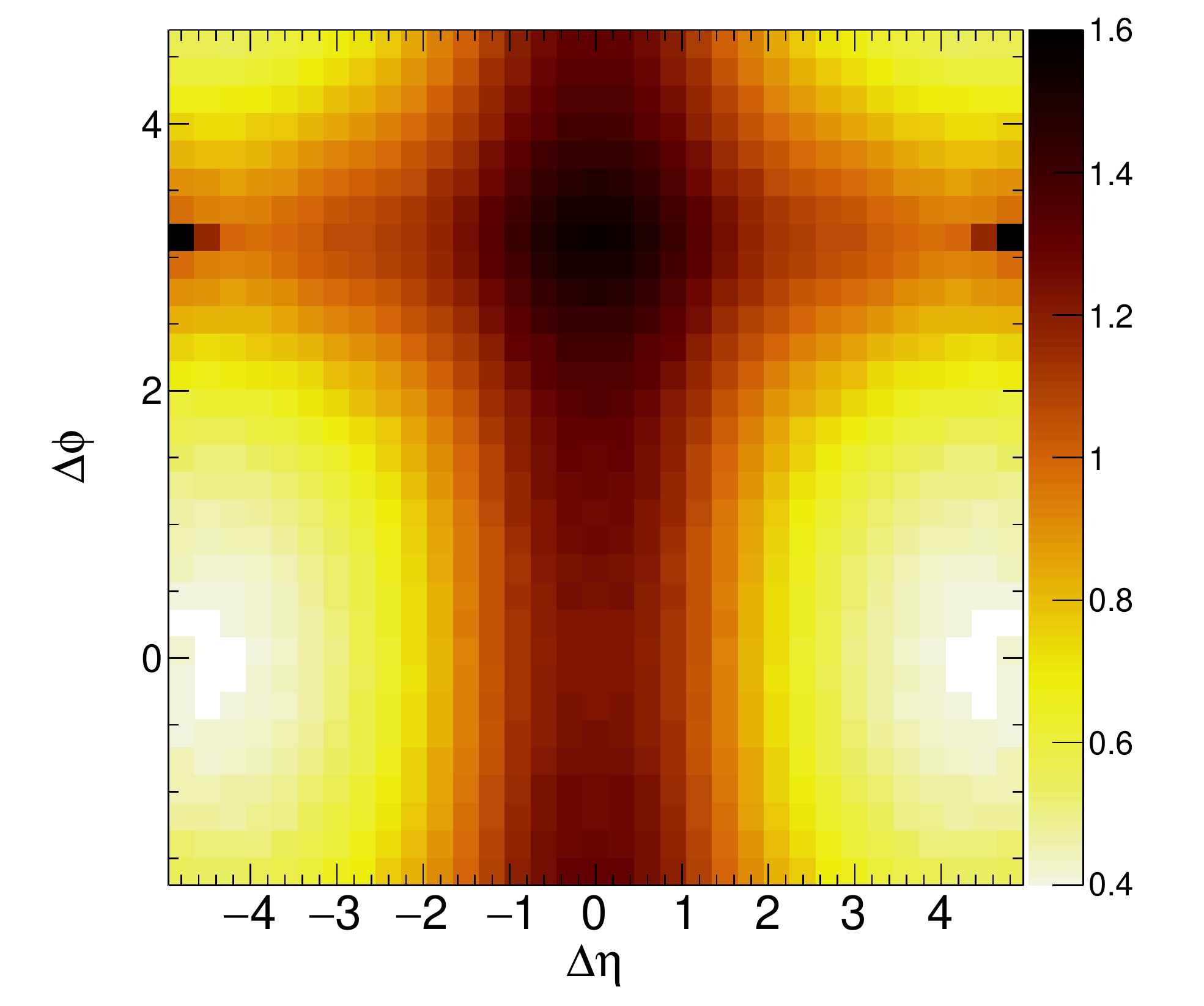} \\ }
\end{minipage}

\caption{ Two-particle correlation functions $C(\Delta\eta,\Delta\phi)$ for different MC models and data at \cmse~=~12.3~GeV. Both particles are required to have \pt~$<$~1.5~GeV. Values of $C(\Delta\eta,\Delta\phi)$ are color-coded.}
\label{Plot:Corr2d:12GeV}
\end{figure}

\begin{figure}[hbtp]

\begin{minipage}[h]{0.45\linewidth}
\center{data (NA61/SHINE)}
\center{\includegraphics[width=1\linewidth]{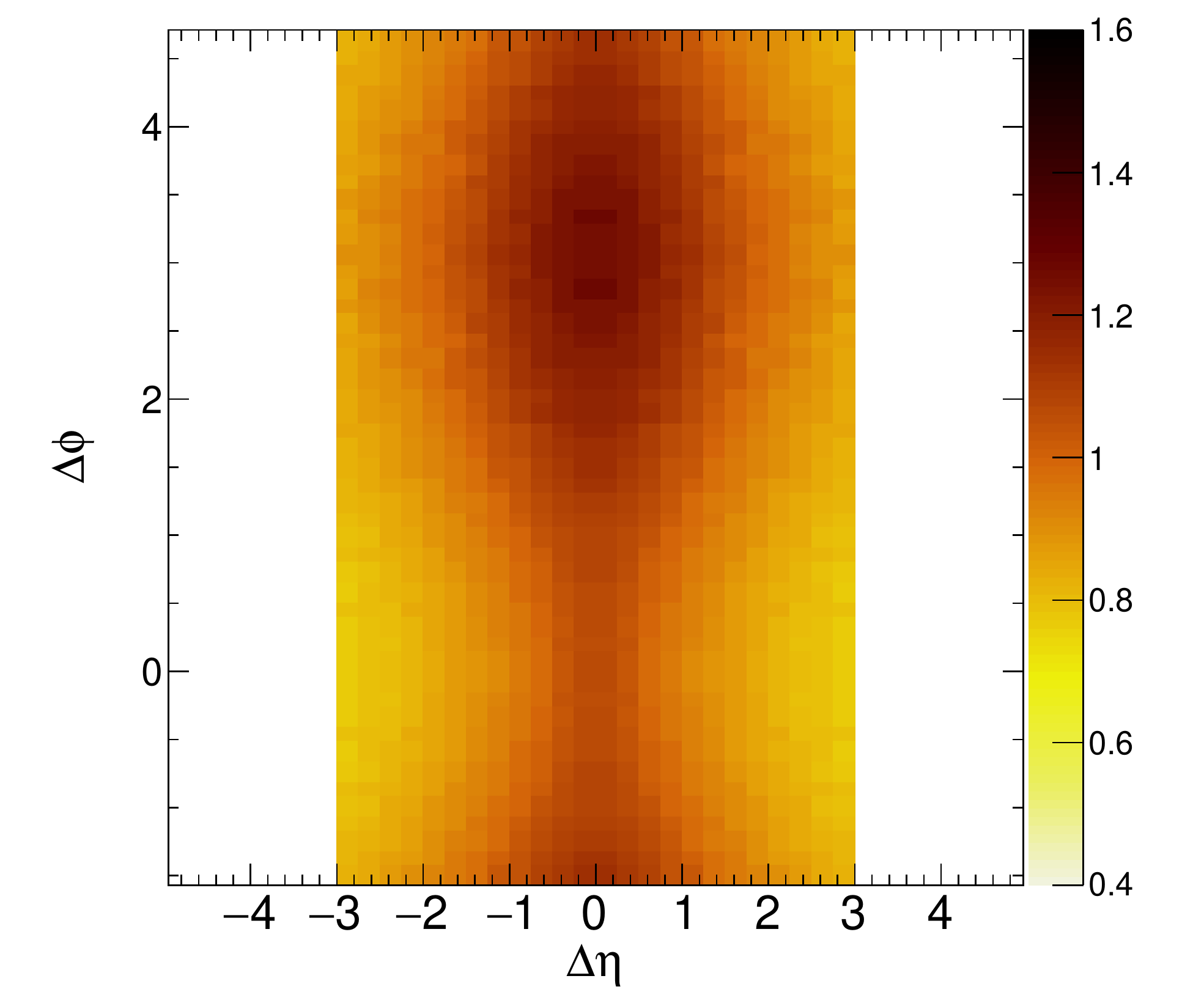} \\ }
\end{minipage}
\begin{minipage}[h]{0.45\linewidth}
\center{\epos 3.4}
\center{\includegraphics[width=1\linewidth]{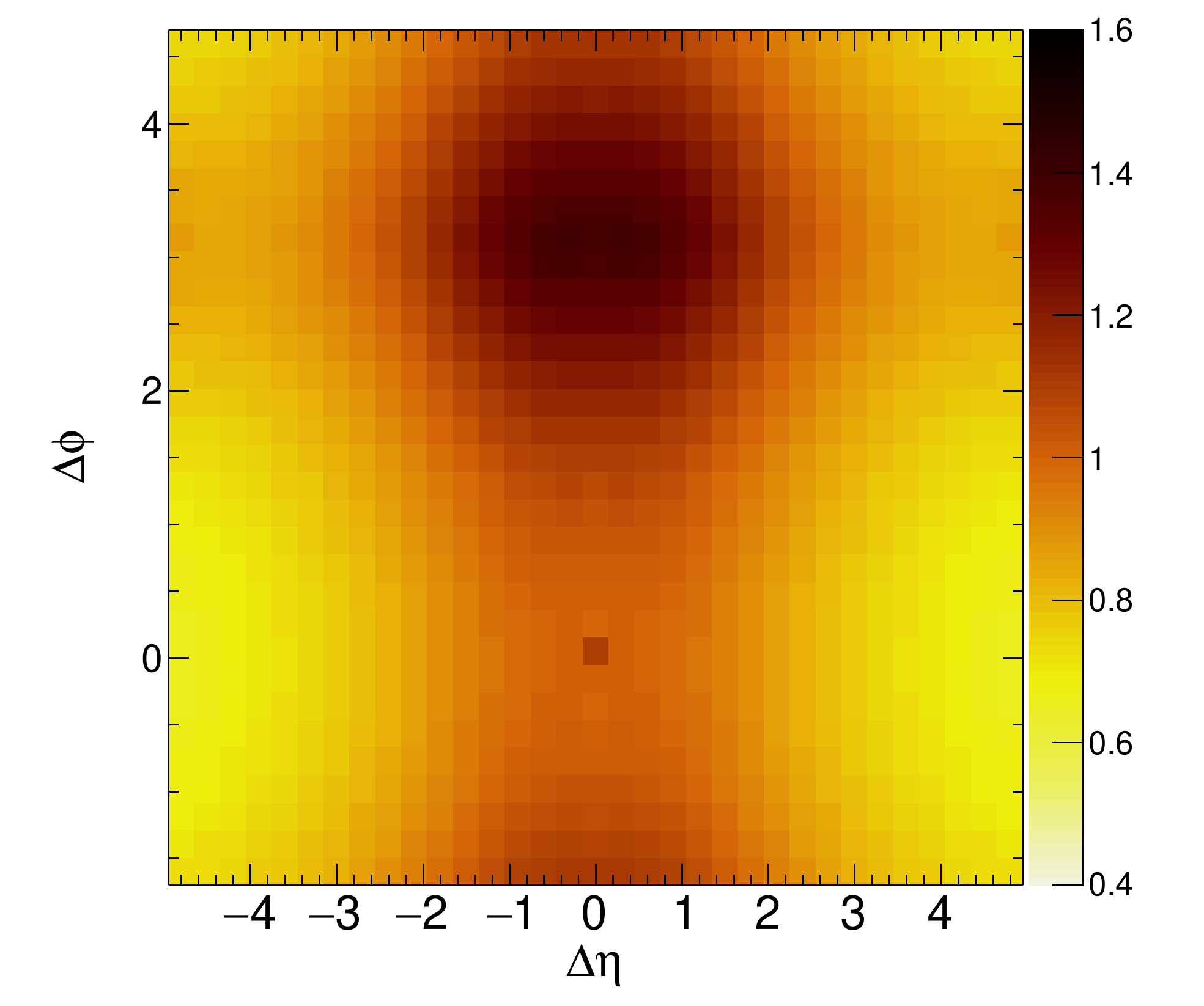} \\ }
\end{minipage}

\begin{minipage}[h]{0.45\linewidth}
\vspace{+0.3cm}
\center{\pythia 8.306}
\center{\includegraphics[width=1\linewidth]{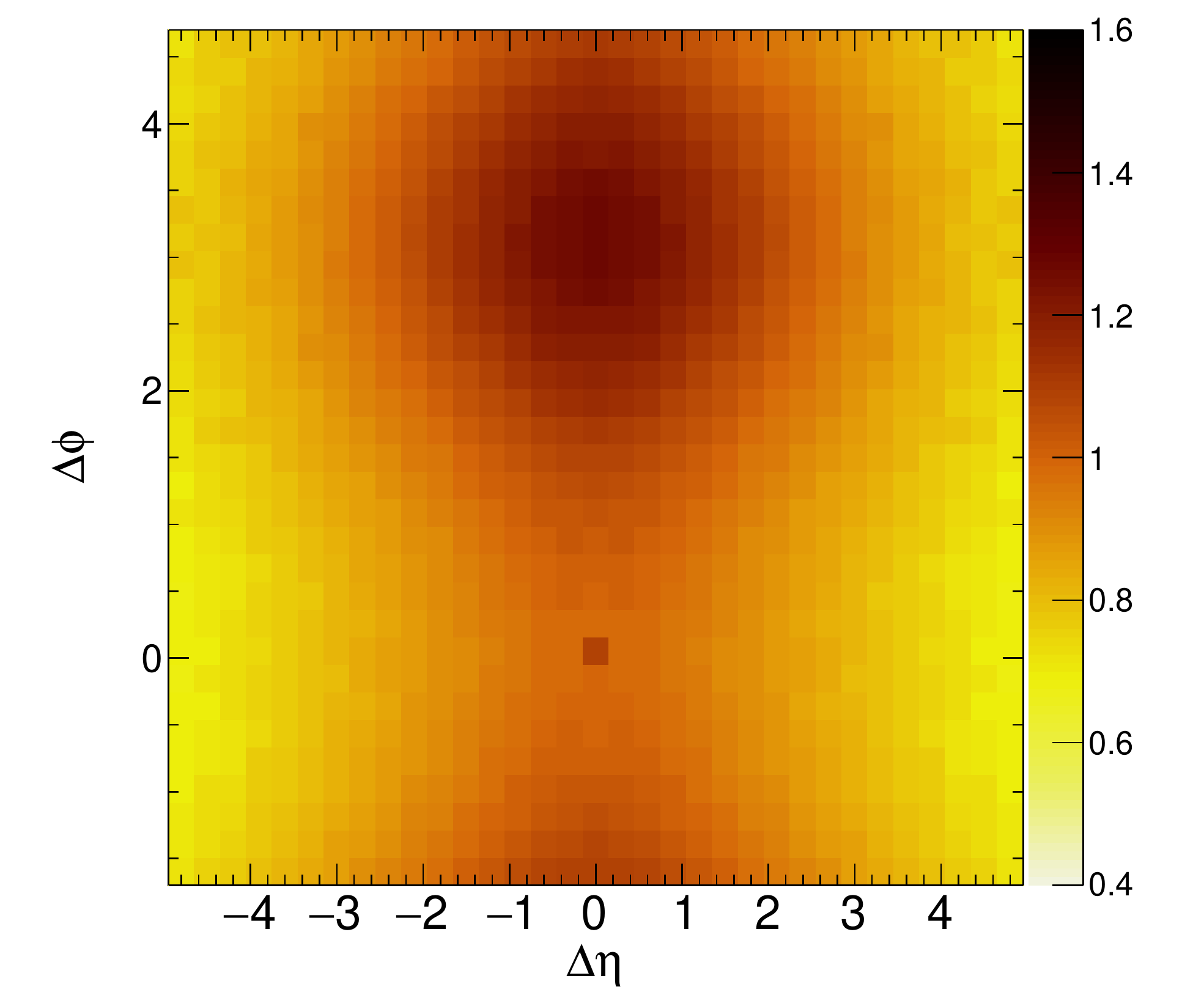} \\ }
\end{minipage}
\begin{minipage}[h]{0.45\linewidth}
\vspace{+0.3cm}
\center{\pythia 8.306 LE1 }
\center{\includegraphics[width=1\linewidth]{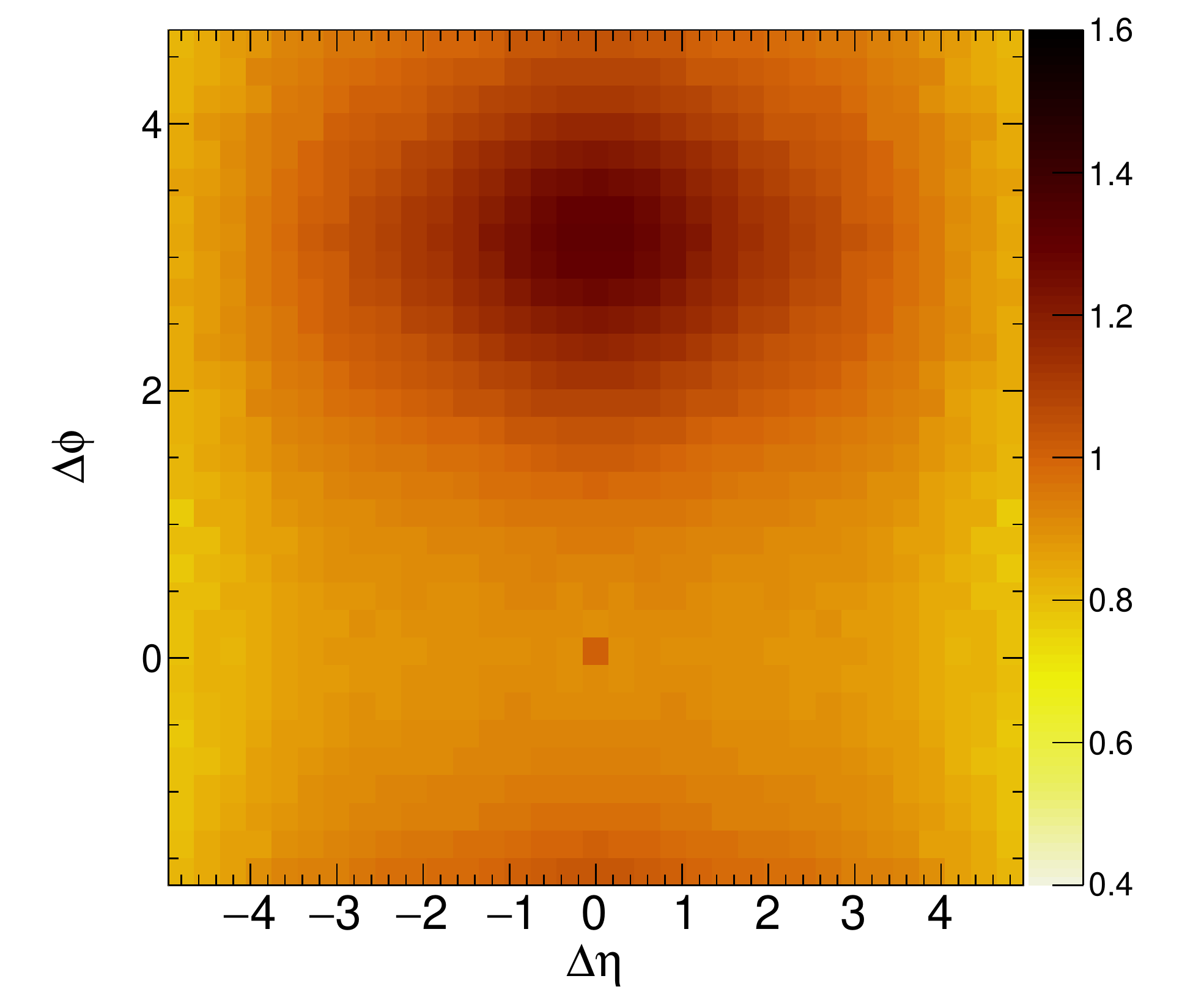} \\ }
\end{minipage}

\begin{minipage}[h]{0.45\linewidth}
\vspace{+0.3cm}
\center{\smashm 2.0.1 }
\center{\includegraphics[width=1\linewidth]{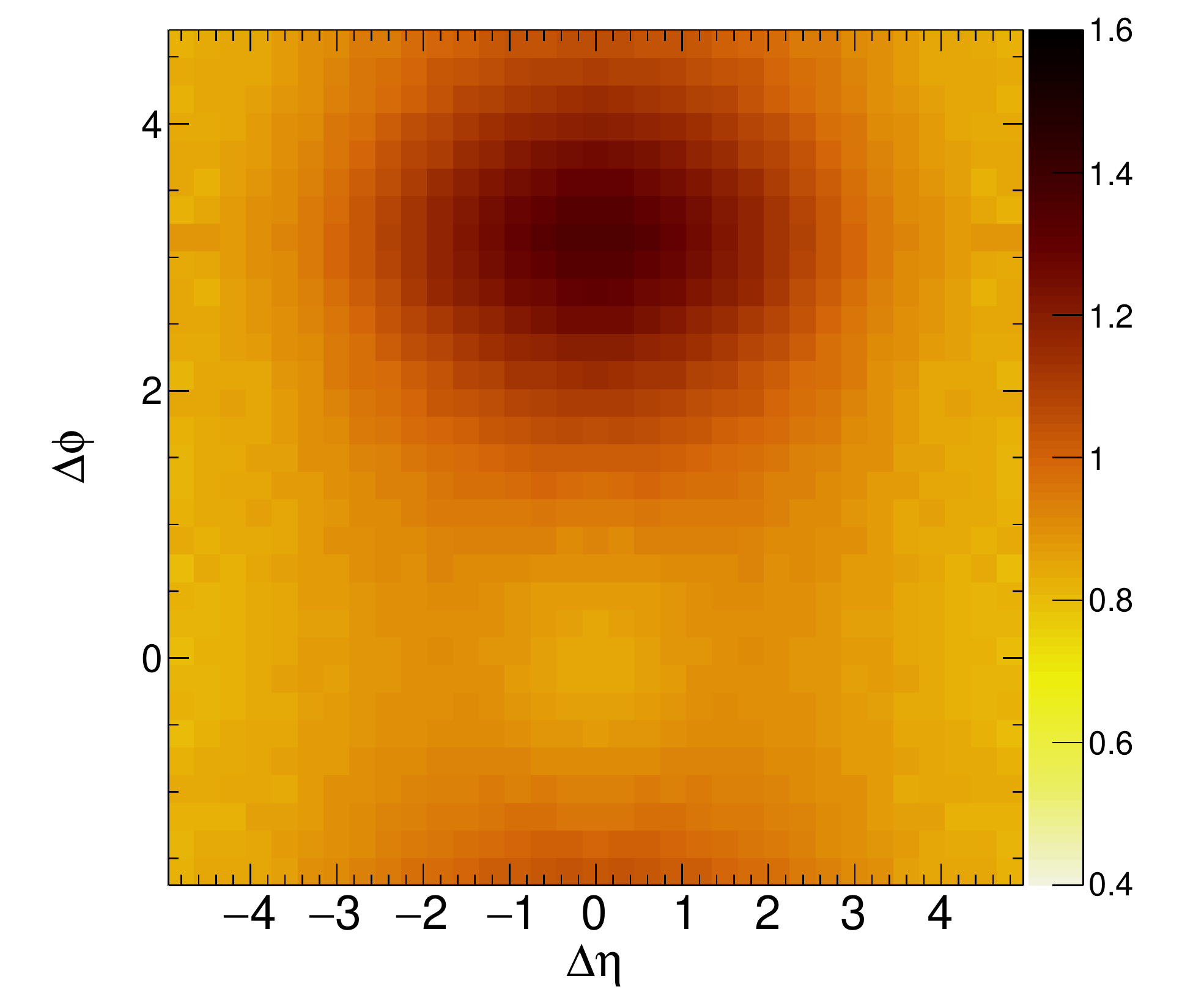} \\ }
\end{minipage}
\begin{minipage}[h]{0.45\linewidth}
\vspace{+0.3cm}
\center{\urqmd 3.4}
\center{\includegraphics[width=1\linewidth]{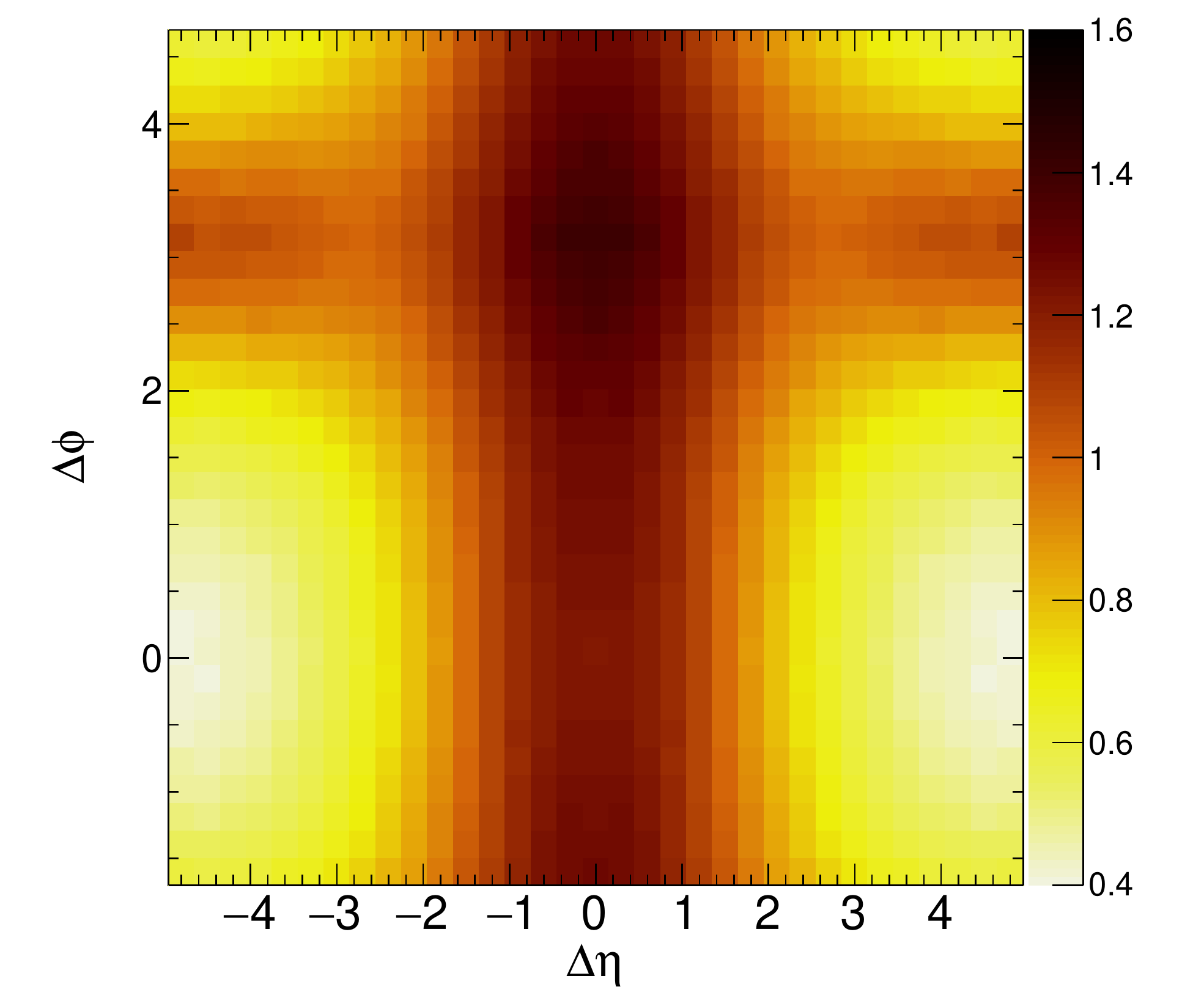} \\ }
\end{minipage}

\caption{ Two-particle correlation functions $C(\Delta\eta,\Delta\phi)$ for different MC models and data at \cmse~=~17.3~GeV. Both particles are required to have \pt~$<$~1.5~GeV. Values of $C(\Delta\eta,\Delta\phi)$ are color-coded.}
\label{Plot:Corr2d:17GeV}
\end{figure}

\FloatBarrier

All in all, no generator can quantitatively describe the data. Some generators fail to describe key correlation regions qualitatively, 
namely and \pythia 8.306, \smashm 2.0.1 can not reproduce the ridge-like structure at $\Delta\eta = 0$ at these collision energies.
Moreover, \urqmd 3.4 and \smashm~2.0.1 produces unnaturally sharp correlations at $\Delta\eta > 3$  at $\Delta\phi = \pi$, 
while \epos 3.4, \pythia 8.306 exhibit a small spike at $\Delta\phi = 0$ and $\Delta\eta = 0$.
As it can be seen tuning of the parameters of \pythia that helped to improve rapidity distributions of p($\bar{\rm p}$), 
$\pi^{\pm}$, $\mathrm{K}^{\pm}$ does not help much (if any) to improve two-particle correlations. 

 Unfortunately, uncertainty (up to 7\% ) of the available experimental data makes 
it hard to use them for quantitative comparison and subsequent tuning. $4 \pi$ geometry and up-to-date apparatus of an MPD detector
at NICA will help to improve data on two-particle correlations at low energies.


\section{Summary and conclusions}
\label{Sec:Summary}

In this paper we have done a pilot review of performance of various MC event generators at NICA energies. 
Namely, we have simulated proton-proton collisions with \epos 3.4,
\pythia 8.306, \urqmd 3.4, \smashm 2.0.1 models and measured mean multiplicities,
mean transverse momenta, and rapidity distributions of $\pi^{\pm}$,  $\mathrm{K}^{\pm}$, p($\bar{\rm p}$) 
and two-particle angular correlations of stable-charged particles in pp collisions with \cmse from $\approx$~5 up to 30~GeV. 
We also suggest a tuned set of parameters for \pythia 8.306, which was obtained at \cmse~=~17.3~GeV. This versions is denoted 
as LE1 that stands for Low Energy 1.  Results of MC simulations are compared to the available data. 

These studies show that  \epos 3.4 decently describes mean multiplicities $\pi^{\pm}$,  $\mathrm{K}^{\pm}$, p($\bar{\rm p}$) 
in the data for studied collision energies.  The standard \pythia 8.306 overestimates the particle multiplicities,
while the tuned version \pythia 8.306 LE1 demonstrates much better description of the data, nearly reaching agreement with the data within uncertainties at \cmse~=~17.3~GeV.
\urqmd 3.4, \smashm 2.0.1 tend to underestimate multiplicities for all the studied meson and antiprotons,
 while overestimating the proton production.

The data of rapidity distributions are best described by \epos 3.4 for all studied particles. 
The standard version of \pythia 8.306 shows higher particle densities over the studied rapidity range, the excess being more prominent at central rapidities. \pythia 8.306 LE1 demonstrates slightly broader rapidity distributions for all studied mesons and antiproton compared to the data. 
The rapidity distribution of protons is still slightly narrower than in the data, but it is a way better than 
for the standard version.
\urqmd 3.4 and \smashm 2.0.1 underestimate particle density of $\pi^{\pm}$ and $\mathrm{K}^{\pm}$ over entire studied rapidity range. 
Production of p($\bar{\rm p}$) is marginally described by \epos 3.4 and \smashm 2.0.1. We also compared results of simulation with
these 5 MC event generators with the available data on multiplicity of  $K^{0}_{\rm S}$, $K^{*}(892)$, $\bar{K}^{*}(892)$, $\phi$,  $\Lambda$,
$\Xi^{-}$, $\overline{\Xi}{}^{-}$  at \cmse~=~17.3~GeV and found that data-MC difference are similar to ones observed for
$\pi^{\pm}$, $\mathrm{K}^{\pm}$, p($\bar{\rm p}$), i.e. the differences are kind of universal for mesons and baryons, respectively.
Also we plotted mean transverse momentum of $\pi^{\pm}$,  $\mathrm{K}^{\pm}$, p($\bar{\rm p}$) for different MC generators.
The spread of results is up 25\%

Two-particle angular correlations revealed that studied MC models show some specific features for each generator and do not
describe the data. Moreover, \epos 3.4, \pythia 8.306, \pythia 8.306 LE1, \urqmd 3.4 exhibit unnaturally sharp structures. 

All-in-all, current understanding of proton-proton collision at this energy range is at best incomplete, and MC models need substantial improvements that poses urgent tasks for the generator developers
in the view of rapidly approaching operation of NICA. 
In this study we suggest a set of tuned parameters for \pythia 8.306 that significantly improves a description 
of the data at \cmse~=~17.3~GeV. This set shows some improvement of description of the data in the entire studied range
of collision energy as well.
Certainly, it is an incomplete tune since it is based on a limited number of observables, and we did not find a single 
solution to address all the observed data-MC discrepancies. 
For instance, we can not make the MC generator fully describe a rapidity distribution of proton without 
distorting the distribution for other particles. 
It is even more complicated and ambitious goal to achieve a better description of two particle correlation. 

It should be noted that there are a lot of studies currently ongoing on this topic. In particular, there are developments
under way in \pythia on implementing hadronic rescattering and low-energy hadron interactions~\cite{Sjostrand:2020gyg} 
which are relevant in the context of NICA studies for both pp and heavy-ion programs. Generator \smashm is being actively
developed. Team in Dubna is developing its cascade model \cite{DCM}.
 
The MPD experiment at NICA  has nearly 4$\pi$ acceptance and will run with up-to-date instrumentation and reconstruction techniques 
thus allowing to improve accuracy of many measured quantities at this energy range and giving the possibility of a much more detailed study
of proton collisions at these energies than ever before.

\section*{Acknowledgments}
This work was supported by the RFBR project 18-02-40131.


\FloatBarrier
\bibliography{GeneratorReview}{}
\bibliographystyle{lucas_unsrt_epjc}
\end{document}